\documentstyle[aps,psfig,prl,twocolumn]{revtex}

\begin{document}
\wideabs{
\title{Thermodynamically consistent mesoscopic fluid particle models for
a van der Waals fluid}
\author{Mar Serrano and Pep Espa\~nol}
\address{Departamento de F\'{\i}sica Fundamental, UNED,
Apartado 60141, 28080 Madrid, Spain}
\date{\today}
\maketitle
\begin{abstract}
The GENERIC structure allows for a unified treatment of different
discrete models of hydrodynamics. We first propose a finite volume
Lagrangian discretization of the continuum equations of hydrodynamics
through the Voronoi tessellation. We then show that a slight
modification of these discrete equations has the GENERIC structure.
The GENERIC structure ensures thermodynamic consistency and allows for
the introduction of correct thermal noise. In this way, we obtain a
consistent discrete model for Lagrangian fluctuating hydrodynamics.
For completeness, we also present the GENERIC versions of the Smoothed
Particle Dynamics model and of the Dissipative Particle Dynamics
model. The thermodynamic consistency endorsed by the GENERIC framework
allows for a coherent discussion of the gas-liquid phase coexistence
of a van der Waals fluid.
\end{abstract}
}
\section{Introduction} 

The behaviour of complex fluids like colloids, emulsions, polymers or
multiphasic fluids is affected by the strong coupling between the
microstructure of these fluids with the macroscopic flow. The
complexity of these systems requires the use of novel computer
simulations techniques and algorithms. Usual macroscopic approaches
that solve partial differential equations with constitutive equations
are not very useful because the basic input, the constitutive equation
is usually not known. Also, these approaches neglect the presence of
thermal noise, which is the responsible for the Brownian motion of
small suspended objects and, therefore, for the diffusive processes
that affect the microstructure of the fluid. In recent years, there has
been a large effort in order to develop mesoscopic techniques in order
to tackle the problems arising in the simulation of
complex fluids.

Dissipative Particle Dynamics is a mesoscopic particle based
simulation method that allows one to model hydrodynamic behavior and
that it consistently includes thermal fluctuations. It was introduced
by Hoogerbrugge and Koelman in 1991 under the motivation of designing
an off-lattice algorithm inspired by the ideas behind the lattice-gas
method \cite{Hoogerbrugge92}. Since then, the model has received a
great deal of attention. From a theoretical point of view, the model
has been given a solid background as a statistical mechanics model
\cite{Espanol95}. The hydrodynamic behavior has been analyzed
\cite{Espanol95,Espanol98} and the methods of kinetic theory have provided
explicit formulae for the transport coefficients in terms of the model
parameters \cite{Marsh97}.  A generalization of DPD has also been
presented in order to conserve energy \cite{Bonet97}. From the side of
applications, the method is very versatile and has proven to be useful
in the simulation of flows in porous media \cite{Koelman93}, colloidal
suspensions \cite{Koelman93,Boek97}, polymer suspensions
\cite{Schlijper95}, microphase separation of copolymers
\cite{Groot97}, multicomponent flows \cite{Coveney96} and thin-film
evolution \cite{Dzwinel99}.

\newpage
The physical picture behind the dissipative particles used in the
model is that they represent mesoscopic portions of real fluid, say
clusters of molecules moving in a coherent and hydrodynamic
fashion. The interaction between these particles are postulated from
simplicity and symmetry principles that ensure the correct
hydrodynamic behavior.  DPD faces, however, a conceptual problem. The
thermodynamic behavior of the model is determined by the conservative
forces introduced in the model. This forces are assumed to be soft
forces in counter distinction to the singular forces of the
Lennard-Jones type used in molecular dynamics. But there is no
well-defined procedure to relate the shape and amplitude of the
conservative forces with a prescribed thermodynamic behavior (although
attempts in that direction have been undertaken, see
Ref. \cite{Groot97}). Also, it is not clear which physical time and
length scales the model actually describes, even though the presence
of thermal noise suggests the foggy area of the mesoscopic realm. We
will see that both problems are closely related.

Dissipative Particle Dynamics is very similar in spirit to the popular
method of Smoothed Particle Dynamics \cite{Espanol98}. The method was
introduced in the context of astrophysics computation in the early
70's \cite{Monaghan92} and very recently it has been applied to the
study of laboratory scale viscous \cite{Takeda94} and thermal flows
\cite{Kum95} in simple geometries. SPD is essentially a Lagrangian
discretization of the Navier-Stokes equations by means of a weight
function. The procedure transforms the partial differential equations
of continuum hydrodynamics into ordinary differential equations. These
equations can be further interpreted as the equations of motion for a
set of particles interacting with prescribed laws of force. The
technique thus allows one to solve PDE's with molecular dynamics
codes. Again, these particles can be understood as physical portions
of the fluid that evolve coherently along the flow. The problem with
Smoothed Particle Hydrodynamics is that it does not include thermal
fluctuations and, therefore, cannot be applied to the study of complex
fluids at mesoscopic scales.

We have recently shown that the conceptual problems in DPD and the
inclusion of thermal fluctuations in SPD can be resolved by
formulating convenient generalizations of both methods under the
general framework of GENERIC \cite{Espanol-prl99}.  In the present
paper, we take a further look at the problem of formulating consistent
models for the simulation of hydrodynamic problems. Our point of view
here is to construct a finite volume algorithm for solving the
Navier-Stokes equations in such a way that thermodynamic consistency
is retained: The resulting algorithm conserves mass, momentum, and
energy, and the entropy is an increasing function of time. Most
important, we show how to include thermal noise in a consistent way,
this is, producing the Einstein distribution function. We end up,
therefore, with an algorithm for simulating fluctuating hydrodynamics
in a Lagrangian way \cite{Eulerian}. This algorithm can be used as the
basis for simulating colloidal suspensions, where the Brownian motion
of the colloidal particles is due to the thermal fluctuations on the
solvent \cite{Bedeaux74}. We show also another potential application
to multiphasic flow of the gas-liquid type.  When the fluid is
described with a van der Waals equation of state, the Einstein
equilibrium distribution allows to discuss the gas-liquid phase
transition in probabilistic terms. Such probabilistic approaches to
equilibrium phase transitions have been studied in the past
\cite{Guemez88}. We should note, however, that the GENERIC finite
volume algorithm should allow us to study fully non-equilibrium
situations created by external boundary conditions that can drive the
system out of equilibrium. Start up of boiling of water in a pot could
be addressed with the proposed GENERIC finite volume algorithm.

Other very recent approaches to the study of the flow of liquid-vapor
coexisting fluids are the Lattice Boltzmann model introduced by Swift,
Osborn and Yeomans \cite{Swift95} and improved in Refs. \cite{Luo98}
in order to have a thermodynamically consistent model for a dense
fluid that may exhibit liquid-vapor coexistence. However, only
isothermal models have been considered up to now. A second very
promising approach is the Direct Simulation Monte Carlo method
\cite{Bird94} that has been conveniently generalized to deal with
dense liquids with liquid-vapor coexistence \cite{Alexander97}.

In order to construct the finite volume algorithm we have been
strongly inspired by the work of Flekkoy and Coveney
\cite{Flekkoy99}. In that paper, the authors present a ``bottom-up''
derivation of Dissipative Particle Dynamics. Physical space is divided
into Voronoi cells and explicit definitions for the mass, momentum and
energy of the cells in terms of the microscopic degrees of freedom
(positions and momenta of the constituent molecules of the fluid) are
given. The time derivatives of these phase functions have the
structure of ``microscopic balance equations'' in a discrete
form. These equations are then divided into ``average'' and
``stochastic'' parts. To further advance into the formulation of a
practical algorithm, the authors then propose {\em phenomenological},
physically sensible, expressions for the average part and require the
fulfillment of the fluctuation-dissipation theorem for the stochastic
part. Because of the use of the phenomenological expressions, we
cannot consider this a ``bottom-up'' approach. Strictly speaking, a
bottom-up approach would require the use of a projection operator
technique or kinetic theory, in order to relate the transport
coefficients with the microscopic dynamics of the system (in the form
of Green-Kubo formulae, for example).

Instead, we propose in this paper a conspicuous ``top-down'' approach
in which the deterministic continuum equations of hydrodynamics are
the starting point. By making intensive use of the smooth Voronoi
tessellation discovered by Flekkoy and Coveney the form of the
discrete equations is dictated by the very structure of the continuum
equations. Our approach is similar to that in
Ref. \cite{Hietel00}. However, we make a further requirement on the
resulting finite volume discretization, which is that they must have
the GENERIC structure. This enforces the addition of a tiny bit into
the momentum equation. Having the GENERIC structure, it is trivial to
obtain the stochastic part, which will be given by the
fluctuation-dissipation theorem.  In the concluding section we will
show the similarities and differences between our equations and those
derived by Flekkoy and Coveney.

The approach presented in this work has also a strong resemblance with
Yuan and Doi simulation method which also uses the Voronoi tessellation
in Lagrangian form \cite{Yuan98} (see also \cite{Yuan93}). They have
applied the method to the simulation of concentrated emulsions under
flow and several other applications to complex fluids are
mentioned. The main difference between our work and that of
Ref. \cite{Yuan98} is the special care we have taken in order to have
thermodynamic consistency through the GENERIC formalism. This allows,
among other things, to include correct thermal noise and describe
diffusive aspects produced by Brownian motion on mesoscopic
objects. Another difference is that we deal with a compressible fluid
in which the pressure is given through the equation of state as a
function of mass and entropy densities, in counterdistinction with
Ref. \cite{Yuan98} where the pressure is obtained by satisfying the
incompressibility condition. Compressibility is necessary if
gas-liquid coexistence is to be described.

\newpage
\section{Review of GENERIC}
\label{review}
For the sake of completeness we review in this section the GENERIC
formalism developed by \"Ottinger and Grmela \cite{generic}.  The
formalism of GENERIC (acronym for General Equation for Non Equilibrium
Reversible Irreversible Coupling) states that all physically sounded
transport equations in non-equilibrium thermodynamics have the same
structure. A large body of evidence confirms this assertion: Linear
irreversible thermodynamics, non-relativistic and relativistic
hydrodynamics, Boltzmann's equation, polymer kinetic theory, and
chemical reactions, just to mention a few, have all the GENERIC
structure \cite{generic,gen-applyed}. The GENERIC formalism is not
only a way of rewriting known transport equations in a physically
transparent way, but it allows us to derive dynamical equations for
new systems not considered so far in an astonishing simple way. The
very structure of the formalism ensures thermodynamic consistency,
energy conservation and positive entropy production.

The essential assumption on which GENERIC is founded is that the
relevant variables $x$ used to describe the system at a certain level
of description evolve in a time scale well-separated from the time
scales of other variables in the system. In other words, the variables
should provide a closed description in which the present state of the
system depends only on the very recent past and memory effects can be
neglected. This is a recurrent theme in non-equilibrium statistical
mechanics since the pioneering works of Zwanzig and Mori
\cite{Zwanzig60}. Actually, the GENERIC structure can be {\em deduced}
from first principles by using standard projection operator formalism
under a Markovian approximation \cite{generic}.

Two basic building blocks in the GENERIC formalism are the energy
$E(x)$ and entropy $S(x)$ functions of the variables $x$ describing
the state of the system at a particular level of description
\cite{generic}. The GENERIC dynamic equations are given then by

\begin{equation}
\frac{dx}{dt}=L\!\cdot\!\frac{\partial E}{\partial x}
+M\!\cdot\!\frac{\partial S}{\partial x}.
\label{gen1}
\end{equation}
The first term in the right hand side is named the {\em reversible}
part of the dynamics and the second term is named the {\em
irreversible} part. The predictive power of GENERIC relies in the fact
that very strong requirements exists on the matrices $L,M$ leaving
small room for the physical input about the system. First, $L$ is
antisymmetric whereas $M$ is symmetric and positive semidefinite. Most
important, the following {\em degeneracy} conditions should hold

\begin{equation}
L\!\cdot\!\frac{\partial S}{\partial x}=0,\quad\quad\quad
M\!\cdot\!\frac{\partial E}{\partial x}=0.
\label{gen2}
\end{equation}
These properties ensure that the energy is a dynamical invariant,
$\dot{E}=0$, and that the entropy is a non-decreasing function of
time, $\dot{S}\ge 0$, as can be proved by a simple application of the
chain rule and the equations of motion (\ref{gen1}). In the case that
other dynamical invariants $I(x)$ exist in the system (as, for
example, linear or angular momentum), then further
conditions must be satisfied by $L,M$. In particular,

\begin{equation}
\frac{\partial I}{\partial x}\!\cdot\!
L\!\cdot\!\frac{\partial E}{\partial x}=0,\quad\quad\quad
\frac{\partial I}{\partial x}\!\cdot\!
M\!\cdot\!\frac{\partial S}{\partial x}=0.
\label{il}
\end{equation}
which ensure that $\dot{I}=0$.

The deterministic equations (\ref{gen1}) are, actually, an
approximation in which thermal fluctuations are neglected. If thermal
fluctuations are not neglected, the dynamics is described by
stochastic differential equations or, equivalently, by a Fokker-Planck
equation that governs the probability distribution function
$\rho=\rho(x,t)$. This FPE has the form \cite{generic}
\begin{equation}
\partial_t\rho =
-\frac{\partial}{\partial x}\!\cdot\!
\left[\rho\left[ 
  L\!\cdot\!\frac{\partial E}{\partial x}
+ M\!\cdot\!\frac{\partial S}{\partial x} \right]
- k_B M\!\cdot\!\frac{\partial \rho}{\partial x}\right],
\label{FPE}
\end{equation}
where $k_B$ is Boltzmann's constant. 

The distribution function of the variables of a given system at
equilibrium is given by the Einstein distribution function. This
assertion can be proved under quite general hypothesis on the mixing
character of the microscopic dynamics of the system \cite{mixing}. If
the microscopic dynamics ensures the existence of dynamical invariants
like the energy $E(x)$ and, perhaps, other invariants $I(x)$, then the
Einstein distribution function takes the form \cite{mixing}

\begin{equation}
\rho^{\rm eq}(x) = g(E(x),I(x))\exp \{ S(x)/k_B\},
\label{einst}
\end{equation}
where the function $g$ is completely determined by the arbitrary
initial distribution of dynamical invariants. For example, if at
an initial time the value of the invariants $E(x),I(x)$ are known
with high precision to be $E_0,I_0$, then the Einstein distribution
function takes the form

\begin{equation}
\rho^{\rm eq}(x)= \frac{\delta(E(x)-E_0)\delta(I(x)-I_0)}
{\Omega(E_0,I_0)}\exp\{k_B^{-1} S(x)\},
\label{ein0}
\end{equation}
where $\Omega(E_0,I_0)$ is the normalization. Given the general
argument behind the Einstein distribution function \cite{mixing}, it
is sensible to demand that the Fokker-Planck equation (\ref{FPE}) has
as its (unique) equilibrium distribution function the Einstein
distribution.  This can be achieved, actually, if the following
further conditions on the form of the matrices $L,M$ hold,

\begin{equation}
\frac{\partial}{\partial x}\!\cdot\!\left[
L \!\cdot\!\frac{\partial E}{\partial x}\right]=0,\quad\quad\quad
M\!\cdot\!\frac{\partial I}{\partial x}=0.
\label{add}
\end{equation}
The first property can be derived independently with projection
operator techniques \cite{generic} whereas the second property implies
that the last equation in (\ref{il}) is automatically satisfied. 

When fluctuations are present, the entropy function $S(x)$ might be a
{\em decreasing} function of time. However, if one considers the entropy
{\em functional}

\begin{equation}
{\cal S}[\rho_t]=\int
S(x)\rho(x,t)dx -k_B\int \rho(x,t)\ln \rho(x,t) dx,
\end{equation}
it is possible to prove by using the Fokker-Planck equation (\ref{FPE}) 
that $\partial_t {\cal S}[\rho_t]\ge 0$. In other words, the entropy
functional  plays the role of a Lyapunov function. 

The stochastic differential equations that
are mathematically equivalent to the above Fokker-Planck equation
are given, with It\^o interpretation, by \cite{Gardiner83}

\begin{equation}
dx = \left[
L\!\cdot\!\frac{\partial E}{\partial x}
+M\!\cdot\!\frac{\partial S}{\partial x}
+k_B\frac{\partial }{\partial x}\!\cdot\!M\right]dt
+d\tilde{x},
\label{sde1}
\end{equation}
to be compared with the deterministic equations (\ref{gen1}).  The
stochastic term $d\tilde{x}$ in Eqn. (\ref{sde1}) is a linear
combination of independent increments of the Wiener process. It
satisfies the mnemotechnical It\^o rule

\begin{equation}
d\tilde{x}d\tilde{x}^T=2k_BM dt,
\label{F-D}
\end{equation}
which means that $d\tilde{x}$ is an infinitesimal of order $1/2$
\cite{Gardiner83}.  Eqn. (\ref{F-D}) is a compact and formal statement of
the fluctuation-dissipation theorem.

When formulating new models it might be convenient to specify
$d\tilde{x}$ directly instead of $M$. This ensures that $M$ through
(\ref{F-D}) automatically satisfies the symmetry and positive definite
character. In order to guarantee that the total energy and dynamical
invariants do not change in time, a strong requirement on the form of
$d\tilde{x}$ holds,
\begin{equation}
\frac{\partial E}{\partial x}\!\cdot\! d\tilde{x}=0,\quad\quad
\frac{\partial I}{\partial x}\!\cdot\! d\tilde{x}=0,
\label{consinv}
\end{equation}
implying the last equations in (\ref{gen2}) and (\ref{add}). The
geometrical meaning of (\ref{consinv}) is clear. The random kicks
produced by $d\tilde{x}$ on the state $x$ are orthogonal to the
gradients of $E,I$. These gradients are perpendicular vectors
(strictly speaking they are one forms) to the hypersurface $E(x)=E_0,
I(x)=I_0$.  Therefore, the kicks let the state $x$ always within the
hypersurface of dynamical invariants.

We finally close this section by noting that, formally, the size of
the fluctuations is governed by the Boltzmann constant $k_B$.  If we
take $k_B\rightarrow 0$, the stochastic differential equation
(\ref{sde1}) becomes the deterministic equations (\ref{gen1}) and the
Fokker-Planck equation (\ref{FPE}) has only first derivatives in the
same way as the Liouville equation. In this limit, the distribution
function $\rho(x,t)$ does not show dispersion and it is essentially a
Dirac delta function evaluated on the solution of the deterministic
equation. In this case, the entropy functional ${\cal S}[\rho]$
reduces to the entropy $S(x)$ and the entropy is a non-decreasing
function of time.

\section{Finite volume method with Voronoi cells}
\label{l-finite-volume}
As a first step for deriving the GENERIC equations for a model of fluid
particles, we consider the method of finite volumes for the numerical
integration of the equations of continuum hydrodynamics. The basic
motivation for this is to have a set of reference equations that serve
as guidelines for the modeling in the GENERIC equations in such a way
that we can consider the GENERIC equations as reasonable
approximations to the equations of continuum hydrodynamics.

The finite volume method consists on integrating the continuum
equations of hydrodynamics in a finite region of space (or {\em finite
volume}) in such a way that ordinary differential equations for the
average fields over the finite regions emerge.  In this section we
present a finite volume method that uses the Voronoi construction as a
conceptually and mathematically elegant method for discretizing the
continuum equations of hydrodynamics.

\begin{figure}[ht] 
\begin{center} 
\psfig{figure=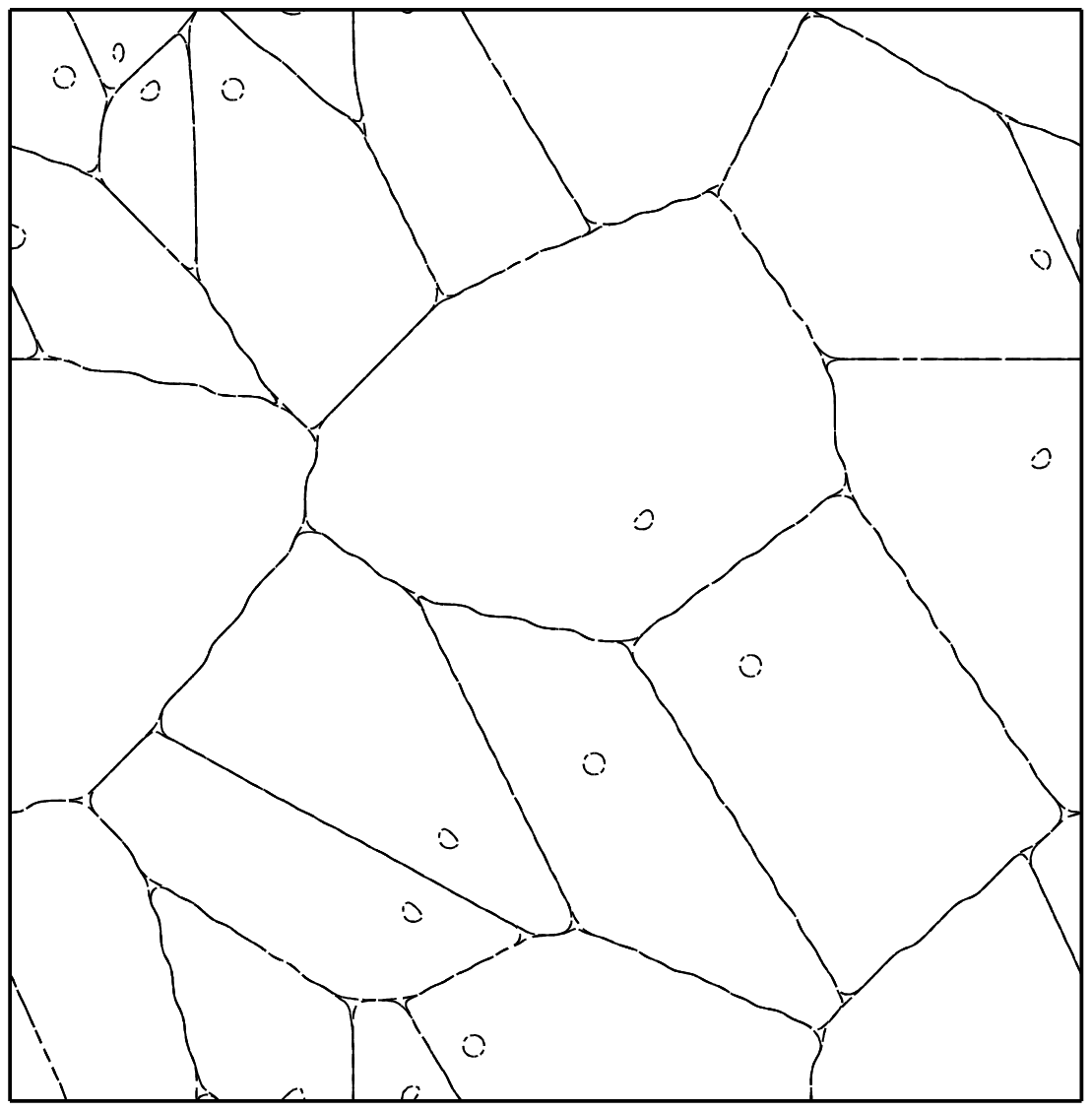,width=10cm,height=10cm}
\caption{\label{tess} Contour line at $\chi_\mu({\bf r})=0.5$ 
for the set of all characteristic functions of the 15 particles 
located at random in a periodic box of size $L=100$. The value 
of $\sigma$ is $ 0.03$.}
\psfig{figure=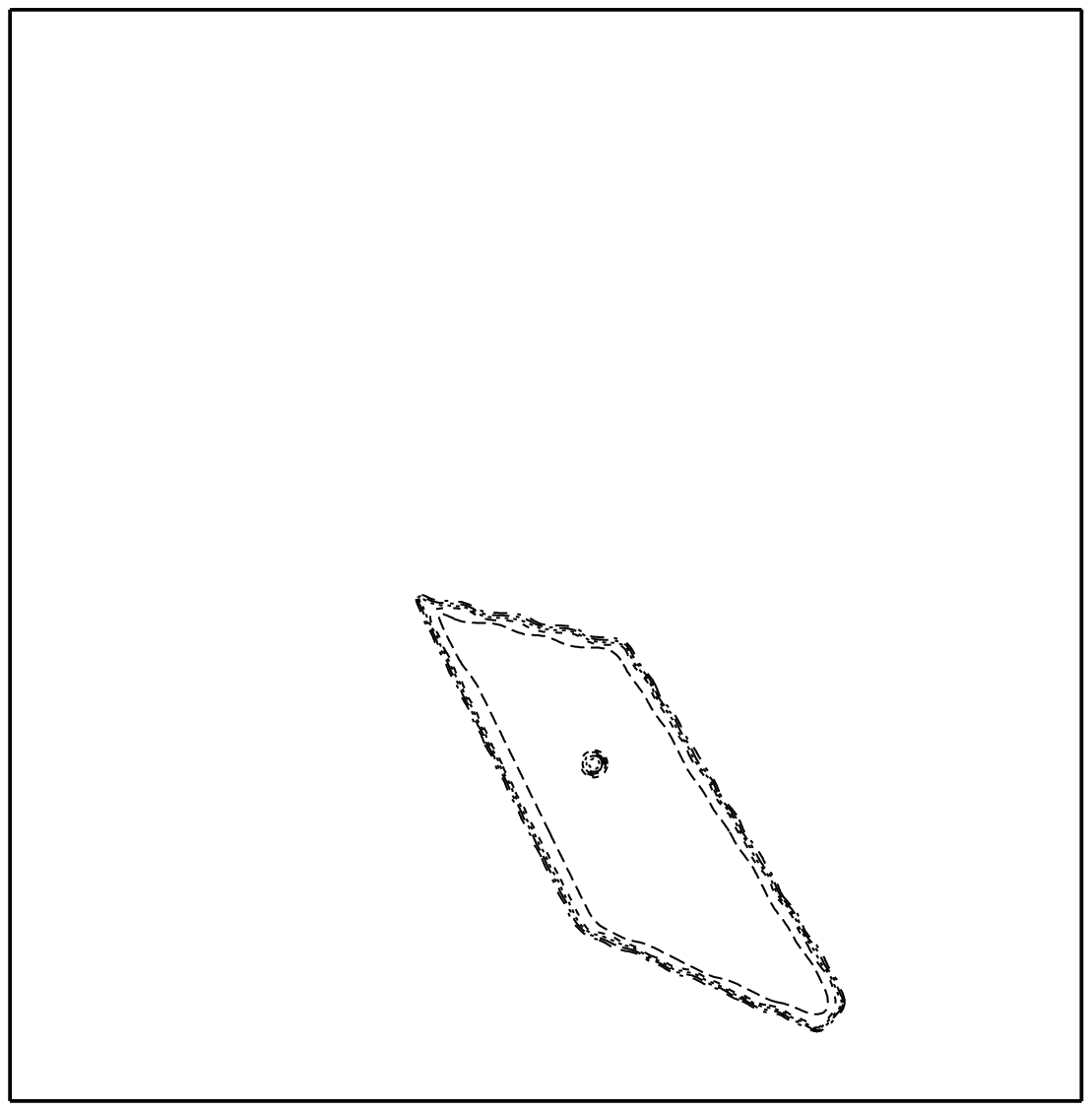,width=10cm,height=10cm}
\caption{\label{tess1}Contour lines of the function $\chi_\mu({\bf r})$  for
particle $\mu=1$ of the previous tessellation. Inside the closed region
the value of $\chi_\mu({\bf r})$ is 1 and outside is zero.}
\end{center}
\end{figure}

Following Flekkoy and Coveney \cite{Flekkoy99}, we introduce
the smoothed characteristic function of the Voronoi cell $\mu$
\begin{equation}
\chi_\mu({\bf r}) = \frac{\Delta(|{\bf r}-{\bf R}_\mu|)}
{\sum_\nu \Delta(|{\bf r}-{\bf R}_\nu|)},
\label{chi}
\end{equation}
where the function $\Delta(r)=\exp\{-r^2/2\sigma^2\}$ is a Gaussian
of width $\sigma$. When $\sigma\rightarrow0$, the smoothed
characteristic function tends to the actual characteristic function
of the Voronoi cell, this is
\begin{equation}
\lim_{\sigma\rightarrow 0}\chi_\mu({\bf r}) = \prod_\nu
\theta(|{\bf r}-{\bf R}_\nu|-|{\bf r}-{\bf R}_\mu|),
\label{chi2}
\end{equation}
where $\theta(x)$ is the Heaviside step function. The Voronoi
characteristic function (\ref{chi2}) takes the value 1 if ${\bf r}$ is
nearer to ${\bf R}_\mu$ than to any other ${\bf R}_\nu$ with
$\nu\neq\mu$. Note that the characteristic function produces a
covering of all space (i.e., a partition of unity), this is,
\begin{equation}
\sum_\mu\chi_\mu({\bf r}) = 1.
\label{cov}
\end{equation}
We introduce the volume of the Voronoi cell through
\begin{equation}
{\cal V}_\mu=\int_{V_T} d{\bf r} \chi_\mu({\bf r}),
\label{volume}
\end{equation}
which satisfies the closure condition
\begin{equation}
\sum_\mu{\cal V}_\mu=V_T,
\label{clo}
\end{equation}
where $V_T$ is the total volume.
In Fig. \ref{tess} we show the Voronoi tessellation corresponding
to 15 particles seeded at random in a periodic box. For an introduction
to the Voronoi tessellation see \cite{Voronoi}.

We mention now some other useful properties of the smoothed characteristic 
function that will be needed later. First, due to the Gaussian form
of $\Delta(r)$, 
\begin{equation}
\nabla \Delta(r)=-\frac{1}{\sigma^2}\Delta(r){\bf r}.
\label{nade}
\end{equation}
Therefore,
\begin{eqnarray}
\frac{\partial }{\partial {\bf r}}\chi_\mu({\bf r})&=&
-\frac{1}{\sigma^2}\chi_\mu({\bf r})({\bf r}-{\bf R}_\mu)
\nonumber\\
&+&\frac{1}{\sigma^2}\chi_\mu({\bf r})
\sum_\nu\chi_\nu({\bf r})({\bf r}-{\bf R}_\nu).
\label{prop1}
\end{eqnarray}
By using the following property
\begin{equation}
\chi_\mu({\bf r})(1- \chi_\mu({\bf r}))=\sum_{\nu\neq\mu}\chi_\mu({\bf r})\chi_\nu({\bf r}),
\label{p2}
\end{equation}
which can be proved by using the definition (\ref{chi}), one
can rewrite Eqn. (\ref{prop1}) as
\begin{equation}
\frac{\partial }{\partial {\bf r}}\chi_\mu({\bf r})=
\frac{1}{\sigma^2}
\sum_\nu\chi_\mu({\bf r})\chi_\nu({\bf r})({\bf R}_\mu-{\bf R}_\nu).
\label{prop1b}
\end{equation}
Another useful relation is,
\begin{eqnarray}
\frac{\partial }{\partial {\bf R}_\nu}\chi_\mu({\bf r})
&=&
\delta_{\mu\nu}\frac{1}{\sigma^2}\chi_\mu({\bf r})({\bf r}-{\bf R}_\mu)
\nonumber\\
&-&\frac{1}{\sigma^2}\chi_\mu({\bf r})\chi_\nu({\bf r})({\bf r}-{\bf R}_\nu).
\label{prop2}
\end{eqnarray}

We now introduce the following two quantities
\begin{eqnarray}
A_{\mu\nu}
&\equiv &R_{\mu\nu}\int \frac{d{\bf r}}{\sigma^2} \chi_\mu({\bf r})\chi_\nu({\bf r}),
\nonumber\\
{\bf c}_{\mu\nu}
&\equiv&\frac{R_{\mu\nu}}{A_{\mu\nu}}\int \frac{d{\bf r}}{\sigma^2}
\chi_\mu({\bf r})\chi_\nu({\bf r}) \left({\bf r}-\frac{{\bf R}_\mu+{\bf R}_\nu}{2}\right),
\label{area}
\end{eqnarray}
where $R_{\mu\nu}=|{\bf R}_\mu-{\bf R}_\nu|$.  In appendix
\ref{ap-vol} we show that in the limit $\sigma\rightarrow 0$ the
quantity $A_{\mu\nu}$ is actually the area of the contact face
$\mu\nu$ between Voronoi cells $\mu$ and $\nu$, whereas the vector
${\bf c}_{\mu\nu}$ is the position of the center of mass of the face
$\mu\nu$ with respect to the ``center'' of the face $({\bf R}_\mu+{\bf
R}_\nu)/2$.

\subsection{Balance equations}

We introduce now the cell average ${[\phi]}_\mu(t)$ over the Voronoi
cell $\mu$ of an arbitrary density field $\phi({\bf r},t)$
\begin{equation}
{[\phi]}_\mu(t) =\frac{1}{{\cal V}_\mu}\int d{\bf r}\phi({\bf r},t)\chi_\mu({\bf r}).
\label{phimu}
\end{equation}
We will refer to ${[\phi]}_\mu (t)$ as a cell variable and we will
see that it is an approximation for the value of the field
at the discrete points given by the cell centers.

In principle, the Voronoi cell centers are allowed to move in an
arbitrary way, this is, ${\bf R}_\mu(t)$ are prescribed functions of
time. The time derivative of the cell averages  is given by
\begin{eqnarray}
\frac{d}{dt} {[\phi]}_\mu (t) &=& -\frac{\dot{\cal V}_\mu}{{\cal V}_\mu}{[\phi]}_\mu (t)
\nonumber\\
&+&\frac{1}{{\cal V}_\mu}\int d{\bf r}\phi({\bf r},t)\frac{d}{dt}\chi_\mu({\bf r})
\nonumber\\
&+&\frac{1}{{\cal V}_\mu}\int d{\bf r}\chi_\mu(t)\partial_t \phi({\bf r},t),
\label{td1}
\end{eqnarray}
where the dot means the time derivative. We see that ${[\phi]}_\mu (t)$ changes
due to both, the motion of the cells and the intrinsic dependence of the
field $\phi({\bf r},t)$ on time.

Now, we will assume that the field $\phi({\bf r},t)$ obeys a 
balance equation, this is 
\begin{equation}
\partial_t\phi({\bf r},t) = -\nabla\!\cdot\!{\bf J}({\bf r},t),
\label{bal}
\end{equation}
where ${\bf J}({\bf r},t)$ is an appropriate current density.  By
integration by parts of the nabla operator and use of
Eqns. (\ref{prop1b}) and (\ref{p2}) one arrives easily at the following
expression

\begin{eqnarray}
\frac{d}{dt}{[\phi]}_\mu (t) &=& -\frac{\dot{\cal V}_\mu(t)}{{\cal V}_\mu}{[\phi]}_\mu (t)
\nonumber\\
&+&\frac{1}{{\cal V}_\mu}\sum_\nu A_{\mu\nu}
{\bf e}_{\mu\nu}\!\cdot\!\left([{\bf J}]_{\mu\nu}
-[\phi]_{\mu\nu}\frac{\dot{{\bf R}}_\mu+\dot{{\bf R}}_\nu}{2}\right)
\nonumber\\
&+&\frac{1}{{\cal V}_\mu}\sum_\nu \frac{A_{\mu\nu}}{R_{\mu\nu}}
[\mbox{\boldmath $\phi$}]^{||}_{\mu\nu}\!\cdot\!\dot{{\bf R}}_{\mu\nu},
\label{l1}
\end{eqnarray}
where 

\begin{eqnarray}
{\bf e}_{\mu\nu} &=& \frac{{\bf R}_{\mu\nu}}{R_{\mu\nu}},
\nonumber\\
{\bf R}_{\mu\nu}&=&{\bf R}_\mu-{\bf R}_\nu,
\nonumber\\
R_{\mu\nu}&=&|{\bf R}_\mu-{\bf R}_\nu|,
\label{defis1}
\end{eqnarray}
and we have introduced the face averages
\begin{eqnarray}
[\cdots]_{\mu\nu}&=& \frac{R_{\mu\nu}}{A_{\mu\nu}}
\int \frac{d{\bf r}}{\sigma^2}\chi_\mu({\bf r})\chi_\nu({\bf r}) \cdots
\nonumber\\
{[\cdots ]}_{\mu\nu}^{||}&=& \frac{R_{\mu\nu}}{A_{\mu\nu}}\int
 \frac{d{\bf r}}{\sigma^2}\chi_\mu({\bf r})\chi_\nu({\bf r}) 
\left({\bf r}-\frac{{\bf R}_\nu+{\bf R}_\mu}{2}\right)\cdots
\label{defis2}
\end{eqnarray}
Note that in the limit of sharp boundaries $\sigma\rightarrow 0$,
$\mbox{\boldmath $\phi$}^{||}_{\mu\nu}$ is a vector which is parallel
to the face $\mu\nu$, whereas ${\bf e}_{\mu\nu}$ is perpendicular to
the face.

We can write Eqn. (\ref{l1}) in the form
\begin{eqnarray}
\frac{d}{dt} \left({\cal V}_\mu{[\phi]}_\mu \right) &=&
\sum_\nu A_{\mu\nu}{\bf e}_{\mu\nu}\left([{\bf J}]_{\mu\nu}-[\phi]_{\mu\nu}
\frac{\dot{{\bf R}}_\mu+\dot{{\bf R}}_\nu}{2}\right)
\nonumber\\
&+&\sum_\nu \frac{A_{\mu\nu}}{R_{\mu\nu}}
[\mbox{\boldmath $\phi$}]^{||}_{\mu\nu}\!\cdot\!\dot{\bf R}_{\mu\nu},
\label{cons}
\end{eqnarray}
which satisfies
\begin{equation}
\frac{d}{dt} \left(\sum_\mu{\cal V}_\mu{[\phi]}_\mu  \right)=0,
\label{cons2}
\end{equation}
due to the symmetries $[\cdots]_{\mu\nu}=[\cdots]_{\nu\mu}$,
$[\cdots]^{||}_{\mu\nu}=[\cdots]^{||}_{\nu\mu}$ of the face averages.
Equation (\ref{cons2}) shows that the Voronoi discretization of the
balance equation (\ref{bal}) conserves {\em exactly} the {\em
extensive} variables (which are of the form density$\times$volume).

\subsection{Finite volumes for an inviscid fluid}
In what follows we will apply the method of finite volumes
to the continuum equations of hydrodynamics. For the sake
of clarity, we first consider the reversible part of
the equations, which correspond the the usual Euler equations
for an inviscid fluid. In the next subsection we consider
the irreversible part.

The Euler equations for an inviscid fluid are the continuity 
equation 
\begin{equation}
\partial_t \rho({\bf r},t) =
 -\nabla\!\cdot\!{\bf g}({\bf r},t),
\label{cont}
\end{equation}
the momentum balance equation
\begin{equation}
\partial_t {\bf g}({\bf r},t) 
= -\nabla P({\bf r},t) - \nabla\!\cdot\!{\bf g}({\bf r},t){\bf v}({\bf r},t),
\label{mombal}
\end{equation}
and the entropy equation
\begin{equation}
\partial_t s({\bf r},t) = -\nabla\!\cdot\!s({\bf r},t){\bf v}({\bf r},t).
\label{content}
\end{equation}
In these equations, $\rho({\bf r},t)$ is the mass density field, ${\bf
g}({\bf r},t)=\rho({\bf r},t){\bf v}({\bf r},t)$ is the momentum
density field, with ${\bf v}({\bf r},t)$ the velocity field, and
$s({\bf r},t)$ is the entropy density field (entropy per unit
volume). The pressure field $P({\bf r},t)$ is given, according to the
local equilibrium assumption, by $P({\bf r},t)=P^{\rm eq}(\rho({\bf
r},t),s({\bf r},t))$ where $P^{\rm eq}(\rho,s)$ is the equilibrium
equation of state that gives the macroscopic pressure in terms of the
mass and entropy densities.

We now write Eqn. (\ref{cons}) for the case that $\phi=\rho$. 
If the Voronoi cells do not move, then $\dot{{\bf R}}_\mu=0$ and
expression (\ref{cons}) simplifies considerably,
\begin{equation}
\frac{d}{dt}M_\mu(t) = \sum_\nu A_{\mu\nu}{\bf e}_{\mu\nu}
\!\cdot\![\rho{\bf v}]_{\mu\nu},
\label{eul}
\end{equation}
where $M_\mu={\cal V}_\mu {[\rho}]_\mu$ is the mass of the Voronoi
cell $\mu$. This corresponds to a {\em Eulerian} discretization of the
continuity equation (\ref{cont}). The physical meaning of
Eqn. (\ref{eul}) is clear: $A_{\mu\nu}{\bf e}_{\mu\nu}\!\cdot\!
[\rho{\bf v}]_{\mu\nu}$ is the total mass per unit time that crosses
face $\mu\nu$ and the rate of change of the mass of the cell $\mu$ is
the sum of this quantity for each face $\mu\nu$.

The {\em Lagrangian} discretization of the continuity equation in
Eqn. (\ref{cont}) is obtained by specifying the motion of the cells
according to
\begin{equation}
\dot{\bf R}_\mu(t)={[{\bf v}]}_{\mu}(t).
\label{dl}
\end{equation}
In this way, the Voronoi cells ``follow'' (the best they can) the
flow field. We can write Eqn. (\ref{cons}) for the case (\ref{cont})
as
\begin{eqnarray}
\left.\dot{M}_\mu\right|_{\rm rev} &=&
\sum_\nu A_{\mu\nu}{\bf e}_{\mu\nu}\!\cdot\!
\left({[\rho{\bf v}]}_{\mu\nu}-
{[\rho]}_{\mu\nu}\frac{{[{\bf v}]}_\mu+{[{\bf v}]}_\nu}{2}\right)
\nonumber\\
&+&\sum_\nu\frac{A_{\mu\nu}}{R_{\mu\nu}}{[\rho]}^{||}_{\mu\nu}\!\cdot\!
({[{\bf v}]}_\mu-{[{\bf v}]}_\nu).
\label{mass1}
\end{eqnarray}
The momentum balance equation (\ref{mombal}) can similarly 
be treated and the following Lagrangian finite volume equation
arise
\begin{eqnarray}
\left.\dot{\bf P}_\mu\right|_{\rm rev} &=&
\sum_\nu A_{\mu\nu}{\bf e}_{\mu\nu}{[P]}_{\mu\nu}
\nonumber\\
&&+\sum_\nu A_{\mu\nu}{\bf e}_{\mu\nu}\!\cdot\!\left(
{[{\bf g}{\bf v}]}_{\mu\nu}-{[{\bf g}]}_{\mu\nu}
\frac{{[{\bf v}]}_\mu+{[{\bf v}]}_\nu}{2}\right)
\nonumber\\
&&+\sum_\nu \frac{A_{\mu\nu}}{R_{\mu\nu}}{[{\bf g}]}^{||}_{\mu\nu}\!\cdot\!({[{\bf v}]}_\mu-{[{\bf v}]}_\nu),
\label{mombal2}
\end{eqnarray}
where we have introduced the total momentum of cell $\mu$ ${\bf
P}_{\mu} = {\cal V}_\mu {\bf g}_\mu$.

Finally, the entropy equation Eqn. (\ref{content}) takes
the following Lagrangian discretization 
\begin{eqnarray}
\left.\dot{S}_\mu\right|_{\rm rev} &=&
\sum_\nu A_{\mu\nu}{\bf e}_{\mu\nu}\!\cdot\!
\left({[s{\bf v}]}_{\mu\nu}-
{[s]}_{\mu\nu}\frac{{[{\bf v}]}_\mu+{[{\bf v}]}_\nu}{2}\right)
\nonumber\\
&+&\sum_\nu\frac{A_{\mu\nu}}{R_{\mu\nu}}{[s]}^{||}_{\mu\nu}\!\cdot\!
({[{\bf v}]}_\mu-{[{\bf v}]}_\nu),
\label{cdent}
\end{eqnarray}
where $S_\mu={\cal V}_\mu s_\mu$ is the total entropy of cell $\mu$.

In all these evolution equations, the subscript $|_{\rm rev}$ denotes
that the equations are actually the reversible part of the dynamics
of a truly viscous fluid.

\subsection{Gradient expansion}
The above equations (\ref{mass1}), (\ref{mombal2}), and (\ref{cdent})
are rigorous and exact and do not depend on the typical size of the
Voronoi cells. By taking sufficiently small cells, we can approximate
the equations and transform them into a {\em closed} set of equations
for the cell variables. In this way, a computationally feasible
algorithm can be proposed for the updating of cell variables.

Let us assume that the hydrodynamic fields have a typical length
scale of variation $\lambda_H$ and that the typical distance
between cell centers is $\lambda_c$. We introduce the {\em resolution}
parameter as $r=\lambda_c/\lambda_H$ and will assume that this
parameter is very small. Therefore the dimensionless quantity
\begin{equation}
\frac{\lambda_c }{\phi({\bf R}_\mu)}\nabla \phi({\bf R}_\mu)\sim r,
\label{re}
\end{equation}
will be very small. By denoting with ${\cal O}(\nabla)$ those terms
which are of relative size $r$, we can write the cell average as

\begin{eqnarray}
{[\phi]}_\mu &\equiv&\frac{1}{{\cal V}_\mu}
\int d{\bf r} \chi_\mu({\bf r}) \phi({\bf r})
\nonumber\\
&=&\frac{1}{{\cal V}_\mu}
\int d{\bf r} \chi_\mu({\bf r}) \phi({\bf r}-{\bf R}_\mu +{\bf R}_\mu)
\nonumber\\
&=&\phi({\bf R}_\mu)+\frac{1}{{\cal V}_\mu}
\int d{\bf r} \chi_\mu({\bf r}) ({\bf r}-{\bf R}_\mu)
\!\cdot\!\nabla\phi({\bf R}_\mu)
+{\cal O}(\nabla^2)
\nonumber\\
&=&\phi({\bf R}_\mu)+{\cal O}(\nabla).
\label{phimub}
\end{eqnarray}
Performing similar Taylor expansions we obtain easily

\begin{equation}
{[\phi]}_{\mu\nu}=\phi\left(\frac{{\bf R}_\mu+{\bf R}_\nu}{2}\right)
+{\cal O}(\nabla).
\label{phimunu}
\end{equation}
Also
\begin{equation}
\phi\left(\frac{{\bf R}_\mu+{\bf R}_\nu}{2}\right)
=\frac{\phi({\bf R}_\mu)+\phi({\bf R}_\nu)}{2}
+{\cal O}(\nabla^2),
\label{phi3}
\end{equation}
and, therefore,
\begin{equation}
{[\phi]}_{\mu\nu}=
\frac{{[\phi]}_{\mu}+{[\phi]}_{\nu}}{2}
+{\cal O}(\nabla).
\label{phi4}
\end{equation}
After some algebra it is easy to show that
\begin{equation}
{[\phi\psi]}_{\mu\nu}={[\phi]}_{\mu\nu}
{[\psi]}_{\mu\nu}+{\cal O}(\nabla^2).
\label{phiprod}
\end{equation}
Finally,
\begin{equation}
{[{\bf \phi}]}^{||}_{\mu\nu}
=\frac{{[\phi]}_{\mu}+{[\phi]}_{\nu}}{2}
{\bf c}_{\mu\nu}+{\cal O}(\nabla).
\label{phipar}
\end{equation}
By using these Taylor approximations in Eqns. (\ref{mass1}),
(\ref{mombal2}), and (\ref{cdent}) we obtain the final Voronoi
finite volume discrete equations for the inviscid fluid,
\begin{eqnarray}
\left.\dot{M}_\mu\right|_{\rm rev}&=&\sum_\nu\frac{A_{\mu\nu}}{R_{\mu\nu}}
\frac{{[\rho]}_{\mu}+{[\rho]}_{\nu}}{2}
{\bf c}_{\mu\nu}\!\cdot\!({[{\bf v}]}_\mu-{[{\bf v}]}_\nu),
\nonumber\\
\left.\dot{\bf P}_\mu\right|_{\rm rev} &=&\sum_\nu A_{\mu\nu}
{\bf e}_{\mu\nu}\frac{{[P]}_{\mu}+{[P]}_{\nu}}{2}
\nonumber\\
&+&\sum_\nu  \frac{A_{\mu\nu}}{R_{\mu\nu}}
\frac{{[\rho]}_{\mu}+{[\rho]}_{\nu}}{2}
\frac{{[{\bf v}]}_{\mu}+{[{\bf v}]}_{\nu}}{2}
{\bf c}_{\mu\nu}\!\cdot\!({[{\bf v}]}_\mu-{[{\bf v}]}_\nu),
\nonumber\\
\left.\dot{S}_\mu\right|_{\rm rev} &=&\sum_\nu\frac{A_{\mu\nu}}{R_{\mu\nu}}
\frac{{[s]}_{\mu}+{[s]}_{\nu}}{2}
{\bf c}_{\mu\nu}\!\cdot\!({[{\bf v}]}_\mu-{[{\bf v}]}_\nu).
\label{fvap}
\end{eqnarray}
These equations become closed equations for $M_\mu,{\bf P}_\mu, S_\mu$
by using
\begin{eqnarray}
{[\rho]}_\mu&=& \frac{M_\mu}{{\cal V}_\mu},
\nonumber\\
{[s]}_\mu&=& \frac{S_\mu}{{\cal V}_\mu},
\nonumber\\
{[P]}_\mu&=& P^{\rm eq}({[\rho]}_\mu,{[s]}_\mu),
\label{rsp}
\end{eqnarray}
where in the last equation for the pressure we have used again a
Taylor expansion and neglected terms of order ${\cal O}(\nabla)$.

\subsection{Finite volumes for a viscous fluid}
Having studied the inviscid fluid, in which there are no
dissipative contributions to the motion of the fluid, we turn
now to the general viscous fluid. The continuum equations
are given by \cite{deGroot84}

\begin{eqnarray}
\partial_t \rho({\bf r},t) &=& -\nabla\!\cdot\!\rho({\bf r},t){\bf v}({\bf r},t),
\nonumber\\
\partial_t {\bf g}({\bf r},t) 
&=& -\nabla P({\bf r},t) - \nabla\!\cdot\!{\bf g}({\bf r},t){\bf v}({\bf r},t)
-\nabla\!\cdot\!(\overline{\bf \Pi}+\Pi {\bf 1}),
\nonumber\\
\partial_t s({\bf r},t) &=& -\nabla\!\cdot\!s({\bf r},t){\bf v}({\bf r},t)
\nonumber\\
&-&\frac{1}{T}\nabla\!\cdot\!{\bf J}^q 
+ \frac{2\eta}{T} \overline{\nabla {\bf v}}:\overline{\nabla {\bf v}}
+\frac{\zeta}{T} (\nabla\!\cdot\!{\bf v})^2,
\label{viscous}
\end{eqnarray}
where, by comparison with Eqns. (\ref{cont}), (\ref{mombal}),
(\ref{content}) we can recognize the purely irreversible terms in the
momentum and entropy equations. Here, $T$ is the temperature field
(which, as the pressure, is a function of $\rho,s$ through the local
equilibrium assumption).  The double dot implies double
contraction. These equations have to be supplemented with the
constitutive equations for the traceless symmetric part $\overline{\bf
\Pi}$ of the viscous stress tensor, the trace $\Pi$ of the viscous
stress tensor, and the heat flux ${\bf J}^q$. They are
\begin{eqnarray}
\overline{\bf \Pi} &=& -2\eta \overline{\nabla {\bf v}},
\nonumber\\
\Pi &=& -\zeta\nabla\!\cdot\!{\bf v},
\nonumber\\
{\bf J}^q &=& -\kappa \nabla T = \kappa T^2\nabla\frac{1}{T},
\label{constitut}
\end{eqnarray}
where the traceless symmetric part of the velocity gradient tensor is
\begin{equation}
\overline{\nabla {\bf v}}
=\frac{1}{2}\left(\nabla {\bf v}+(\nabla {\bf v})^T\right)
-\frac{1}{D}\nabla\!\cdot\!{\bf v}.
\label{on}
\end{equation}
Here, $D$ is the dimension of physical space.
In principle, the shear viscosity $\eta$, the bulk viscosity $\zeta$
and the thermal conductivity $\kappa$ might depend on the state of the
fluid through $\rho,s$.

Following identical steps as for the inviscid fluid, we see that the
viscous (irreversible) term in the momentum balance equation
translates into
\begin{equation}
\left.\dot{\bf P}_\mu \right|_{\rm irr}=
 \sum_\nu A_{\mu\nu}{\bf e}_{\mu\nu}\!\cdot\![\eta \overline{\nabla {\bf v}}]_{\mu\nu}
+\sum_\nu A_{\mu\nu}{\bf e}_{\mu\nu}[\zeta \nabla \!\cdot\!{\bf v}]_{\mu\nu}.
\label{irrp}
\end{equation}
We now consider the gradient expansion on each term. 
For example,

\begin{eqnarray}
\left[\eta \overline{\nabla {\bf v}}\right]_{\mu\nu}
&=&\frac{\left[\eta \overline{\nabla {\bf v}}\right]_{\mu}
  +\left[\eta \overline{\nabla {\bf v}}\right]_{\nu}}{2}+{\cal O}(\nabla),
\nonumber\\
\left[\zeta \nabla \!\cdot\!{\bf v}\right]_{\mu\nu}
&=&\frac{[\zeta \nabla \!\cdot\!{\bf v}]_{\mu}+[\zeta \nabla \!\cdot\!{\bf v}]_{\nu}}{2}+{\cal O}(\nabla).
\label{t1}
\end{eqnarray}
Therefore, Eqn. (\ref{irrp}) becomes

\begin{equation}
\left.\dot{\bf P}_\mu \right|_{\rm irr}=
 \sum_\nu {\bf \Omega}_{\mu\nu}\!\cdot\![\eta \overline{\nabla {\bf v}}]_{\nu}
+\sum_\nu {\bf \Omega}_{\mu\nu}[\zeta \nabla \!\cdot\!{\bf v}]_{\nu}+{\cal O}(\nabla),
\label{irrp2}
\end{equation}
where we have introduced
\begin{equation}
{\bf \Omega}_{\mu\nu} = \frac{1}{2}A_{\mu\nu}{\bf e}_{\mu\nu}
\label{aeom}
\end{equation}
Note that
for any quantity $\phi$ we have
\begin{equation}
{[\nabla \phi]}_\mu = -\frac{1}{{\cal V}_\mu}
\sum_\nu {\bf \Omega}_{\mu\nu} {[\phi]}_\nu+{\cal O}(\nabla)
\label{gdisc}
\end{equation}
Therefore, we see that ${\bf \Omega}_{\mu\nu}$ is a sort of discrete
version of the gradient operator. This discrete gradient satisfies
\begin{equation}
\sum_{\nu\neq\mu}{\bf \Omega}_{\mu\nu}
=\sum_{\nu\neq\mu} \frac{1}{2}A_{\mu\nu}{\bf e}_{\mu\nu} = 0
\label{green}
\end{equation}
which is essentially the statement of the divergence theorem as can
be shown from the identity
\begin{eqnarray}
0&=&\int d{\bf r} \chi_\mu({\bf r}) \frac{\partial}{\partial {\bf r}}1
=-\int d{\bf r} \frac{\partial}{\partial {\bf r}}\chi_\mu({\bf r}) 
\nonumber\\
&=&-\sum_\nu A_{\mu\nu}{\bf e}_{\mu\nu}
\label{diverthe}
\end{eqnarray}
where Eqn. (\ref{prop1b}) has been used in the last equality.

The $\alpha,\beta$ component of the tensor in the first
term in the lhs of (\ref{irrp2}) becomes
\begin{eqnarray}
[\eta\overline{\nabla {\bf v}}^{\alpha\beta}]_{\mu} &=&
[\eta]_{\mu}[\overline{\nabla {\bf v}}^{\alpha\beta}]_{\mu} +{\cal O}(\nabla)
\nonumber\\
&=&
-\frac{{[\eta]}_\mu}{{\cal V}_\mu}\left(
\frac{1}{2}\sum_\nu \left({\bf \Omega}_{\mu\nu}^\alpha{[{\bf v}^\beta]}_{\nu}
+{\bf \Omega}_{\mu\nu}^\beta{[{\bf v}^\alpha]}_{\nu}\right)\right.
\nonumber\\
&-&\left.\frac{1}{D}\delta^{\alpha\beta}\sum_\nu
{\bf \Omega}_{\mu\nu}\!\cdot\!{[{\bf v}]}_{\nu}\right) +{\cal O}(\nabla).
\label{gradterm}
\end{eqnarray}
In a similar way,
\begin{equation}
[\zeta\nabla \!\cdot\!{\bf v}]_\mu =
-\frac{{[\zeta]}_\mu}{{\cal V}_\mu}
\sum_\nu {\bf \Omega}_{\mu\nu}\!\cdot\!{[{\bf v}]}_{\nu}.
\label{divterm}
\end{equation}
By introducing the following discrete versions of the quantities
$\overline{\bf \Pi}, \Pi, \overline{\nabla {\bf v}},\nabla\!\cdot\!{\bf v}$
in Eqns. (\ref{constitut})

\begin{eqnarray}
\overline{\bf \Pi}_\mu &=& -\frac{2{[\eta]}_\mu}{{\cal V}_\mu}
\overline{\bf G}_\mu,
\quad\quad\quad\quad
\Pi_\mu = -\frac{{[\zeta]}_\mu}{{\cal V}_\mu}
D_\mu,
\nonumber\\
\overline{\bf G}_\mu^{\alpha\beta} &= &-
\left[
\frac{1}{2}\sum_\nu[
{\bf \Omega}_{\mu\nu}^{\alpha}{\bf v}_\nu^\beta+
{\bf \Omega}_{\mu\nu}^{\beta}{\bf v}_\nu^\alpha]-
\frac{1}{D}\delta^{\alpha\beta}\sum_\nu{\bf \Omega}_{\mu\nu}\!\cdot\!{\bf v}_\nu\right],
\nonumber\\
D_\mu &=&-\sum_\nu{\bf \Omega}_{\mu\nu}\!\cdot\!{\bf v}_\nu,
\label{grad}
\end{eqnarray}
we can write the irreversible part of the momentum equation
as

\begin{equation}
\left.\dot{\bf P}_\mu \right|_{\rm irr}=
 \sum_\nu {\bf \Omega}_{\mu\nu}\!\cdot\!({\bf \Pi}_\nu+\Pi_\nu {\bf 1})
+{\cal O}(\nabla).
\label{irrmom}
\end{equation}

After a very similar procedure, the irreversible part
of the dynamics in the entropy equation can be cast also
in the form

\begin{eqnarray}
T_\mu\left.\dot{S}_\mu\right|_{\rm irr} &=& 
\sum_\nu{\bf \Omega}_{\mu\nu}\!\cdot\!{[{\bf J}^q]}_\nu
\nonumber\\
&+&\frac{2\eta_\mu}{{\cal V}_\mu}
\overline{{\bf G}}_{\mu}:\overline{{\bf G}}_{\mu}
+\frac{\zeta_\mu}{{\cal V}_\mu}D^2_\mu,
\label{irrent}
\end{eqnarray}
where 
\begin{equation}
{[{\bf J}^q]}_\mu = -\frac{{[\kappa]}_\mu}{{\cal V}_\mu}T_\mu^2
\sum_\nu{\bf \Omega}_{\mu\nu}\frac{1}{T_\nu}.
\label{hf}
\end{equation}

Addition of the irreversible parts Eqn. (\ref{irrmom}), (\ref{irrent})
to the reversible part Eqns. (\ref{fvap}) leads to the final
Lagrangian finite volume discretization of continuum hydrodynamics.
The numerical solution of the resulting set of ordinary differential
equations would produce results which are accurate to order $r$.

Admittedly, there are many other possibilities for approximating
equation (\ref{irrp}) to first order in gradients. The presentation
above is just a convenient form which has the particular GENERIC
structure, as will be shown in the following sections.

\section{GENERIC model of fluid particles}
\label{GEN-model}
We have presented in Ref. \cite{Espanol-prl99} a description of a Newtonian
fluid in terms of discrete {\em fluid particles}.  In this section, we
present a slight modification of the fluid particle model presented in
Ref. \cite{Espanol-prl99} inspired by the results of the previous section on
the finite Voronoi volume discretization.

In the fluid particle model of Ref. \cite{Espanol-prl99}, the fluid particles
are understood as thermodynamic subsystems which move with the
flow. The state of the system was described by the set of variables
$x=\{{\bf R}_\mu, {\bf P}_\mu,{\cal V}_\mu,S_\mu,\;\;i=1,\ldots,M\}$,
where $M$ is the number of fluid particles and ${\bf R}_\mu$ is the
position, ${\bf P}_\mu$ is the momentum, ${\cal V}_\mu$ is the volume,
and $S_\mu$ is the entropy of the $\mu$-th fluid particle.  Because
each fluid particle is understood as a thermodynamic subsystem, it has
a well-defined thermodynamic fundamental equation. The fundamental
equation relates the internal energy ${\cal E}_\mu$ of the fluid
particle with its mass $M_\mu$, volume ${\cal V}_\mu$ and entropy
$S_\mu$, this is ${\cal E}_\mu = {\cal E}(M_\mu,{\cal V}_\mu,S_\mu)$.
The {\em local equilibrium hypothesis} assumes that the fundamental
equation for the fluid particles has the same functional form as the
fundamental equation for the whole system at equilibrium.

The volume ${\cal V}_\mu$ was considered in Ref. \cite{Espanol-prl99} as an
independent variable to be included in $x$.  In the appendix
\ref{ap-dep} we show that, despite our intention of considering the
volume as an independent variable, due to the particular form of the
matrix $L$ selected in Ref. \cite{Espanol-prl99}, the volume is
actually a function of the positions. For this reason, in this paper
we consider from the very beginning that the volume ${\cal V}_\mu$ of
the fluid particles is a function of the positions of the particles,
i.e., ${\cal V}_\mu={\cal V}_\mu({\bf R}_1,\ldots,{\bf R}_M)$. We will
actually assume that the volume of particle $\mu$ is the volume of the
Voronoi cell of this particle, this is, Eqn. (\ref{volume}).

Unfortunately, the fact that the volume is not an actual independent
thermodynamic variable limits the applicability of the model in
Ref. \cite{Espanol-prl99}. In order to recover the thermodynamic versatility
required, it is necessary to define a model in which the {\em mass} of
the particles changes. From the point of view of the finite volume
method of the previous section, it is fairly clear that the mass of
the Voronoi cells should be taken as a dynamical variable.

The state of the fluid is given, therefore, by $x=\{{\bf R}_\mu,{\bf
P}_\mu,M_\mu,S_\mu\}$.  The energy and entropy functions are
postulated to have the form

\begin{eqnarray}
E(x) &=& \sum_\mu\frac{{\bf P}_\mu^2}{2M_\mu}
+{\cal E}(M_\mu,S_\mu,{\cal V}_\mu),
\nonumber\\
S(x) &=& \sum_\mu S_\mu.
\label{ES}
\end{eqnarray}
where ${\cal V}_\mu$ is an implicit function of the positions of 
the fluid particles.
Regarding the dynamical invariants of the system, we require that the
total mass $M(x)=\sum_\mu M_\mu$ and total momentum ${\bf P}(x)=
\sum_\mu{\bf P}_\mu$ are the only dynamical invariants. Conservation
of angular momentum would require the introduction of spin variables
in this discrete model \cite{Espanol98}.

The gradients of energy and entropy are given by

\begin{equation}
\frac{\partial E}{\partial x} =
\left(
\begin{array}{c} -\sum_\gamma \frac{\partial {\cal V}_\gamma}{\partial {\bf R}_\nu} P_\gamma
\\
\\{\bf v}_\nu
\\
\\-\frac{{\bf v}_\nu^2}{2}+\mu_\nu 
\\
\\T_\mu
\end{array}\right),
\quad\quad\quad
\frac{\partial S}{\partial x} =
\left(
\begin{array}{c} {\bf 0}\\ \\ {\bf 0}\\ \\ 0\\ \\ 1
\end{array}\right),
\label{derES}
\end{equation}
where we have introduced the velocity, pressure, chemical potential per unit mass, and temperature 
according to the usual definitions,
\begin{eqnarray}
{\bf v}_\nu&=&\frac{{\bf P}_\nu}{M_\nu},
\nonumber\\
-P_{\nu}&=&\frac{\partial {\cal E}_\nu}{\partial {\cal V}_\nu},
\nonumber\\
\mu_{\nu}&=&\frac{\partial {\cal E}_\nu}{\partial M_\nu},
\nonumber\\
T_{\nu}&=&\frac{\partial {\cal E}_\nu}{\partial S_\nu}.
\label{intdef}
\end{eqnarray}

\section{Reversible dynamics}
\label{reversible}

In this section we consider the reversible part of the dynamics for
the fluid particle model. The
matrix $L$ is made of $M\times M$ blocks ${\bf L}_{\mu\nu}$ of size
$8\times 8$. The antisymmetry of $L$ translates into ${\bf
L}_{\mu\nu}=-{\bf L}_{\nu\mu}$. 

We have a first strong requirements for the form of $L$. We wish that the
reversible part of the dynamics produces the following equations of
motion for the positions 
\begin{equation}
\dot{\bf R}_\mu = {\bf v}_\mu.
\label{require}
\end{equation}
The simplest non-trivial reversible part that produces the
above equation has the following form

\begin{equation}
\left(
\begin{array}{c} 
\dot{\bf R}_\mu\\
\\
\dot{\bf P}_\mu\\
\\
\dot{M}_\mu\\
\\
\dot{S}_\mu
\end{array}\right)
=\sum_\nu {\bf L}_{\mu\nu}
\left(
\begin{array}{c} -\sum_\gamma \frac{\partial {\cal V}_\gamma}{\partial {\bf R}_\nu} P_\gamma
\\
\\{\bf v}_\nu
\\
\\-\frac{{\bf v}_\nu^2}{2}+\mu_\nu 
\\
\\T_\nu
\end{array}\right),
\label{revgen}
\end{equation}
where the block ${\bf L}_{\mu\nu}$ has the structure
\begin{equation}
{\bf L}_{\mu\nu} = 
\left(
\begin{array}{ccccccccc}
 {\bf 0} && {\bf 1}\delta_{\mu\nu} && {\bf 0} && {\bf 0}  \\
\\
-{\bf 1}\delta_{\mu\nu} && {\bf \Lambda}_{\mu\nu} && {\bf \Delta}_{\mu\nu} && {\bf \Gamma}_{\mu\nu} \\
\\
{\bf 0} &&-{\bf \Delta}_{\nu\mu}&& 0 && 0 \\
\\
{\bf 0} &&-{\bf \Gamma}_{\nu\mu} && 0 && 0 \\
\end{array}\right).
\label{lij}
\end{equation}
The first row of ${\bf L}_{\mu\nu}$ ensures the equation of motion
(\ref{require}). The first column is fixed by antisymmetry of
$L$. Note that in order to have antisymmetry of ${\bf L}_{\mu\nu}$
(which, in turn, ensures energy conservation), it is necessary that
${\bf \Lambda}^T_{\mu\nu}=-{\bf \Lambda}_{\nu\mu}$. Performing the
matrix multiplication in Eqn. (\ref{revgen}), the reversible part of
the dynamics takes the form

\begin{eqnarray}
\dot{\bf R}_\mu &=& {\bf v}_\mu,
\nonumber\\
\dot{\bf P}_\mu &=& 
\sum_\nu \frac{\partial {\cal V}_\nu}{\partial {\bf R}_\mu} P_\nu
+\sum_\nu {\bf \Lambda}_{\mu\nu}\!\cdot\!{\bf v}_\nu 
\nonumber\\
&+&\sum_\nu{\bf \Delta}_{\mu\nu}\left(-\frac{{\bf v}_\nu^2}{2}+\mu_\nu\right)
+{\bf \Gamma}_{\mu\nu} T_\nu,
\nonumber\\
\dot{M}_\mu &=&- \sum_\nu {\bf \Delta}_{\nu\mu} {\bf v}_\nu,
\nonumber\\
\dot{S}_\mu &=&- \sum_\nu {\bf \Gamma}_{\nu\mu} {\bf v}_\nu.
\label{revdes}
\end{eqnarray}
We now develop the pressure term by using Eqn. (\ref{fin1}) of
appendix \ref{ap-vol}
\begin{eqnarray}
\sum_{\nu} \frac{\partial {\cal V}_\nu}{\partial {\bf R}_\mu} P_\nu
&=&\sum_{\nu\neq\mu} \frac{\partial {\cal V}_\nu}{\partial {\bf R}_\mu} (P_\nu-P_\mu)
\nonumber\\
&=&\sum_{\nu\neq\mu} 
\int \frac{d{\bf r}}{\sigma^2}\chi_\mu({\bf r})\chi_\nu({\bf r})({\bf r}-{\bf R}_\mu)(P_\mu-P_\nu)
\nonumber\\
&=&\sum_{\nu\neq\mu} A_{\mu\nu}{\bf e}_{\mu\nu}\frac{P_\mu+P_\nu}{2}
\nonumber\\
&&+\sum_{\nu\neq\mu} \frac{A_{\mu\nu}}{R_{\mu\nu}}{\bf c}_{\mu\nu}(P_\mu-P_\nu),
\label{volpres}
\end{eqnarray}
where use has been made of the property (\ref{green}).

The momentum equation becomes
\begin{eqnarray}
\dot{{\bf P}}_\mu&=&
\sum_{\nu} A_{\mu\nu}{\bf e}_{\mu\nu}\frac{P_\mu+P_\nu}{2}
+\sum_\nu\left({\bf \Lambda}_{\mu\nu}\!\cdot\!{\bf v}_\nu-
{\bf \Delta}_{\mu\nu}\frac{{\bf v}_\nu^2}{2}\right)
\nonumber\\
&+&
\sum_\nu\left(
 \frac{A_{\mu\nu}}{R_{\mu\nu}}{\bf c}_{\mu\nu}(P_\mu-P_\nu)
+{\bf \Delta}_{\mu\nu}\mu_\nu +{\bf \Gamma}_{\mu\nu}T_\nu\right).
\label{m2}
\end{eqnarray}

Now, the basic question to answer is, What forms for ${\bf
\Lambda}_{\mu\nu}$, ${\bf \Delta}_{\mu\nu}$ and ${\bf
\Gamma}_{\mu\nu}$ should we use in order to consider
Eqns. (\ref{revdes}) as a discrete version of hydrodynamics?  In what
follows we will propose forms for these quantities in such a way that
Eqns. (\ref{revdes}) and (\ref{mass1}), (\ref{mombal2}), (\ref{cdent})
coincide as much as possible.

The vectors ${\bf \Delta}_{\mu\nu}$ and ${\bf \Gamma}_{\mu\nu}$ are
easily identified by comparing the mass and entropy equations in
(\ref{fvap}) and (\ref{revdes}). The matrix ${\bf \Lambda}_{\mu\nu}$
is obtained by inspection from the comparison between the momentum
equation in (\ref{fvap}) and (\ref{revdes}). Our proposals are

\begin{eqnarray}
{\bf\Delta}_{\mu\nu} &=&
\frac{A_{\mu\nu}}{R_{\mu\nu}}\frac{\rho_\mu+\rho_\nu}{2}{\bf
c}_{\mu\nu} -\delta_{\mu\nu}\sum_\sigma
\frac{A_{\mu\sigma}}{R_{\mu\sigma}}\frac{\rho_\mu+\rho_\sigma}{2}{\bf
c}_{\mu\sigma},
\nonumber\\
{\bf\Gamma}_{\mu\nu} &=&
\frac{A_{\mu\nu}}{R_{\mu\nu}}\frac{s_\mu+s_\nu}{2}{\bf
c}_{\mu\nu} -\delta_{\mu\nu}\sum_\sigma
\frac{A_{\mu\sigma}}{R_{\mu\sigma}}\frac{s_\mu+s_\sigma}{2}{\bf
c}_{\mu\sigma},
\nonumber\\
{\bf \Lambda}_{\mu\nu}&=&
\frac{A_{\mu\nu}}{R_{\mu\nu}}\frac{\rho_\mu+\rho_\nu}{2}
\left[\frac{{\bf v}_\mu+{\bf v}_\nu}{2}{\bf c}_{\mu\nu}
-{\bf c}_{\mu\nu}\frac{{\bf v}_\mu+{\bf v}_\nu}{2}\right]
\nonumber\\
&-&\delta_{\mu\nu}\sum_\sigma
\frac{A_{\mu\sigma}}{R_{\mu\sigma}}\frac{\rho_\mu+\rho_\sigma}{2}
\left[\frac{{\bf v}_\mu+{\bf v}_\sigma}{2}{\bf c}_{\mu\sigma}
-{\bf c}_{\mu\sigma}\frac{{\bf v}_\mu+{\bf v}_\sigma}{2}\right].
\nonumber\\
&&\label{proposal}
\end{eqnarray}
Note that ${\bf \Lambda}_{\mu\nu}^T=-{\bf \Lambda}_{\mu\nu}
={\bf \Lambda}_{\nu\mu}$ and,
therefore, the antisymmetry of $L$ is ensured.  Note also that
$\sum_\nu{\bf \Gamma}_{\mu\nu}=0$ and, therefore, the degeneracy
condition $L\!\cdot\!\partial S/\partial x=0$ is satisfied.

By substitution of these forms into the mass and entropy equations
in (\ref{revdes}) one obtains the mass and entropy equations
obtained in the finite volume method Eqns. (\ref{fvap}). Substitution
into the momentum equation (\ref{m2}) leads to the finite volume momentum
equation (\ref{fvap}), with an additional term which is

\begin{eqnarray}
&&
\sum_\nu
 \frac{A_{\mu\nu}}{R_{\mu\nu}}{\bf c}_{\mu\nu}
\left((P_\mu-P_\nu)-\frac{\rho_\mu+\rho_\nu}{2}(\mu_\mu -\mu_\nu )\right.
\nonumber\\
&&-\left.\frac{s_\mu+s_\nu}{2}(T_\mu- T_\nu)\right).
\label{gd}
\end{eqnarray}
This term is strongly reminiscent of the Gibbs-Duhem relation
which, in differential forms is $dP-\rho d\mu-sdT=0$. For this
reason, we expect that this term, although not exactly zero, will
be very small.

In summary, the proposed GENERIC equations for the reversible part of the
evolution of the variables ${\bf R}_\mu,{\bf P}_\mu,M_\mu,S_\mu$ are

\begin{eqnarray}
\dot{\bf R}_\mu &=& {\bf v}_\mu,
\nonumber\\
\dot{\bf P}_\mu &=&\sum_\nu A_{\mu\nu}{\bf e}_{\mu\nu}\frac{P_{\mu}+P_{\nu}}{2}
\nonumber\\
&+&\sum_\nu  \frac{A_{\mu\nu}}{R_{\mu\nu}}\frac{\rho_{\mu}+\rho_{\nu}}{2}
\frac{{\bf v}_{\mu}+{\bf v}_{\nu}}{2}
{\bf c}_{\mu\nu}\!\cdot\!({\bf v}_\mu-{\bf v}_\nu)
\nonumber\\
&+&\sum_\nu
 \frac{A_{\mu\nu}}{R_{\mu\nu}}{\bf c}_{\mu\nu}
\left((P_\mu-P_\nu)-\frac{\rho_\mu+\rho_\nu}{2}(\mu_\mu -\mu_\nu )\right.
\nonumber\\
&&-\left.\frac{s_\mu+s_\nu}{2}(T_\mu- T_\nu)\right),
\nonumber\\
\dot{M}_\mu &=&\sum_\nu\frac{A_{\mu\nu}}{R_{\mu\nu}}
\frac{\rho_{\mu}+\rho_{\nu}}{2}
{\bf c}_{\mu\nu}\!\cdot\!({\bf v}_\mu-{\bf v}_\nu),
\nonumber\\
\dot{S}_\mu &=&\sum_\nu\frac{A_{\mu\nu}}{R_{\mu\nu}}
\frac{s_{\mu}+s_{\nu}}{2}
{\bf c}_{\mu\nu}\!\cdot\!({\bf v}_\mu-{\bf v}_\nu).
\label{REVER}
\end{eqnarray}
Here, ${\bf v}_\mu = {\bf P}_\mu/M_\mu$, $\rho_\mu = M_\mu/{\cal
V}_\mu$ and $s_\mu = S_\mu/{\cal V}_\mu$. These GENERIC equations for
the reversible part of the dynamics Eqns. (\ref{revdes}) are identical
(except for the small Gibbs-Duhem term) to the finite volume
discretization of the continuum equations of inviscid hydrodynamics
Eqns. (\ref{fvap}) and can, therefore, be considered as a proper
discretization of the continuum equations of hydrodynamics.  Total
mass, momentum, and energy are conserved exactly and the total entropy
does not change in time due to this reversible motion.

\section{Irreversible dynamics}
\label{irreversible}

In this section we consider the irreversible part of the dynamics
$M\!\cdot\!\partial S/\partial x$. We will postulate the random terms
$d\tilde{x}$ for the discrete equations and will construct, through
the fluctuation-dissipation theorem (\ref{F-D}), the matrix $M$ and
the irreversible part of the dynamics. If we guess correctly the
random terms, the resulting discrete equations should consistently
produce the correct dissipative part of the dynamics.

Thermal fluctuations are introduced into the continuum equations of
hydrodynamics through the divergence of a random stress tensor and a
random heat flux \cite{Landau59},\cite{Espanol-PA98}. In principle,
one could think about a random {\em mass} flux that would, according
to the fluctuation-dissipation theorem, produce an irreversible term
in the mass balance equation. Such term is absent in simple fluids but
not in mixtures (it produces the diffusion terms). The reason why
there is no such a random mass flux in the continuous description of a
simple fluid can be understood with the method of projection
operators. As it is well-known, the method produces Green-Kubo
expressions for the transport coefficients. This Green-Kubo forms
involve the correlation of the {\em projected} currents.  Since, in
the continuum case, the time derivative of the microscopic density
field is precisely (minus) the divergence of the microscopic momentum
density field (which is itself a relevant variable), it turns out that
the projected current vanish exactly and there is no Green-Kubo
transport coefficient in the mass equation. The situation is different
when one considers discrete variables. The discrete variables are the
mass, momentum and internal energy of the Voronoi cells as functions
of the position and momenta of the fluid molecules. Even though a
projection operator derivation of the equations of motion for these
variables is extremely involved, it is possible to show that the
projected mass current does not strictly vanish. This amounts to
accept that the mass in a given cell fluctuates not only due to the
indirect action of the random stress and heat flux but also through
the direct effect of a random mass flux. For the time being and for
the sake of simplicity, however, we assume that this random mass flux
can be neglected. In this case, the noise term in the equation of
motion (\ref{sde1}) has the form $d\tilde{x}^T\rightarrow
\left( {\bf 0},d\tilde{\bf P}_\mu,0,d\tilde{S}_\mu\right)$.

In the following subsections we consider two different implementations
of the noise terms. The first one, through a random stress tensor and
random heat flux, can be considered as the natural way of constructing
discrete equations that, in the continuum limit, converge towards the
equations of continuum hydrodynamics. The second implementation is a
cartoon of the first one and leads to the Dissipative Particle
Dynamics algorithm.

\subsection{Finite volume hydrodynamic}
\label{fin-vol}
By analogy with the continuum fluctuating hydrodynamics we construct
the random terms $d\tilde{\bf P}_\mu,d\tilde{S}_\mu $ as the discrete
divergences of a random flux

\begin{eqnarray}
d\tilde{\bf P}_\mu &=& 
\sum_\nu{\bf \Omega}_{\mu\nu}\!\cdot\!d\tilde{\mbox{\boldmath $\sigma$}}_\nu,
\nonumber\\
d\tilde{S}_\mu &=& 
\frac{1}{T_\mu}\sum_\nu{\bf \Omega}_{\mu\nu}\!\cdot\!d\tilde{\bf J}^q_\nu
-\frac{1}{T_\mu}d\tilde{\mbox{\boldmath $\sigma$}}_\mu
:\sum_\nu{\bf \Omega}_{\nu\mu}{\bf v}_\nu^T.
\label{ran2}
\end{eqnarray}
We will select the form (\ref{aeom}) for ${\bf \Omega}_{\mu\nu}$,
but the particular form is not important for the time being.
The random stress $d\tilde{\mbox{\boldmath $\sigma$}}_\mu
$ and random heat flux $d\tilde{\bf J}^q_\mu $ are defined by

\begin{eqnarray}
d\tilde{\mbox{\boldmath $\sigma$}}_\mu &=&
a_\mu\overline{d{\bf W}}^{S}_{\mu}
+b_\mu\frac{{\bf 1}}{D}{\rm tr}[d{\bf W}_{\mu}],
\nonumber\\
d\tilde{\bf J}^q_\mu &=& c_\mu d{\bf V}_\mu.
\label{ran3}
\end{eqnarray}
The coefficients $a_\mu,b_\mu,c_\mu$ are given by
\begin{eqnarray}
a_\mu &=& \left(4k_BT_\mu\frac{\eta_\mu}{{\cal V}_\mu}\right)^{1/2},
\nonumber\\
b_\mu &=& \left(2Dk_BT_\mu\frac{\zeta_\mu}{{\cal V}_\mu}\right)^{1/2},
\nonumber\\
c_\mu &=& T_\mu\left(2k_B\frac{\kappa_\mu}{{\cal V}_\mu}\right)^{1/2}.
\label{abc}
\end{eqnarray}
Here, $D$ is the physical dimension of space, $\eta_\mu$ is the shear
viscosity, $\zeta_\mu$ is the bulk viscosity, and $\kappa_\mu$ is the
thermal conductivity. These transport coefficient might depend in
general on the thermodynamic state of the fluid particle $\mu$. The
particular form of the coefficients in Eqn. (\ref{abc}) might appear
somehow arbitrary. Actually, it is only after writing up the final
discrete equations and comparing them with the finite volume equations
(\ref{irrmom}), (\ref{irrent}) that we could extract the particular
functional form of these coefficients.

The traceless symmetric
random matrix $\overline{d{\bf W}}^{S}_\mu$ is given by
\begin{equation}
\overline{d{\bf W}}^S_{\mu}=
\frac{1}{2}\left[d{\bf W}_{\mu}+d{\bf W}^T_{\mu}\right]
-\frac{1}{D}{\rm tr}[d{\bf W}_{\mu}]{\bf 1}.
\label{decomp}
\end{equation}
$d{\bf W}_{\mu}$ is a matrix
of independent Wiener increments.  The vector $d{\bf V}_\mu$ is also a
vector of independent Wiener increments.  They satisfy the It\^o
mnemotechnical rules
\begin{eqnarray}
d{\bf W}^{ii'}_{\mu}d{\bf W}^{jj'}_{\nu}&=&
\delta_{\mu\nu}\delta_{ij}\delta_{i'j'}dt,
\nonumber\\
d{\bf V}^{i}_\mu d{\bf V}^{j}_\nu&=&\delta_{\mu\nu}\delta_{ij}dt,
\nonumber\\
d{\bf V}^{i}_\mu d{\bf W}^{jj'}_\nu&=&0,
\label{ran3b}
\end{eqnarray}
where latin indices denote tensorial components.
Note that the postulated forms for $d\tilde{{\bf P}}_\mu,d\tilde{S}_\mu$ in
Eqn. (\ref{ran2}) satisfy 
\begin{eqnarray}
\sum_\mu{\bf
v}_\mu\!\cdot\!d\tilde{{\bf P}}_\mu+T_\mu d\tilde{S}_\mu&=&0,
\nonumber\\
\sum_\mu d\tilde{{\bf P}}_\mu&=&0,
\label{de0}
\end{eqnarray}
 and, therefore, Eqns. (\ref{consinv}) are
satisfied. This means that the postulated noise terms conserve
momentum and energy exactly. It is now a matter of algebra to
construct the dyadic $d\tilde{x}d\tilde{x}^T$ and from
Eqn. (\ref{F-D}) extract the matrix $M$. 
The procedure is rather cumbersome but standard.

Once $M$ is constructed, the terms $M\!\cdot\!\partial S/\partial x$
in the equation of motion (\ref{gen1}) can be written up. By
assuming that the transport coefficients do not depend on the entropy
density (but they might depend on the mass density), the resulting
equations of motion are

\begin{eqnarray}
\left.d{\bf P}_\mu\right|_{\rm irr} &=& 
\sum_\nu{\bf \Omega}_{\mu\nu}\!\cdot\!
(\mbox{\boldmath $\Pi$}_\nu +\Pi_\nu{\bf 1})dt 
+ d\tilde{\bf P}_\mu,
\nonumber\\
\left.T_\mu d{S}_\mu \right|_{\rm irr}&=& \left(1-\frac{k_B}{C_{Vi}}\right)
\left[\frac{2\eta_\mu}{{\cal V}_\mu}
\overline{\bf G}_\mu:\overline{\bf G}_\mu 
+\frac{\zeta_\mu}{{\cal V}_\mu} D_\mu^2 \right]dt
\nonumber\\
&+&\sum_\nu{\bf \Omega}_{\mu\nu} \!\cdot\!{\bf J}^q_\nu dt
\nonumber\\
&-&\frac{k_B}{T_\mu C_\mu}\sum_\nu{\bf \Omega}_{\mu\nu}^2
\frac{\kappa_\nu}{{\cal V}_\nu}T_\nu^2 dt
\nonumber\\
&-&\frac{k_BT_\mu}{m}
\left(\left(\frac{D^2+D-2}{2D}\right)
\frac{2\eta_\mu}{{\cal V}_\mu}+\frac{\zeta_\mu}{{\cal V}_\mu}\right)
\sum_\nu{\bf \Omega}_{\nu\mu}^2dt
\nonumber\\
&+&T_\mu d\tilde{S}_\mu.
\label{eqmot}
\end{eqnarray}
In these equations, we have introduced the same quantities
as in Eqn. (\ref{grad}). The heat flux ${\bf J}^q_\mu$ is defined by
\begin{equation}
{\bf J}^q_\mu=-T_\mu^2\frac{\kappa_\mu}{{\cal V}_\mu}\sum_\nu{\bf \Omega}_{\nu\mu}
\frac{1}{T_\nu}\left(1-\frac{k_B}{C_\nu}\right).
\label{disc}
\end{equation}
Finally, the heat capacity at constant volume of particle $\mu$ is defined by
\begin{equation}
C_\mu=T_\mu\left(\frac{\partial T_\mu}{\partial s_\mu}\right)_{\cal V}^{-1}.
\label{cv}
\end{equation}

We observe that, quite remarkably, the above equations are in the
limit $k_B\rightarrow0$ identical to the irreversible part of the
particular finite volume discretization of the continuum hydrodynamic
equations presented in section \ref{l-finite-volume}.  We have,
therefore, shown that these equations (\ref{eqmot}) are a proper
discretization of the irreversible part of hydrodynamics with thermal
noise included consistently.

By collecting the reversible part $L\!\cdot\!\partial E/\partial x$ in
Eqn. (\ref{revgen}) and the irreversible part $M\!\cdot\!\partial
S/\partial x$ in Eqn. (\ref{eqmot}), the final equations of motion for
the discrete hydrodynamic variables could be finally written.

\subsection{Irreversible part of DPD}
\label{dpd}

In this section we show how, by postulating a different form for the
noise $d\tilde{x}$, one can obtain an irreversible part which is
closely related to the irreversible part of the Dissipative Particle
Dynamics model. The Dissipative Particle Dynamics model that we
present here, then, is a natural generalization of the classical DPD
model \cite{Hoogerbrugge92} in which not only an internal energy (or
entropy) variable is included as in Refs. \cite{Bonet97} but also a
mass density variable is introduced. From the GENERIC point of view,
this DPD model differs from the finite volume hydrodynamics in section
\ref{fin-vol} model only in the form of the dissipative and random
terms.

Instead of (\ref{ran2}) the postulated structure of the random terms
is $d\tilde{x}\rightarrow({\bf 0}, d\tilde{\bf
P}_\mu,0,d\tilde{S}_\mu)$ with the following definitions

\begin{eqnarray}
d\tilde{\bf P}_\mu&=&\sum_\nu {\bf B}_{\mu\nu} dW_{\mu\nu},
\nonumber\\
d\tilde{S}_\mu &=& -\frac{1}{2T_\mu}\sum_\nu {\bf B}_{\mu\nu}\!\cdot\!{\bf v}_{\mu\nu}dW_{\mu\nu}
+\frac{1}{T_\mu}\sum_\nu A_{\mu\nu}dV_{\mu\nu},
\label{noisedpd}
\end{eqnarray}
where ${\bf B}_{\mu\nu}=-{\bf B}_{\nu\mu}$, ${\bf B}_{\mu\mu}=0$, and
$A_{\mu\nu}=A_{\nu\mu}$ are suitable functions of position and,
perhaps, other state variables.  The independent Wiener processes
satisfy $dW_{\mu\nu}=dW_{\nu\mu}$ and $dV_{\mu\nu}=-dV_{\nu\mu}$ and
the following It\^o mnemotechnical rules
\begin{eqnarray}
dW_{\mu\mu'}dW_{\nu\nu'} &=& [\delta_{\mu\nu}\delta_{\mu'\nu'}+\delta_{\mu\nu'}\delta_{\mu'\nu}]dt,
\nonumber\\
dV_{\mu\mu'}dV_{\nu\nu'} &=& [\delta_{\mu\nu}\delta_{\mu'\nu'}-\delta_{\mu\nu'}\delta_{\mu'\nu}]dt,
\nonumber\\
dW_{\mu\mu'}dV_{\nu\nu'} &=& 0.
\label{wie}
\end{eqnarray}
Note that the noise terms in Eqn. (\ref{noisedpd}) satisfy the requirements
(\ref{consinv}) which take the form
\begin{eqnarray}
\sum_\mu d\tilde{\bf P}_\mu&=&0,
\nonumber\\
\sum {\bf v}_\mu\!\cdot\!d\tilde{\bf P}_\mu+T_\mu d\tilde{S}_\mu&=&0.
\label{dpds}
\end{eqnarray}
The first equation ensures momentum conservation while the second
equation ensures energy conservation. Note that the
random force $d\tilde{\bf P}_\mu$ provides ``kicks'' to the particles
along the line joining the particles, and it satisfies Newton's third
law. The first term of the random term $d\tilde{S}_\mu$ is suggested by
the last equation in Eqn. (\ref{dpds}) whereas the last
term is dictated by our wish of modeling heat conduction
\cite{Bonet97}.

We note that the DPD noise terms (\ref{noisedpd}) can be viewed as a
cartoon of the noise terms (\ref{ran2}) in the finite Voronoi volume
model. For this reason, we will assume the following form for
$A_{\mu\nu},{\bf B}_{\mu\nu}$ to remain as close as possible to
(\ref{ran2}) while retaining the correct symmetries for
$A_{\mu\nu},{\bf B}_{\mu\nu}$,

\begin{eqnarray}
A_{\mu\nu} &=& \left(2k_BT_\mu T_\nu\frac{\kappa}{\overline{\cal V}}
\right)^{1/2}|{\bf \Omega}_{\mu\nu}|,
\nonumber\\
{\bf B}_{\mu\nu}&=&\left(2k_B\frac{T_\mu T_\nu}{T_\mu+T_\nu}
\frac{\gamma}{\overline{\cal V}}
\right)^{1/2}\mbox{\boldmath $\Omega$}_{\mu\nu}.
\label{bij}
\end{eqnarray}
We have introduced $\kappa$ as a parameter with dimensions of a
thermal conductivity and a coefficient $\gamma$ with dimensions of a
viscosity. This DPD model has only a single viscosity, instead of two
viscosities (shear and bulk) appearing in the finite volume model of
the previous subsection. We denote with ${\overline {\cal V}}=V_T/M$
the average volume per particle.

It is now a matter of simple algebra to construct the irreversible
matrix $M$ for the DPD algorithm with the
fluctuation-dissipation theorem (\ref{F-D}). The final irreversible
part of the equations of motion are
\begin{eqnarray}
\left. d{\bf P}_\mu\right|_{\rm irr} &=&
-\sum_\nu\frac{\gamma_{\mu\nu}}{\overline{\cal V}}
({\bf v}_{\mu\nu}\!\cdot\!{\bf \Omega}_{\mu\nu}){\bf \Omega}_{\mu\nu}dt
+ d\tilde{\bf P}_\mu,
\nonumber\\
\left.T_\mu d{S}_\mu\right|_{\rm irr} &=& 
\sum_\nu\frac{\gamma_{\mu\nu}}{2\overline{\cal V}}
({\bf \Omega}_{\mu\nu}\!\cdot\!{\bf v}_{\mu\nu})^2dt
\nonumber\\
&+&\frac{\kappa}{\overline{\cal V}}
\sum_\nu {\bf \Omega}^2_{\mu\nu}(T_\nu-T_\mu)dt
\nonumber\\
&-&\frac{\gamma}{2\overline{\cal V}}
\sum_\nu\frac{T_\nu}{T_\mu+T_\nu}\frac{k_B}{C_\mu}
({\bf \Omega}_{\mu\nu}\!\cdot\!{\bf v}_{\mu\nu})^2dt
\nonumber\\
&-&\frac{2\gamma k_B T_\mu}{m\overline{\cal V}}
\sum_\nu\frac{T_\nu}{T_\mu+T_\nu}{\bf \Omega}_{\mu\nu}^2 dt
\nonumber\\
&-&\frac{k_B}{C_{\mu}}\frac{\kappa}{\overline{\cal V}}
\sum_{\nu\neq \mu}{\bf \Omega}^2_{\mu\nu}T_\nu dt
+T_\mu d\tilde{S}_\mu.
\label{eqmotdpd}
\end{eqnarray}
we have defined the pair viscosity as
\begin{equation}
\gamma_{\mu\nu}=\gamma
\left(1-
\frac{T_\mu T_\nu}{(T_\mu+T_\nu)^2}
\left(\frac{k_B}{C_\mu}+\frac{k_B}{C_\nu}\right)\right).
\label{gij}
\end{equation}
We discuss each term of these equations now. The momentum of
a particle changes irreversibly due to a friction force that
depends on the velocity differences between particles and
due to a random noise. This is the conventional form of
the DPD forces except for the fact that the friction coefficient
depends on the temperatures of the particles (although with a 
small prefactor of order $k_B/C_\mu$). 
Concerning the equation for the evolution of the entropy, the first
term models the process of viscous heating, the fact that the motion
of the particles creates an internal friction that increases the
internal energy of the particles. This term is dictated basically by
the requirement of energy conservation. The second term takes into
account the process of heat conduction and it can be understood as a
simple discretization of the heat conduction equation. Note again that
this form of the conduction term is physically more reasonable that
the expressions given in \cite{Bonet97}. The next three terms
are proportional to $k_B$ and ensure the exact conservation of energy,
as can be seen by explicitly computing $dE=\frac{\partial E}{\partial x}dx
+\frac{1}{2}\frac{\partial^2 E}{\partial x\partial x}d\tilde{x}d\tilde{x}$.

At this point, we would like to make a close comparison between the
model presented in Ref. \cite{Flekkoy99} and the one presented in this
paper. In Ref. \cite{Flekkoy99} the variables used to describe the
system of fluid particles are the positions, momentum and energy of
the particles. The mass is assumed to be a constant. We will focuse on
the momentum equation because in order to compare the energy equation
in both works we should transform our system of variables with the
entropy to a system with the energy as independent variable. The
momentum equation in \cite{Flekkoy99} is (changing their notation
to our notation)
\begin{eqnarray}
\dot{\bf P}_\mu&=&
-\sum_\nu A_{\mu\nu}\left\{
\frac{P_\mu-P_\nu}{2}{\bf e}_{\mu\nu}\right.
\nonumber\\
&-&\left.\frac{\eta}{R_{\mu\nu}}[{\bf v}_{\mu\nu}+
({\bf v}_{\mu\nu}\!\cdot\!{\bf e}_{\mu\nu}){\bf e}_{\mu\nu}]\right\}
+M_\mu {\bf g}+\tilde{\bf F}_\mu
\label{mofle}
\end{eqnarray}
where ${\bf g}$ is a body force like gravity and $\tilde{\bf F}_\mu$
is a stochastic force with a form fixed by the fluctuation-dissipation
theorem.  We note that the reversible part of this equation, the
pressure term, is identical to ours except for the small Gibbs-Duhem
term required for thermodynamic consistency when the mass and entropy
are not conserved by the reversible part of the dynamics as happens in
our case. The pressure difference instead of the sum is used in
Eqn. (\ref{mofle}), but both are equivalent in view of
Eqn. (\ref{diverthe}). The irreversible term proportional to the
``viscosity'' $\eta$, is of a form similar to the DPD irreversible
term $d{\bf P}_\mu|_{\rm irr}$ in Eqns. (\ref{eqmotdpd}), with ${\bf
\Omega}_{\mu\nu}$ given by Eqn. (\ref{aeom}). A term proportional to
${\bf v}_{\mu\nu}$, which breaks angular momentum conservation is
included \cite{Espanol98} but poses no conceptual problems. However,
we note that dimensionally $\eta$ is not a true viscosity, a volume
factor is missing. Actually, we expect that a simulation of
Eqns. (\ref{mofle}) with a fixed value of $\eta$ at different
resolutions (this is, the same external length scale, channel width
for example, and different number of fluid particles which have,
consequently, different typical volumes) would produce different
values for the actual viscosity of the fluid being modeled.

\section{The SPH model}
\label{sphmodel}
In this section we show that the Smoothed Particle Hydrodynamics
algorithm for an inviscid fluid is a particular case of the GENERIC
Eqs. (\ref{revdes}). For a viscous fluid, the SPH algorithm involves
the irreversible part (\ref{eqmot}) with no fluctuation effects,
i.e. $k_B=0$.

In SPH, a density variable $d_\mu$ associated to each fluid particle is
defined through
\begin{equation}
d_\mu =\sum_\nu W(|{\bf R}_\mu-{\bf R}_\nu|),
\label{density}
\end{equation}
where the weight function $ W(r)$ is a bell-shaped function with
finite support $h$ and normalized to unity. If particles are close
together, then the density defined by Eqn. (\ref{density}) is higher
in that region of space. From the density one can define a volume
associated to each fluid particle

\begin{equation}
{\cal V}_\mu({\bf R}_1,\ldots,{\bf R}_M) = \frac{1}{d_\mu}.
\label{volsph}
\end{equation}
The SPH algorithm for an inviscid fluid is obtained from Eqns. (\ref{revdes})
by using ${\bf
\Lambda}_{\mu\nu}$= ${\bf \Delta}_{\mu\nu}={\bf
\Gamma}_{\mu\nu}=0$. The derivative of the volume (\ref{volsph}) 
with respect to the positions of the particles is easily 
computed as
\begin{equation}
\frac{\partial {\cal V}_\nu}{\partial {\bf R}_\mu} =
-\frac{1}{d^2_\nu}(\mbox{\boldmath $\omega$}_{\mu\nu}+
\delta_{\mu\nu}\sum_k\mbox{\boldmath $\omega$}_{\mu\gamma}),
\label{omsph}
\end{equation}
where 
\begin{equation}
\mbox{\boldmath $\omega$}_{\mu\nu}= W'(r_{\mu\nu}){\bf e}_{\mu\nu}.
\end{equation}
The prime here denotes the derivative and ${\bf e}_{\mu\nu}$ is
the unit vector joining particles $i,j$. 

Substitution of Eqn. (\ref{omsph}) into the reversible
equations (\ref{revdes}) leads to

\begin{eqnarray}
\dot{\bf R}_\mu &=& {\bf v}_\mu,
\nonumber\\
M_\mu\dot{\bf v}_\mu &=& -\sum_\nu
\mbox{\boldmath $\omega$}_{\mu\nu}\left[
\frac{P_\nu}{d_\nu^2}+\frac{P_\mu}{d_\mu^2}\right],
\nonumber\\
\dot{M}_\mu &=& 0,
\nonumber\\
\dot{S}_\mu &=& 0.
\label{revsph}
\end{eqnarray}
This is the form of the discretization preferred by Monagan
\cite{Monaghan92}. It is remarkable that this form is forced solely by
the selection of the volume function in Eqn. (\ref{volsph}) and the
GENERIC structure. For a viscous fluid, the irreversible part of the
SPH equations are simply obtained from Eqns. (\ref{eqmot}) by using as
${\bf \Omega}_{\mu\nu}$ the function $-\partial {\cal V}_\nu/\partial
{\bf R}_\mu$ given in Eqn. (\ref{omsph}).

In the SPH equations (\ref{revsph}), the mass of the particles is a
constant. We will show in section \ref{equilibrium}, when we discuss
the equilibrium distribution function, that this constancy makes the
algorithm unsuitable for studying gas-liquid coexistence. On the other
hand, the SPH algorithm suffers from an unphysical feature: if the
pressures of the neighbors of particle $\mu$ are equal to the pressure
of this particle $\mu$, there is still a remnant force on this
particle, which is physically unacceptable.  A possible correction of
the above defect is as follows.

Instead of the volume defined as in (\ref{volsph}) we define the
volume as

\begin{equation}
{\cal V}_\mu = \frac{1}{d_\mu}\frac{{\cal V}_T}{\sum_k d_k^{-1}}
\label{volsphcorr}
\end{equation}
which satisfies $\sum_\mu {\cal V}_\mu ={\cal V}_T$. Note that the
correction factor ${\cal V}_T/\sum_k d_k^{-1}$ is expected to be close
to 1. Note also that now the volume of a given particle depends on the
coordinates of all the particles in the system. This breaks the local
definition of the volume of a particle in terms of the positions of
the neighbours.

Now, by using this expression for the corrected volume of a particle
 we obtain

\begin{eqnarray}
{\bf \Omega}_{\mu\nu} &=& \left[
\frac{{\cal V}_\nu}{d_\nu}(\delta_{\mu\nu}\sum_\gamma
\mbox{\boldmath $\omega$}_{\mu\gamma}+\mbox{\boldmath $\omega$}_{\mu\nu})
\right.
\nonumber\\
&-&\left.\frac{{\cal V}_\nu}{{\cal V}_T}\sum_\gamma
\left(\frac{{\cal V}_\nu}{d_\nu}+\frac{{\cal V}_\mu}{d_\mu}\right)
\mbox{\boldmath $\omega$}_{\mu\gamma}\right],
\end{eqnarray}
and, therefore,
\begin{equation}
M_\mu\dot{\bf v}_\mu = 
-\sum_\nu\left[(P_\mu -\overline{P})\frac{{\cal V}_\mu}{d_\mu}
+(P_\nu -\overline{P})\frac{{\cal V}_\nu}{d_\nu}\right]
\mbox{\boldmath $\omega$}_{\mu\nu},
\label{neq}
\end{equation}
where $\overline{P}$ is a sort of ``spatial average'' of the
pressures of the particles, i.e.,
\begin{equation}
\overline{P}=\frac{1}{{\cal V}_T}\sum_\mu {\cal V}_\mu P_\mu.
\label{spave}
\end{equation}
Equation (\ref{neq}) has the following good features: Total momentum
and total volume are conserved variables. If the pressure of all the
particles is exactly the same, there is no force on the particles. And
the following drawbacks: The force on particle $\mu$ depends on the
state of {\em all} the particles of the systems. Therefore, the change
in momentum of a bulk of fluid particles takes place non-locally, not
through the neighbourhood of this bulk. This non-local effect breaks
the local transport of momentum and therefore the macroscopic
hydrodynamic behaviour. We expect that this effect is small,
particularly when the range of the weight function is much larger than
the interparticle distance. The second drawback is that the remnant
force is zero only if the pressures of {\em all} the particles of the
system are equal. If in a local region of the system the pressures are
equal for the particles in that region but different to the pressure
of other regions, then there still exists a remnant force on the
particles of that region.

A final word on the {\em overlapping coefficient} is in order.  We
define the overlapping coefficient $s$ as the ratio of the range $h$
of the weight function $W(r)$ to the typical interparticle distance
$\lambda$ between fluid particles, this is $s=h/\lambda$. When $s\gg
1$ a given particle has typically many neigbhours. It is clear that if
the overlapping coefficient is much larger than 1 then, for
homogeneously distributed particles all the volumes of all the
particles will be typically the same.  In the limit $s$ large, then,
the pressures of the particles, which depend only on the volumes of
the particles, will be the same and the forces on the particles will
be negligible. In this limit, the equations of motion (\ref{revsph})
lose their sense. Actually, we expect that the volume defined through
(\ref{volsphcorr}) will have sense only if the overlapping coefficient
is slightly larger than 1.  However, the usual derivations of SPH by
convoluting the continuum equations of hydrodynamics with the weight
function $W(r)$ make sense only in the limit of large overlapping, in
such a way that the integrals can be reasonably approximated by
sums. Because of this inconsistency and the problems discussed above,
we prefer the formulation of discrete hydrodynamics in terms of finite
Voronoi volumes rather than in terms of weight functions as it is done
in SPH. From a computational point of view, we note that the large cpu
time required in order to update the Voronoi mesh (i.e. compute the
quantities $A_{\mu\nu},{\bf c}_{\mu\nu}$) may be compensated by the
fact that typically only six neighbours need to be considered in the
finite Voronoi volume simulation whereas 30-40 neighbours are needed
in a SPH simulation in 2D.

We would like to comment finally on the approach proposed in
Ref. \cite{Hietel00} where a finite volume approach similar to the one
presented here is advocated. However, these authors do not consider
the Voronoi limit $\sigma\rightarrow0$ but rather take a finite width
for the function $\Delta(r)$, providing an overlapping coefficient of
typically 1.4. Because they do not consider the Voronoi construction,
they encounter the difficulty of evaluating integrals similar to those
in Eqns. (\ref{area}). In one dimension, as is the case considered in
Ref. \cite{Hietel00}, this can be achieved with a numerical integration
method, but this becomes readily infeasible in higher dimensions.

\section{Equilibrium distribution function}
\label{equilibrium}
In this section we discuss the equilibrium distribution function
$\rho^{\rm eq}(x)$ corresponding to the equations of motion
(\ref{REVER}) plus the irreversible terms (\ref{eqmot}). Note that the
equilibrium distribution function of Eqns. (\ref{eqmot}) and
(\ref{eqmotdpd}) is the same, irrespective of the actual form of the
irreversible part of the dynamics. This is because both sets of
equations have the GENERIC structure. The GENERIC structure of the
equations of motion ensures that the equilibrium distribution function
for these variables is given by Einstein distribution function in the
presence of dynamical invariants, Eqn. (\ref{einst}). Because total
mass $M(x)$ total energy $E_0$ and momentum ${\bf P}_0$ are conserved
by the dynamics, the Einstein distribution function will be given by
\begin{eqnarray}
\rho^{\rm eq}(x)&=& 
\frac{1}{\Omega}
\delta(M(x)-{\cal M}_0)\delta(E(x)-E_0)\delta({\bf P}(x)-{\bf P}_0),
\nonumber\\
&\times &\exp\{k_B^{-1} S(x)\},
\label{ein}
\end{eqnarray}
where we have assumed that we know with absolute precision the values
of the total mass ${\cal M}_0$, energy $E_0$ and momentum ${\bf P}_0$
at the initial time. This is the situation in a computer
simulation. $\Omega$ is a normalization factor that ensures the
normalization of $\rho^{\rm eq}(x)$. A word is in order about the
total volume. We note that due to the definition of the volume in
Eqn. (\ref{volume}), {\em any} configuration of positions ${\bf
R}_\mu$ of the particles gives that the total volume has the same
value $\sum_\mu {\cal V}_\mu = {\cal V}_0$. Therefore, even though
total volume is conserved, it does not produce a restriction in the
form of a delta function in Eqn. (\ref{ein}).

\subsection{The most probable state}
The most probable state $x^*$ at equilibrium according to (\ref{ein})
is the one that maximizes the entropy $S(x)$ subject to the
constraints $M(x)=M_0$, $E(x)=E_0$, and ${\bf P}(x)= {\bf P}_0$. By
introducing Lagrange multipliers $\beta,\lambda$ and ${\bf V}$,
the most probable state $x^*$ is the state that maximizes
$k_B^{-1}S(x) - \beta (E(x) - \mbox{\boldmath $V$}\!\cdot\!{\bf
P}(x)-\lambda M(x))$ without constraints. By equating the partial
derivatives with respect to every variable to zero, one obtains the
following implicit equations for the most probable values $x^*=\{{\bf
R}_\mu^*,{\bf P}_\mu^*,M^*_\mu,S^*_\mu\}$
\begin{eqnarray}
\sum_\nu\frac{\partial {\cal V}_{\nu}}{\partial {\bf R}_\mu}P_\nu(x^*)&=&0,
\nonumber\\
\frac{{\bf P}^*_\mu}{M^*_\mu}&=&\mbox{\boldmath $V$},
\nonumber\\
\mu_\mu(x^*)&=&\lambda+\frac{3}{2}{\bf V}^2,
\nonumber\\
T_\mu(x^*)&=&\frac{1}{k_B\beta }.
\label{*}
\end{eqnarray}
The second equation states that in the most probable state all
particles move at the same velocity $\mbox{\boldmath $V$}$ which might
be set to zero without loss of generality.  The two last equations
state, then, that the temperature and chemical potential per unit mass
of all the fluid particles are equal at the most probable value of the
discrete hydrodynamic variables. This implies that the pressure is
also the same for all the fluid particles (in a simple fluid the
intensive parameters are not independent \cite{callen}). The first
equation is, therefore, trivially satisfied (because $\sum_\nu{\cal
V}_\nu={\rm ctn}$).

\subsection{Marginal distribution functions}

In this subsection we will integrate out the momentum variables in
Eqn. (\ref{ein}) in order to have more specific information about the
distribution of different variables at equilibrium. To this end, we
denote the state $x=(y,\{{\bf P}\})$ where $y=(\{{\bf
R}\},\{M\},\{S\})$ is the set of positions, masses, and entropies of
all particles. Note that the total entropy and internal energy in
Eqn. (\ref{ES}) do not depend on momentum variables.

By integrating the distribution function $\rho^{\rm eq}(x)$
over momenta we will have the probability $\rho^{\rm eq}(y)$ 
of a realization of  $y$

\begin{eqnarray}
\rho^{\rm eq}(y)&=&\exp \left\{S(y)/k_B\right\}
\frac{1}{\Omega_0}\delta\left(\sum^M_\mu M_\mu - {\cal M}_0\right)
\nonumber\\
&\times&
\int d{\bf P}_1\ldots d{\bf P}_M
\delta\left(\sum_\mu^M{\bf P}_\mu-{\bf P}_0\right)
\nonumber\\
&\times&
\delta\left(\sum_\mu^M
\left(\frac{{\bf P}^2_\mu}{2M_\mu}+{\cal E}_\mu(y)\right)-E_0\right)
\nonumber\\
&=&\exp \left\{S(y)/k_B\right\}
\frac{1}{\Omega_0}\delta\left(\sum^M_\mu M_\mu - {\cal M}_0\right)
\nonumber\\
&\times&
\prod_\mu^M(2M_\mu)^{D/2}\frac{\omega_{D(M-1)}}{2}
\left[E_0-\sum_\mu{\cal E}_\mu(y)\right]^{\frac{D(M-1)}{2}-1},
\nonumber\\
\label{pne}
\end{eqnarray}
where we have used Eqn. (\ref{11}) of appendix \ref{ap-mol}.

We find now a convenient approximation to Eqn. (\ref{pne}) by noting
that this probability is expected to be highly peaked around the most
probable state.  Therefore, for those values of the variables
${\cal E}_\mu(y)$ for which $\rho^{\rm eq}(y)$ is appreciably different
from zero we can approximate

\begin{eqnarray}
&&\left[E_0-\sum_\mu{\cal E}_\mu\right]^P
= \left[E_0-\sum_\mu{\cal E}_\mu^*\right]^P
\nonumber\\
&\times&\left[1+\frac{1}{P}\beta^*\sum_\mu({\cal E}_\mu^*-{\cal E}_\mu)\right]^P
\nonumber\\
&\approx&
\left[E_0-\sum_\mu{\cal E}_\mu^*\right]^P
\exp\{\beta^*\sum_\mu \left({\cal E}_\mu^*-{\cal E}_\mu\right)\}
\nonumber\\
&=&{\rm ctn.} \exp \{-\beta^*\sum_\mu{\cal E}_\mu\},
\label{approx}
\end{eqnarray}
where ${\cal E}_\mu^*$ is the most probable value of ${\cal E}_\mu$
and $P=D(M-1)/2-1$ is a very large number and we have introduced
\begin{equation}
\beta^*=\frac{D(M-1)/2-1}{E_0-\sum_\nu{\cal E}_\nu^*}\approx
\frac{DM/2}{E_0-\sum_\nu{\cal E}_\nu^*}.
\label{beta}
\end{equation}

Finally, we can write Eqn. (\ref{pne}) as
\begin{eqnarray}
\rho^{\rm eq}(y)
&=&\frac{1}{\Omega'_0}
\exp \left\{S(y)/k_B-\beta^*\sum^M_\mu{\cal E}_\mu(y)\right\}
\nonumber\\
&\times&\delta\left(\sum^M_\mu M_\mu - {\cal M}_0\right),
\label{pne2}
\end{eqnarray}
where $\Omega'_0$ is the corresponding normalization function.
In Eqn. (\ref{pne2}) we have neglected a term
$\sum_\mu \log M_\mu$ in front of $\sum_\mu{\cal E}_\mu (y)$. Note
that ${\cal E}_\mu$ is a first order function of its arguments
$S_\mu,M_\mu,{\cal V}_\mu$ and, therefore, it is of order $M_\mu$.

Changing back to the original notation we write Eqn. (\ref{pne2}) in
the form

\begin{eqnarray}
\rho^{\rm eq}(\{{\bf R},M,S\})
&=&\frac{1}{\Omega'_0}
\exp \left\{\sum^M_\mu S_\mu/k_B-\beta^*{\cal E}_\mu(M_\mu,S_\mu,{\cal V}_\mu)\right\}
\nonumber\\
&\times&\delta\left(\sum^M_\mu M_\mu - {\cal M}_0\right).
\label{pne3}
\end{eqnarray}
We see, therefore, that by integrating the momenta the
``microcanonical'' form Eqn. (\ref{ein}) becomes the ``canonical''
form (\ref{pne3}).  

We are interested now on the distribution function $P(M_\mu,S_\mu)$ that
the particular cell $\mu$ has the values $M_\mu,S_\mu$ for its mass
and entropy, irrespective of the values of the rest of the variables
in the system. We integrate (\ref{pne3}) over the variables
of all cells except $M_\mu,S_\mu$. 

\begin{eqnarray}
P(M_\mu,S_\mu) &=&
\frac{1}{\Omega'_0} \int d\{{\bf R}\}d^{(M-1)}\{M\}d^{(M-1)}\{S\}
\nonumber\\
&\times&
\exp \left\{\sum^M_{\nu} S_\nu/k_B
-\beta^*{\cal E}_\nu(M_\nu,S_\nu,{\cal V}_\nu)\right\}
\nonumber\\ 
&\times&
\delta\left(\sum^M_\mu M_\mu - {\cal M}_0\right)
\nonumber\\
&=&
\frac{{\cal V}_0^M}{\Omega'_0} \int d\{V\}d^{(M-1)}\{M\}d^{(M-1)}\{S\}
\nonumber\\
&\times&
\exp \left\{\sum^M_{\nu} S_\nu/k_B
-\beta^*{\cal E}_\nu(M_\nu,S_\nu, V_\nu)\right\}
\nonumber\\ 
&\times&
\delta\left(\sum^M_\mu M_\mu - {\cal M}_0\right)F(V_1,\cdots,V_M)
\nonumber\\ 
\label{pms}
\end{eqnarray}
where we have introduced the identity
\begin{equation}
\int d\{V\}\prod_\mu^M\delta(V_\mu-{\cal V}_\mu(\{{\bf R}\}))=1,
\label{iden}
\end{equation}
and the function
\begin{equation}
F(V_1,\ldots,V_M)=\frac{1}{V_T^M}
\int d\{{\bf R}\}\prod_\mu^M\delta(V_\mu-{\cal V}_\mu(\{{\bf R}\})).
\label{fvv}
\end{equation}
This function is the probability density that the the particles have
the particular distribution $V_1,\ldots,V_M$ of volumes provided that
the distribution function of the positions is uniform.  The
calculation of this function is difficult and we do not attempt to do
it. Rather, we will assume that this function is highly peaked around
$(\overline{V},\cdots, \overline{V})$ in such a way that all the cells
have approximately the same volume $\overline{V}={V}_T/M$. Under this
approximation, Eqn. (\ref{pms}) becomes

\begin{eqnarray}
P(M_\mu,S_\mu) 
&=&\frac{1}{\Omega'_0}
\exp \left\{S_\mu/k_B-\beta^*{\cal E}_\mu(M_\mu,S_\mu, \overline{V})\right\}
\nonumber\\
&\times&\Phi({\cal M}_0-M_\mu),
\label{pm3}
\end{eqnarray}
where we have introduced the function

\begin{eqnarray}
\Phi(X)&=&\int d^{(M-1)}\{M\}d^{(M-1)}\{S\}
\nonumber\\
&\times&
\exp \left\{\sum^M_{\nu\neq\mu} S_\nu/k_B
-\beta^*{\cal E}_\nu(M_\nu,S_\nu, \overline{V})\right\}
\nonumber\\ 
&\times&
\delta\left(\sum^M_{\nu\neq\mu} M_\nu - X\right).
\nonumber\\ 
\label{phix}
\end{eqnarray}
The functional form of $\Phi(X)$ is very well approximated
by an exponential. This can be seen by taking the derivative

\begin{eqnarray}
\Phi'(X)&=&-\int d^{(M-1)}\{M\}d^{(M-1)}\{S\}
\nonumber\\
&\times&
\exp \left\{\sum^M_{\nu\neq\mu} S_\nu/k_B
-\beta^*{\cal E}_\nu(M_\nu,S_\nu, \overline{V})\right\}
\nonumber\\ 
&\times&
\frac{\partial}{\partial M_\sigma}\delta\left(\sum^M_{\nu\neq\mu} M_\nu - X)\right)
\nonumber\\ 
&=&-\beta^*\int d^{(M-1)}\{M\}d^{(M-1)}\{S\}\mu(M_\sigma,S_\sigma, \overline{V})
\nonumber\\
&\times&
\exp \left\{\sum^M_{\nu\neq\mu} S_\nu/k_B
-\beta^*{\cal E}_\nu(M_\nu,S_\nu, \overline{V})\right\}
\nonumber\\ 
&\times&
\delta\left(\sum^M_{\nu\neq\mu} M_\nu - X)\right),
\nonumber\\ 
\label{phix2}
\end{eqnarray}
where we have integrated by parts in the second equality and $\sigma$
is the label of any cell except $\mu$.  The most probable value of the
integrand in Eqn. (\ref{phix2}) is the solution of
\begin{equation}
\mu(M_\nu,S_\nu,\overline{V}) = \overline{\lambda},
\end{equation}
where the chemical potential per unit mass equates $\overline{\lambda}$, a
suitable Lagrange multiplier that accounts for the mass conserving
delta function.  We expect that when the number of variables $M$ is
very large, the integrand in Eqn. (\ref{phix2}) becomes highly peaked
around this most probable value. In this case, we have

\begin{equation}
\Phi'(X)\approx -\beta^*\overline{\lambda} \Phi(X),
\end{equation}
and therefore $\Phi(X)$ is the exponential
$\exp\{-\beta^*\overline{\lambda}X\}$.

Returning back to Eqn. (\ref{pm3}) we have finally,
\begin{equation}
P(M_\mu,S_\mu) 
=\frac{1}{Z}
\exp \left\{S_\mu/k_B-\beta^*\left({\cal E}(M_\mu,S_\mu, \overline{V})
-\overline{\lambda} M_\mu\right)\right\},
\label{pmsap}
\end{equation}
where $Z$ is the appropriate normalization.  Note that the argument of
the exponential is the Gibbs free energy with fixed values for the
inverse temperature $\beta^*$ and chemical potential per unit
mass $\overline{\lambda}$.

\subsection{van der Waals fluid}
\label{f=0}
We will particularize the discussion of $P(M_\mu,S_\mu)$ by
considering the fundamental equation ${\cal E}(M_\mu,S_\mu,
\overline{V})$ for a van der Waals fluid.  First we note that the
internal energy is a first order function of its variables and,
therefore,
\begin{equation}
{\cal E}(M_\mu,S_\mu, \overline{V})
=\overline{V}\epsilon(n_\mu,s_\mu),
\label{for}
\end{equation}
where $n_\mu = M_\mu/(m_0\overline{V})$ is the number density,
$s_\mu=S_\mu/\overline{V}$ is the entropy density, and $\epsilon={\cal
E}/\overline{V}$ is the internal energy density of cell
$\mu$. Therefore, from Eqn. (\ref{pmsap}) we can obtain the
probability density that cell $\mu$ has the values $n,s$ for its
number density and entropy density. It is given by 
\begin{equation}
P(n,s) 
=\frac{1}{Z}
\exp \overline{V}\left\{s/k_B-\beta^*(\epsilon(n,s)
-\overline{\lambda} m_0 n)\right\}.
\label{pden}
\end{equation}

It is convenient to use reduced units for the van der Waals fluid (see
appendix \ref{ap-vdW} for details of notation). In reduced units the
distribution function becomes
\begin{equation}
P(\tilde{n},\tilde{s})=
\frac{1}{Z}
\exp \{\tilde{V}\left\{
\tilde{s}-\tilde{\beta}^{\rm ext}(\tilde{\epsilon}(\tilde{n},\tilde{s})-
\tilde{\mu}^{\rm ext}\tilde{n})\right\}
\label{ptil}
\end{equation}

The most probable value of this distribution function occurs
at $\tilde{n}^*,\tilde{s}^*$ which are solutions of the equations
\begin{eqnarray}
\frac{\partial\tilde{\epsilon}(\tilde{n}^*,\tilde{s}^*)}{\partial \tilde{n}}
&=& \tilde{\mu}(\tilde{n}^*,\tilde{s}^*) = \tilde{\mu}^{\rm ext}
\nonumber\\
\frac{\partial\tilde{\epsilon}(\tilde{n}^*,\tilde{s}^*)}{\partial \tilde{s}}
&=& \tilde{T}(\tilde{n}^*,\tilde{s}^*) =\frac{1}{\tilde{\beta}^{\rm ext}}
\label{**}
\end{eqnarray}
For fixed $\tilde{n}$, the relation between $\tilde{T}$ and 
$\tilde{s}$ is monotonic, whereas the chemical potential
has the form 
\begin{equation}
\tilde{\mu}=\tilde{T}
\left(\ln\left(\frac{\tilde{n}}{3-\tilde{n}}\right)+\frac{\tilde{n}}{3-\tilde{n}}\right)
-\frac{9}{4}\tilde{n} 
-\frac{D}{2}\tilde{T}\ln\left(\frac{\tilde {T}}{c}\right).
\label{mu}
\end{equation}

\begin{figure}[ht]
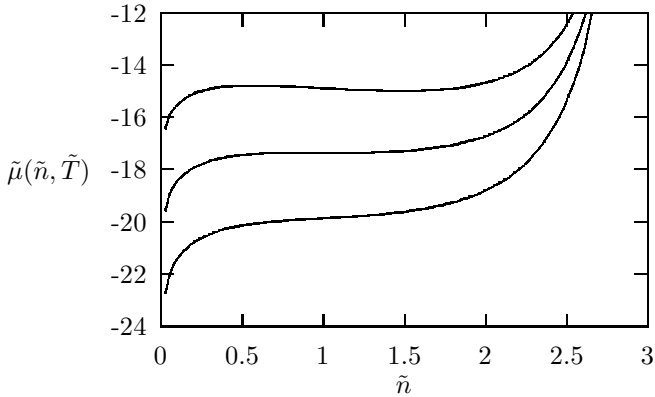

\begin{center}
\setlength{\unitlength}{0.240900pt}
\ifx\plotpoint\undefined\newsavebox{\plotpoint}\fi
\sbox{\plotpoint}{\rule[-0.200pt]{0.400pt}{0.400pt}}%


\caption{\label{figmu} Chemical potential
$\tilde{\mu}(\tilde{n},\tilde{T})$ as a function of $\tilde{n}$ for
different $\tilde{T}$. In descending order, $T=0.85,1.0, 1.15$. Observe
that for $\tilde{T}<1$ the equation
$\tilde{\mu}(\tilde{n},\tilde{T})=\tilde{\mu}^{\rm ext}$ might have
three solutions for $\tilde{n}$, depending on the actual value of
$\tilde{\mu}^{\rm ext}$. }
\end{center}
\end{figure}
\begin{figure}[ht]
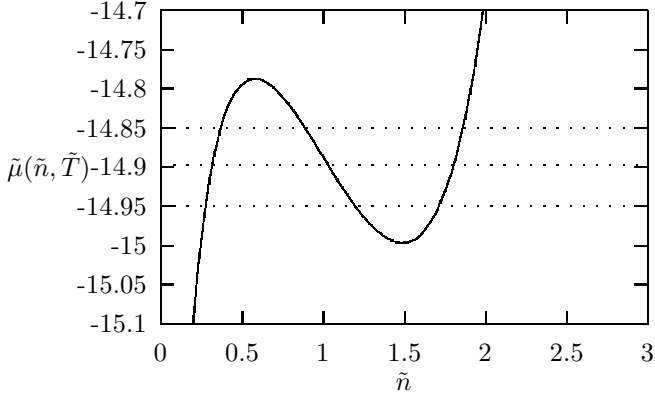

\begin{center}
\setlength{\unitlength}{0.240900pt}
\ifx\plotpoint\undefined\newsavebox{\plotpoint}\fi


\caption{\label{figmu1} 
Zoom of the previous figure for the isotherm $T=0.85$. The equal
area construction gives a 
value $\tilde{\mu}^{\rm ext}=-14.8971$ which produces two humps of
equal height in $P(\tilde{n},\tilde{T})$.}
\end{center}
\end{figure}
In Fig. \ref{figmu} we show the chemical potential for different
values of the temperature. We observe that for $\tilde{T}<1$ there are
three solutions for $n^*$ in the first equation in (\ref{**}).  This
means that depending on $\tilde{\beta}^{\rm ext}$ and
$\tilde{\mu}^{\rm ext}$, the distribution function
$P(\tilde{n},\tilde{s})$ can present a bimodal form.

Due to the particular functional form of the fundamental equation for
the van der Waals gas, it is more convenient to study the distribution
function $P(\tilde{n},\tilde{T})$ instead of $P(\tilde{n},\tilde{s})$.
This function is computed in the appendix \ref{ap-vdW}.

\begin{figure}[ht] 
\begin{center} 
\psfig{figure=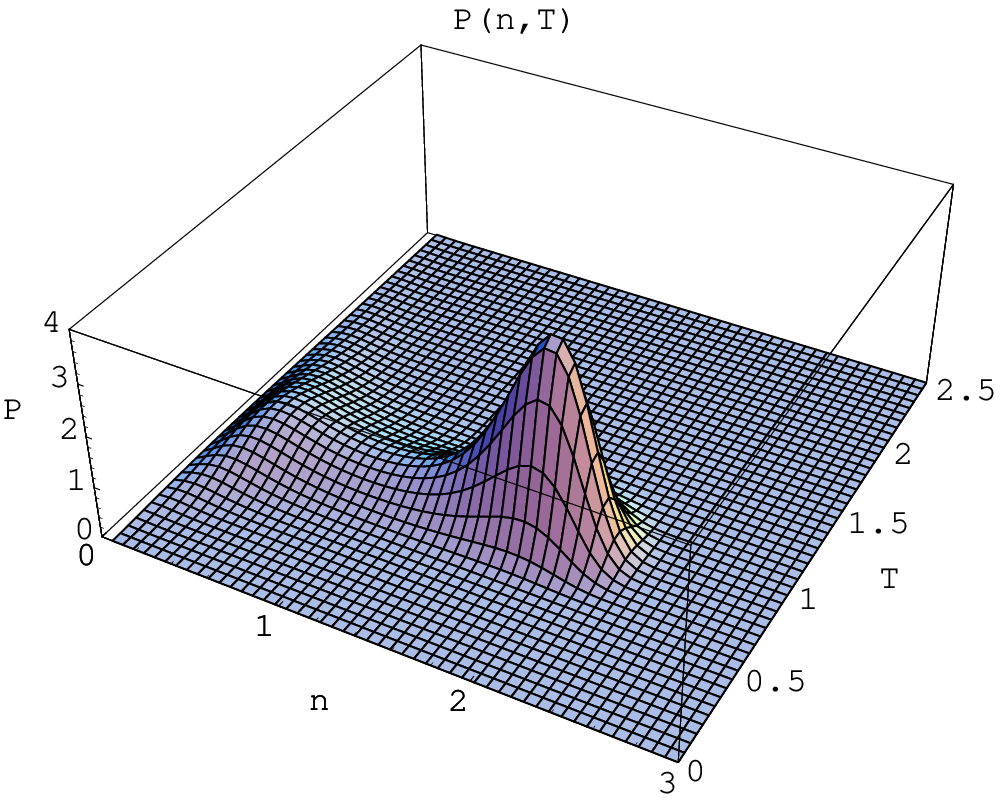,width=8cm,height=8cm}
\caption{\label{bimodal1} The distribution function $P(\tilde{n},\tilde{T})$
for an external temperature $1/\tilde{\beta}^{\rm ext}=0.85$. The external
chemical potential is $\tilde{\mu}^{\rm ext}=-14.85$. The typical volume
of the cells is $\tilde{V}=20$.}
\psfig{figure=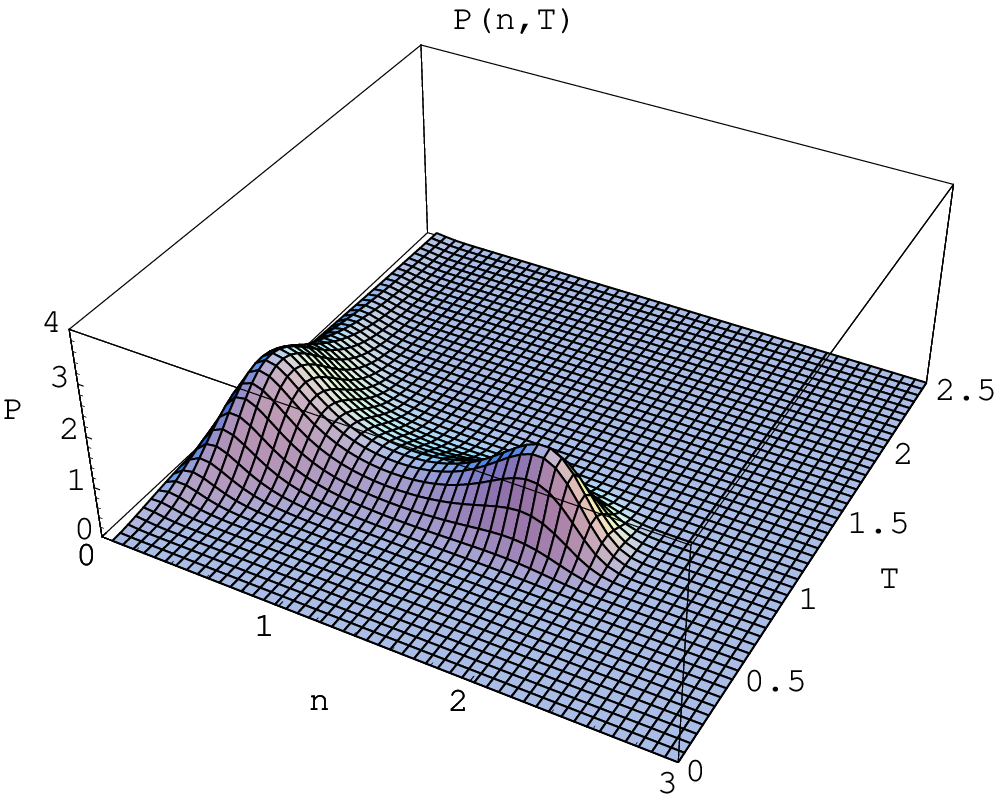,width=8cm,height=8cm}
\caption{\label{bimodal2} The distribution function $P(\tilde{n},\tilde{T})$
for an external temperature $1/\tilde{\beta}^{\rm ext}=0.85$. The external
chemical potential is $\tilde{\mu}^{\rm ext}=-14.8971$. The typical volume
of the cells is $\tilde{V}=20$. The two maxima have equal height.}
\psfig{figure=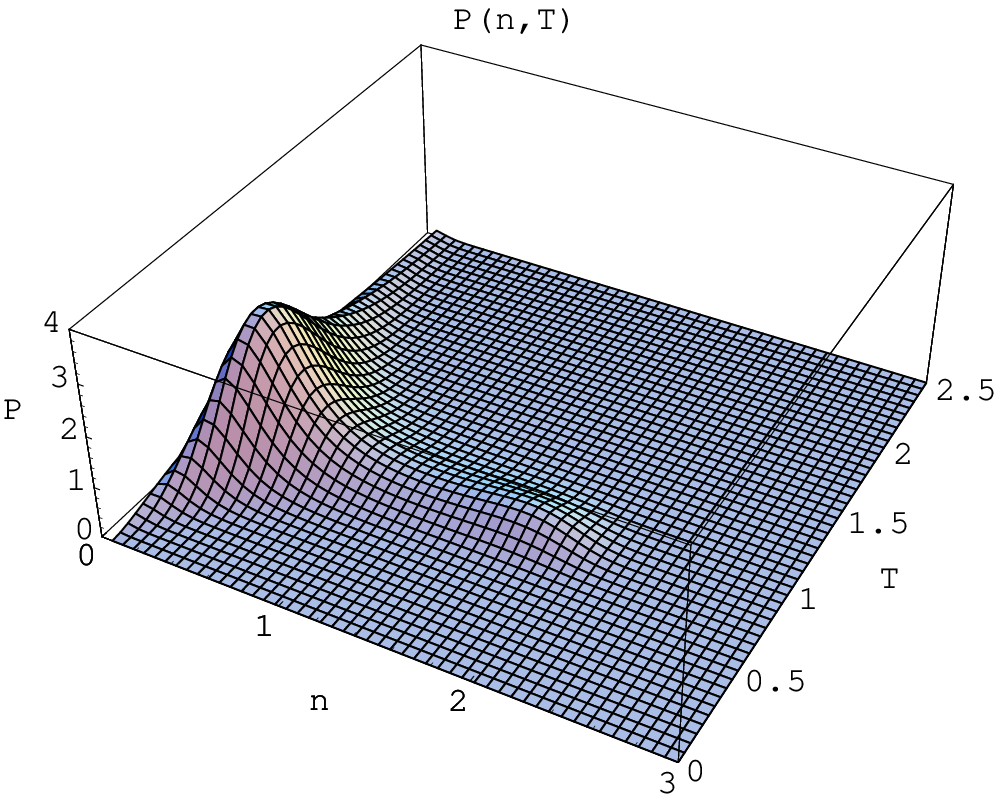,width=8cm,height=8cm}
\caption{\label{bimodal3} The distribution function $P(\tilde{n},\tilde{T})$
for an external temperature $1/\tilde{\beta}^{\rm ext}=0.85$. The external
chemical potential is $\tilde{\mu}^{\rm ext}=-14.95$. The typical volume
of the cells is $\tilde{V}=20$.}
\end{center}
\end{figure}

In Figs. \ref{bimodal1},\ref{bimodal2},\ref{bimodal3} we show the
distribution of number density and temperature
$P(\tilde{n},\tilde{T})$ for a value of the external temperature
$1/\tilde{\beta}^{\rm ext}=0.85$ (below the unit critical temperature)
and three different values of the external chemical potential
$\tilde{\mu}^{\rm ext}= -14.85,-14.8971,-14.95$. We observe the
presence of two humps at the same value of the temperature
$\tilde{T}=0.85$ and different values of the density $\tilde{n}$.  The
existence of a bimodal structure in the distribution function
$P(\tilde{n},\tilde{s})$ is a reflection of the gas-liquid transition,
because a cell of size $\tilde{V}$ can have a non-vanishing
probability of having two different values of the density.  The
external chemical potential $\tilde{\mu}^{\rm ext}$ controls the
relative magnitude of the two maxima. For the particular value
$\tilde{\mu}^{\rm ext}=-14.8971$, at this value of the external
temperature, the two heights are equal. This value is called the
chemical potential of coexistence and satisfies the equal area rule,
see Fig. \ref{figmu1}.

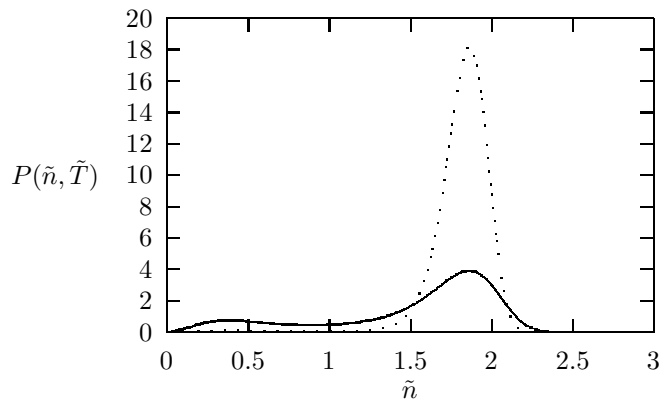
\begin{figure}[ht]
\begin{center}
\setlength{\unitlength}{0.240900pt}
\ifx\plotpoint\undefined\newsavebox{\plotpoint}\fi
\begin{picture}(1049,629)(0,0)
\font\gnuplot=cmr10 at 10pt
\gnuplot
\sbox{\plotpoint}{\rule[-0.200pt]{0.400pt}{0.400pt}}%
\put(220.0,113.0){\rule[-0.200pt]{184.288pt}{0.400pt}}
\put(220.0,113.0){\rule[-0.200pt]{0.400pt}{118.764pt}}
\put(220.0,113.0){\rule[-0.200pt]{4.818pt}{0.400pt}}
\put(198,113){\makebox(0,0)[r]{0}}
\put(965.0,113.0){\rule[-0.200pt]{4.818pt}{0.400pt}}
\put(220.0,162.0){\rule[-0.200pt]{4.818pt}{0.400pt}}
\put(198,162){\makebox(0,0)[r]{2}}
\put(965.0,162.0){\rule[-0.200pt]{4.818pt}{0.400pt}}
\put(220.0,212.0){\rule[-0.200pt]{4.818pt}{0.400pt}}
\put(198,212){\makebox(0,0)[r]{4}}
\put(965.0,212.0){\rule[-0.200pt]{4.818pt}{0.400pt}}
\put(220.0,261.0){\rule[-0.200pt]{4.818pt}{0.400pt}}
\put(198,261){\makebox(0,0)[r]{6}}
\put(965.0,261.0){\rule[-0.200pt]{4.818pt}{0.400pt}}
\put(220.0,310.0){\rule[-0.200pt]{4.818pt}{0.400pt}}
\put(198,310){\makebox(0,0)[r]{8}}
\put(965.0,310.0){\rule[-0.200pt]{4.818pt}{0.400pt}}
\put(220.0,360.0){\rule[-0.200pt]{4.818pt}{0.400pt}}
\put(198,360){\makebox(0,0)[r]{10}}
\put(965.0,360.0){\rule[-0.200pt]{4.818pt}{0.400pt}}
\put(220.0,409.0){\rule[-0.200pt]{4.818pt}{0.400pt}}
\put(198,409){\makebox(0,0)[r]{12}}
\put(965.0,409.0){\rule[-0.200pt]{4.818pt}{0.400pt}}
\put(220.0,458.0){\rule[-0.200pt]{4.818pt}{0.400pt}}
\put(198,458){\makebox(0,0)[r]{14}}
\put(965.0,458.0){\rule[-0.200pt]{4.818pt}{0.400pt}}
\put(220.0,507.0){\rule[-0.200pt]{4.818pt}{0.400pt}}
\put(198,507){\makebox(0,0)[r]{16}}
\put(965.0,507.0){\rule[-0.200pt]{4.818pt}{0.400pt}}
\put(220.0,557.0){\rule[-0.200pt]{4.818pt}{0.400pt}}
\put(198,557){\makebox(0,0)[r]{18}}
\put(965.0,557.0){\rule[-0.200pt]{4.818pt}{0.400pt}}
\put(220.0,606.0){\rule[-0.200pt]{4.818pt}{0.400pt}}
\put(198,606){\makebox(0,0)[r]{20}}
\put(965.0,606.0){\rule[-0.200pt]{4.818pt}{0.400pt}}
\put(220.0,113.0){\rule[-0.200pt]{0.400pt}{4.818pt}}
\put(220,68){\makebox(0,0){0}}
\put(220.0,586.0){\rule[-0.200pt]{0.400pt}{4.818pt}}
\put(348.0,113.0){\rule[-0.200pt]{0.400pt}{4.818pt}}
\put(348,68){\makebox(0,0){0.5}}
\put(348.0,586.0){\rule[-0.200pt]{0.400pt}{4.818pt}}
\put(475.0,113.0){\rule[-0.200pt]{0.400pt}{4.818pt}}
\put(475,68){\makebox(0,0){1}}
\put(475.0,586.0){\rule[-0.200pt]{0.400pt}{4.818pt}}
\put(603.0,113.0){\rule[-0.200pt]{0.400pt}{4.818pt}}
\put(603,68){\makebox(0,0){1.5}}
\put(603.0,586.0){\rule[-0.200pt]{0.400pt}{4.818pt}}
\put(730.0,113.0){\rule[-0.200pt]{0.400pt}{4.818pt}}
\put(730,68){\makebox(0,0){2}}
\put(730.0,586.0){\rule[-0.200pt]{0.400pt}{4.818pt}}
\put(858.0,113.0){\rule[-0.200pt]{0.400pt}{4.818pt}}
\put(858,68){\makebox(0,0){2.5}}
\put(858.0,586.0){\rule[-0.200pt]{0.400pt}{4.818pt}}
\put(985.0,113.0){\rule[-0.200pt]{0.400pt}{4.818pt}}
\put(985,68){\makebox(0,0){3}}
\put(985.0,586.0){\rule[-0.200pt]{0.400pt}{4.818pt}}
\put(220.0,113.0){\rule[-0.200pt]{184.288pt}{0.400pt}}
\put(985.0,113.0){\rule[-0.200pt]{0.400pt}{118.764pt}}
\put(220.0,606.0){\rule[-0.200pt]{184.288pt}{0.400pt}}
\put(45,359){\makebox(0,0){$P(\tilde{n},\tilde{T})$}}
\put(602,23){\makebox(0,0){$\tilde{n}$}}
\put(220.0,113.0){\rule[-0.200pt]{0.400pt}{118.764pt}}
\put(228,114){\usebox{\plotpoint}}
\put(228,113.67){\rule{1.686pt}{0.400pt}}
\multiput(228.00,113.17)(3.500,1.000){2}{\rule{0.843pt}{0.400pt}}
\put(235,115.17){\rule{1.700pt}{0.400pt}}
\multiput(235.00,114.17)(4.472,2.000){2}{\rule{0.850pt}{0.400pt}}
\put(243,117.17){\rule{1.700pt}{0.400pt}}
\multiput(243.00,116.17)(4.472,2.000){2}{\rule{0.850pt}{0.400pt}}
\put(251,119.17){\rule{1.700pt}{0.400pt}}
\multiput(251.00,118.17)(4.472,2.000){2}{\rule{0.850pt}{0.400pt}}
\multiput(259.00,121.61)(1.355,0.447){3}{\rule{1.033pt}{0.108pt}}
\multiput(259.00,120.17)(4.855,3.000){2}{\rule{0.517pt}{0.400pt}}
\put(266,124.17){\rule{1.700pt}{0.400pt}}
\multiput(266.00,123.17)(4.472,2.000){2}{\rule{0.850pt}{0.400pt}}
\put(274,126.17){\rule{1.700pt}{0.400pt}}
\multiput(274.00,125.17)(4.472,2.000){2}{\rule{0.850pt}{0.400pt}}
\put(282,127.67){\rule{1.927pt}{0.400pt}}
\multiput(282.00,127.17)(4.000,1.000){2}{\rule{0.964pt}{0.400pt}}
\put(290,128.67){\rule{1.686pt}{0.400pt}}
\multiput(290.00,128.17)(3.500,1.000){2}{\rule{0.843pt}{0.400pt}}
\put(297,129.67){\rule{1.927pt}{0.400pt}}
\multiput(297.00,129.17)(4.000,1.000){2}{\rule{0.964pt}{0.400pt}}
\put(336,129.67){\rule{1.927pt}{0.400pt}}
\multiput(336.00,130.17)(4.000,-1.000){2}{\rule{0.964pt}{0.400pt}}
\put(344,128.67){\rule{1.686pt}{0.400pt}}
\multiput(344.00,129.17)(3.500,-1.000){2}{\rule{0.843pt}{0.400pt}}
\put(305.0,131.0){\rule[-0.200pt]{7.468pt}{0.400pt}}
\put(359,127.67){\rule{1.927pt}{0.400pt}}
\multiput(359.00,128.17)(4.000,-1.000){2}{\rule{0.964pt}{0.400pt}}
\put(367,126.67){\rule{1.927pt}{0.400pt}}
\multiput(367.00,127.17)(4.000,-1.000){2}{\rule{0.964pt}{0.400pt}}
\put(351.0,129.0){\rule[-0.200pt]{1.927pt}{0.400pt}}
\put(382,125.67){\rule{1.927pt}{0.400pt}}
\multiput(382.00,126.17)(4.000,-1.000){2}{\rule{0.964pt}{0.400pt}}
\put(375.0,127.0){\rule[-0.200pt]{1.686pt}{0.400pt}}
\put(398,124.67){\rule{1.686pt}{0.400pt}}
\multiput(398.00,125.17)(3.500,-1.000){2}{\rule{0.843pt}{0.400pt}}
\put(390.0,126.0){\rule[-0.200pt]{1.927pt}{0.400pt}}
\put(421,123.67){\rule{1.927pt}{0.400pt}}
\multiput(421.00,124.17)(4.000,-1.000){2}{\rule{0.964pt}{0.400pt}}
\put(405.0,125.0){\rule[-0.200pt]{3.854pt}{0.400pt}}
\put(467,123.67){\rule{1.927pt}{0.400pt}}
\multiput(467.00,123.17)(4.000,1.000){2}{\rule{0.964pt}{0.400pt}}
\put(429.0,124.0){\rule[-0.200pt]{9.154pt}{0.400pt}}
\put(490,124.67){\rule{1.927pt}{0.400pt}}
\multiput(490.00,124.17)(4.000,1.000){2}{\rule{0.964pt}{0.400pt}}
\put(498,125.67){\rule{1.927pt}{0.400pt}}
\multiput(498.00,125.17)(4.000,1.000){2}{\rule{0.964pt}{0.400pt}}
\put(475.0,125.0){\rule[-0.200pt]{3.613pt}{0.400pt}}
\put(514,126.67){\rule{1.686pt}{0.400pt}}
\multiput(514.00,126.17)(3.500,1.000){2}{\rule{0.843pt}{0.400pt}}
\put(521,128.17){\rule{1.700pt}{0.400pt}}
\multiput(521.00,127.17)(4.472,2.000){2}{\rule{0.850pt}{0.400pt}}
\put(529,129.67){\rule{1.927pt}{0.400pt}}
\multiput(529.00,129.17)(4.000,1.000){2}{\rule{0.964pt}{0.400pt}}
\put(537,130.67){\rule{1.927pt}{0.400pt}}
\multiput(537.00,130.17)(4.000,1.000){2}{\rule{0.964pt}{0.400pt}}
\put(545,132.17){\rule{1.500pt}{0.400pt}}
\multiput(545.00,131.17)(3.887,2.000){2}{\rule{0.750pt}{0.400pt}}
\put(552,134.17){\rule{1.700pt}{0.400pt}}
\multiput(552.00,133.17)(4.472,2.000){2}{\rule{0.850pt}{0.400pt}}
\multiput(560.00,136.61)(1.579,0.447){3}{\rule{1.167pt}{0.108pt}}
\multiput(560.00,135.17)(5.579,3.000){2}{\rule{0.583pt}{0.400pt}}
\put(568,139.17){\rule{1.500pt}{0.400pt}}
\multiput(568.00,138.17)(3.887,2.000){2}{\rule{0.750pt}{0.400pt}}
\multiput(575.00,141.61)(1.579,0.447){3}{\rule{1.167pt}{0.108pt}}
\multiput(575.00,140.17)(5.579,3.000){2}{\rule{0.583pt}{0.400pt}}
\multiput(583.00,144.60)(1.066,0.468){5}{\rule{0.900pt}{0.113pt}}
\multiput(583.00,143.17)(6.132,4.000){2}{\rule{0.450pt}{0.400pt}}
\multiput(591.00,148.61)(1.579,0.447){3}{\rule{1.167pt}{0.108pt}}
\multiput(591.00,147.17)(5.579,3.000){2}{\rule{0.583pt}{0.400pt}}
\multiput(599.00,151.59)(0.710,0.477){7}{\rule{0.660pt}{0.115pt}}
\multiput(599.00,150.17)(5.630,5.000){2}{\rule{0.330pt}{0.400pt}}
\multiput(606.00,156.60)(1.066,0.468){5}{\rule{0.900pt}{0.113pt}}
\multiput(606.00,155.17)(6.132,4.000){2}{\rule{0.450pt}{0.400pt}}
\multiput(614.00,160.59)(0.821,0.477){7}{\rule{0.740pt}{0.115pt}}
\multiput(614.00,159.17)(6.464,5.000){2}{\rule{0.370pt}{0.400pt}}
\multiput(622.00,165.59)(0.671,0.482){9}{\rule{0.633pt}{0.116pt}}
\multiput(622.00,164.17)(6.685,6.000){2}{\rule{0.317pt}{0.400pt}}
\multiput(630.00,171.59)(0.710,0.477){7}{\rule{0.660pt}{0.115pt}}
\multiput(630.00,170.17)(5.630,5.000){2}{\rule{0.330pt}{0.400pt}}
\multiput(637.00,176.59)(0.671,0.482){9}{\rule{0.633pt}{0.116pt}}
\multiput(637.00,175.17)(6.685,6.000){2}{\rule{0.317pt}{0.400pt}}
\multiput(645.00,182.59)(0.671,0.482){9}{\rule{0.633pt}{0.116pt}}
\multiput(645.00,181.17)(6.685,6.000){2}{\rule{0.317pt}{0.400pt}}
\multiput(653.00,188.59)(0.581,0.482){9}{\rule{0.567pt}{0.116pt}}
\multiput(653.00,187.17)(5.824,6.000){2}{\rule{0.283pt}{0.400pt}}
\multiput(660.00,194.59)(0.821,0.477){7}{\rule{0.740pt}{0.115pt}}
\multiput(660.00,193.17)(6.464,5.000){2}{\rule{0.370pt}{0.400pt}}
\multiput(668.00,199.59)(0.821,0.477){7}{\rule{0.740pt}{0.115pt}}
\multiput(668.00,198.17)(6.464,5.000){2}{\rule{0.370pt}{0.400pt}}
\multiput(676.00,204.61)(1.579,0.447){3}{\rule{1.167pt}{0.108pt}}
\multiput(676.00,203.17)(5.579,3.000){2}{\rule{0.583pt}{0.400pt}}
\put(684,207.17){\rule{1.500pt}{0.400pt}}
\multiput(684.00,206.17)(3.887,2.000){2}{\rule{0.750pt}{0.400pt}}
\put(506.0,127.0){\rule[-0.200pt]{1.927pt}{0.400pt}}
\put(699,207.17){\rule{1.700pt}{0.400pt}}
\multiput(699.00,208.17)(4.472,-2.000){2}{\rule{0.850pt}{0.400pt}}
\multiput(707.00,205.93)(0.821,-0.477){7}{\rule{0.740pt}{0.115pt}}
\multiput(707.00,206.17)(6.464,-5.000){2}{\rule{0.370pt}{0.400pt}}
\multiput(715.00,200.93)(0.581,-0.482){9}{\rule{0.567pt}{0.116pt}}
\multiput(715.00,201.17)(5.824,-6.000){2}{\rule{0.283pt}{0.400pt}}
\multiput(722.00,194.93)(0.494,-0.488){13}{\rule{0.500pt}{0.117pt}}
\multiput(722.00,195.17)(6.962,-8.000){2}{\rule{0.250pt}{0.400pt}}
\multiput(730.59,185.51)(0.488,-0.626){13}{\rule{0.117pt}{0.600pt}}
\multiput(729.17,186.75)(8.000,-8.755){2}{\rule{0.400pt}{0.300pt}}
\multiput(738.59,175.21)(0.485,-0.721){11}{\rule{0.117pt}{0.671pt}}
\multiput(737.17,176.61)(7.000,-8.606){2}{\rule{0.400pt}{0.336pt}}
\multiput(745.59,165.30)(0.488,-0.692){13}{\rule{0.117pt}{0.650pt}}
\multiput(744.17,166.65)(8.000,-9.651){2}{\rule{0.400pt}{0.325pt}}
\multiput(753.59,154.51)(0.488,-0.626){13}{\rule{0.117pt}{0.600pt}}
\multiput(752.17,155.75)(8.000,-8.755){2}{\rule{0.400pt}{0.300pt}}
\multiput(761.59,144.72)(0.488,-0.560){13}{\rule{0.117pt}{0.550pt}}
\multiput(760.17,145.86)(8.000,-7.858){2}{\rule{0.400pt}{0.275pt}}
\multiput(769.59,135.69)(0.485,-0.569){11}{\rule{0.117pt}{0.557pt}}
\multiput(768.17,136.84)(7.000,-6.844){2}{\rule{0.400pt}{0.279pt}}
\multiput(776.00,128.93)(0.671,-0.482){9}{\rule{0.633pt}{0.116pt}}
\multiput(776.00,129.17)(6.685,-6.000){2}{\rule{0.317pt}{0.400pt}}
\multiput(784.00,122.94)(1.066,-0.468){5}{\rule{0.900pt}{0.113pt}}
\multiput(784.00,123.17)(6.132,-4.000){2}{\rule{0.450pt}{0.400pt}}
\multiput(792.00,118.95)(1.579,-0.447){3}{\rule{1.167pt}{0.108pt}}
\multiput(792.00,119.17)(5.579,-3.000){2}{\rule{0.583pt}{0.400pt}}
\put(800,115.17){\rule{1.500pt}{0.400pt}}
\multiput(800.00,116.17)(3.887,-2.000){2}{\rule{0.750pt}{0.400pt}}
\put(807,113.67){\rule{1.927pt}{0.400pt}}
\multiput(807.00,114.17)(4.000,-1.000){2}{\rule{0.964pt}{0.400pt}}
\put(815,112.67){\rule{1.927pt}{0.400pt}}
\multiput(815.00,113.17)(4.000,-1.000){2}{\rule{0.964pt}{0.400pt}}
\put(691.0,209.0){\rule[-0.200pt]{1.927pt}{0.400pt}}
\put(823.0,113.0){\rule[-0.200pt]{37.099pt}{0.400pt}}
\put(228,113){\usebox{\plotpoint}}
\put(228.00,113.00){\usebox{\plotpoint}}
\multiput(235,113)(20.756,0.000){0}{\usebox{\plotpoint}}
\put(248.76,113.00){\usebox{\plotpoint}}
\multiput(251,113)(20.756,0.000){0}{\usebox{\plotpoint}}
\multiput(259,113)(20.547,2.935){0}{\usebox{\plotpoint}}
\put(269.44,114.00){\usebox{\plotpoint}}
\multiput(274,114)(20.595,2.574){0}{\usebox{\plotpoint}}
\multiput(282,115)(20.756,0.000){0}{\usebox{\plotpoint}}
\put(290.13,115.02){\usebox{\plotpoint}}
\multiput(297,116)(20.756,0.000){0}{\usebox{\plotpoint}}
\put(310.82,116.00){\usebox{\plotpoint}}
\multiput(313,116)(20.756,0.000){0}{\usebox{\plotpoint}}
\multiput(320,116)(20.756,0.000){0}{\usebox{\plotpoint}}
\put(331.57,116.00){\usebox{\plotpoint}}
\multiput(336,116)(20.595,-2.574){0}{\usebox{\plotpoint}}
\multiput(344,115)(20.756,0.000){0}{\usebox{\plotpoint}}
\put(352.27,115.00){\usebox{\plotpoint}}
\multiput(359,115)(20.756,0.000){0}{\usebox{\plotpoint}}
\put(372.98,114.25){\usebox{\plotpoint}}
\multiput(375,114)(20.756,0.000){0}{\usebox{\plotpoint}}
\multiput(382,114)(20.756,0.000){0}{\usebox{\plotpoint}}
\put(393.72,114.00){\usebox{\plotpoint}}
\multiput(398,114)(20.756,0.000){0}{\usebox{\plotpoint}}
\multiput(405,114)(20.756,0.000){0}{\usebox{\plotpoint}}
\put(414.47,114.00){\usebox{\plotpoint}}
\multiput(421,114)(20.756,0.000){0}{\usebox{\plotpoint}}
\put(435.23,114.00){\usebox{\plotpoint}}
\multiput(436,114)(20.756,0.000){0}{\usebox{\plotpoint}}
\multiput(444,114)(20.756,0.000){0}{\usebox{\plotpoint}}
\put(455.98,114.00){\usebox{\plotpoint}}
\multiput(460,114)(20.756,0.000){0}{\usebox{\plotpoint}}
\multiput(467,114)(20.756,0.000){0}{\usebox{\plotpoint}}
\put(476.74,114.00){\usebox{\plotpoint}}
\multiput(483,114)(20.756,0.000){0}{\usebox{\plotpoint}}
\put(497.49,114.00){\usebox{\plotpoint}}
\multiput(498,114)(20.756,0.000){0}{\usebox{\plotpoint}}
\multiput(506,114)(20.595,2.574){0}{\usebox{\plotpoint}}
\put(518.19,115.00){\usebox{\plotpoint}}
\multiput(521,115)(20.756,0.000){0}{\usebox{\plotpoint}}
\multiput(529,115)(20.595,2.574){0}{\usebox{\plotpoint}}
\put(538.86,116.23){\usebox{\plotpoint}}
\multiput(545,117)(20.547,2.935){0}{\usebox{\plotpoint}}
\put(559.44,118.93){\usebox{\plotpoint}}
\multiput(560,119)(20.136,5.034){0}{\usebox{\plotpoint}}
\multiput(568,121)(19.077,8.176){0}{\usebox{\plotpoint}}
\put(578.88,125.94){\usebox{\plotpoint}}
\multiput(583,128)(16.604,12.453){0}{\usebox{\plotpoint}}
\put(595.34,138.34){\usebox{\plotpoint}}
\multiput(599,142)(11.143,17.511){0}{\usebox{\plotpoint}}
\put(607.26,155.21){\usebox{\plotpoint}}
\put(616.79,173.62){\usebox{\plotpoint}}
\put(624.23,192.97){\usebox{\plotpoint}}
\multiput(630,211)(4.307,20.304){2}{\usebox{\plotpoint}}
\multiput(637,244)(4.070,20.352){2}{\usebox{\plotpoint}}
\multiput(645,284)(3.483,20.461){2}{\usebox{\plotpoint}}
\multiput(653,331)(2.718,20.577){3}{\usebox{\plotpoint}}
\multiput(660,384)(2.988,20.539){3}{\usebox{\plotpoint}}
\multiput(668,439)(3.098,20.523){2}{\usebox{\plotpoint}}
\multiput(676,492)(3.796,20.405){2}{\usebox{\plotpoint}}
\put(685.49,540.12){\usebox{\plotpoint}}
\put(692.08,559.14){\usebox{\plotpoint}}
\put(702.79,546.75){\usebox{\plotpoint}}
\multiput(707,532)(3.098,-20.523){3}{\usebox{\plotpoint}}
\multiput(715,479)(2.008,-20.658){3}{\usebox{\plotpoint}}
\multiput(722,407)(2.118,-20.647){4}{\usebox{\plotpoint}}
\multiput(730,329)(2.261,-20.632){4}{\usebox{\plotpoint}}
\multiput(738,256)(2.487,-20.606){2}{\usebox{\plotpoint}}
\multiput(745,198)(3.975,-20.371){2}{\usebox{\plotpoint}}
\multiput(753,157)(6.563,-19.690){2}{\usebox{\plotpoint}}
\multiput(761,133)(11.513,-17.270){0}{\usebox{\plotpoint}}
\put(771.30,119.36){\usebox{\plotpoint}}
\multiput(776,116)(20.136,-5.034){0}{\usebox{\plotpoint}}
\put(790.68,113.16){\usebox{\plotpoint}}
\multiput(792,113)(20.756,0.000){0}{\usebox{\plotpoint}}
\multiput(800,113)(20.756,0.000){0}{\usebox{\plotpoint}}
\put(811.43,113.00){\usebox{\plotpoint}}
\multiput(815,113)(20.756,0.000){0}{\usebox{\plotpoint}}
\multiput(823,113)(20.756,0.000){0}{\usebox{\plotpoint}}
\put(832.18,113.00){\usebox{\plotpoint}}
\multiput(838,113)(20.756,0.000){0}{\usebox{\plotpoint}}
\put(852.94,113.00){\usebox{\plotpoint}}
\multiput(854,113)(20.756,0.000){0}{\usebox{\plotpoint}}
\multiput(861,113)(20.756,0.000){0}{\usebox{\plotpoint}}
\put(873.69,113.00){\usebox{\plotpoint}}
\multiput(877,113)(20.756,0.000){0}{\usebox{\plotpoint}}
\multiput(885,113)(20.756,0.000){0}{\usebox{\plotpoint}}
\put(894.45,113.00){\usebox{\plotpoint}}
\multiput(900,113)(20.756,0.000){0}{\usebox{\plotpoint}}
\multiput(908,113)(20.756,0.000){0}{\usebox{\plotpoint}}
\put(915.21,113.00){\usebox{\plotpoint}}
\multiput(923,113)(20.756,0.000){0}{\usebox{\plotpoint}}
\put(935.96,113.00){\usebox{\plotpoint}}
\multiput(939,113)(20.756,0.000){0}{\usebox{\plotpoint}}
\multiput(946,113)(20.756,0.000){0}{\usebox{\plotpoint}}
\put(956.72,113.00){\usebox{\plotpoint}}
\multiput(962,113)(20.756,0.000){0}{\usebox{\plotpoint}}
\multiput(970,113)(20.756,0.000){0}{\usebox{\plotpoint}}
\put(977,113){\usebox{\plotpoint}}
\end{picture}

\caption{\label{voldep1} The distribution function
$P(\tilde{n},\tilde{T})$ along the isotherm $\tilde{T}=0.85$ for two
different values of $\tilde{V}=20$ (solid line) and $60$ (dotted
line). Here, the external chemical potential is $\tilde{\mu}^{\rm
ext}=-14.85$. Observe that the distribution becomes more peaked as the
typical volume $\tilde{V}$ of the cells increases. The relative height
increases and, eventually, in the thermodynamic limit
$\tilde{V}\rightarrow\infty$ only one peak survives.}
\end{center}
\end{figure}

\begin{figure}[ht]
\begin{center}
\setlength{\unitlength}{0.240900pt}
\ifx\plotpoint\undefined\newsavebox{\plotpoint}\fi
\begin{picture}(1049,629)(0,0)
\font\gnuplot=cmr10 at 10pt
\gnuplot
\sbox{\plotpoint}{\rule[-0.200pt]{0.400pt}{0.400pt}}%
\put(220.0,113.0){\rule[-0.200pt]{184.288pt}{0.400pt}}
\put(220.0,113.0){\rule[-0.200pt]{0.400pt}{118.764pt}}
\put(220.0,113.0){\rule[-0.200pt]{4.818pt}{0.400pt}}
\put(198,113){\makebox(0,0)[r]{0}}
\put(965.0,113.0){\rule[-0.200pt]{4.818pt}{0.400pt}}
\put(220.0,175.0){\rule[-0.200pt]{4.818pt}{0.400pt}}
\put(198,175){\makebox(0,0)[r]{0.5}}
\put(965.0,175.0){\rule[-0.200pt]{4.818pt}{0.400pt}}
\put(220.0,236.0){\rule[-0.200pt]{4.818pt}{0.400pt}}
\put(198,236){\makebox(0,0)[r]{1}}
\put(965.0,236.0){\rule[-0.200pt]{4.818pt}{0.400pt}}
\put(220.0,298.0){\rule[-0.200pt]{4.818pt}{0.400pt}}
\put(198,298){\makebox(0,0)[r]{1.5}}
\put(965.0,298.0){\rule[-0.200pt]{4.818pt}{0.400pt}}
\put(220.0,360.0){\rule[-0.200pt]{4.818pt}{0.400pt}}
\put(198,360){\makebox(0,0)[r]{2}}
\put(965.0,360.0){\rule[-0.200pt]{4.818pt}{0.400pt}}
\put(220.0,421.0){\rule[-0.200pt]{4.818pt}{0.400pt}}
\put(198,421){\makebox(0,0)[r]{2.5}}
\put(965.0,421.0){\rule[-0.200pt]{4.818pt}{0.400pt}}
\put(220.0,483.0){\rule[-0.200pt]{4.818pt}{0.400pt}}
\put(198,483){\makebox(0,0)[r]{3}}
\put(965.0,483.0){\rule[-0.200pt]{4.818pt}{0.400pt}}
\put(220.0,544.0){\rule[-0.200pt]{4.818pt}{0.400pt}}
\put(198,544){\makebox(0,0)[r]{3.5}}
\put(965.0,544.0){\rule[-0.200pt]{4.818pt}{0.400pt}}
\put(220.0,606.0){\rule[-0.200pt]{4.818pt}{0.400pt}}
\put(198,606){\makebox(0,0)[r]{4}}
\put(965.0,606.0){\rule[-0.200pt]{4.818pt}{0.400pt}}
\put(220.0,113.0){\rule[-0.200pt]{0.400pt}{4.818pt}}
\put(220,68){\makebox(0,0){0}}
\put(220.0,586.0){\rule[-0.200pt]{0.400pt}{4.818pt}}
\put(348.0,113.0){\rule[-0.200pt]{0.400pt}{4.818pt}}
\put(348,68){\makebox(0,0){0.5}}
\put(348.0,586.0){\rule[-0.200pt]{0.400pt}{4.818pt}}
\put(475.0,113.0){\rule[-0.200pt]{0.400pt}{4.818pt}}
\put(475,68){\makebox(0,0){1}}
\put(475.0,586.0){\rule[-0.200pt]{0.400pt}{4.818pt}}
\put(603.0,113.0){\rule[-0.200pt]{0.400pt}{4.818pt}}
\put(603,68){\makebox(0,0){1.5}}
\put(603.0,586.0){\rule[-0.200pt]{0.400pt}{4.818pt}}
\put(730.0,113.0){\rule[-0.200pt]{0.400pt}{4.818pt}}
\put(730,68){\makebox(0,0){2}}
\put(730.0,586.0){\rule[-0.200pt]{0.400pt}{4.818pt}}
\put(858.0,113.0){\rule[-0.200pt]{0.400pt}{4.818pt}}
\put(858,68){\makebox(0,0){2.5}}
\put(858.0,586.0){\rule[-0.200pt]{0.400pt}{4.818pt}}
\put(985.0,113.0){\rule[-0.200pt]{0.400pt}{4.818pt}}
\put(985,68){\makebox(0,0){3}}
\put(985.0,586.0){\rule[-0.200pt]{0.400pt}{4.818pt}}
\put(220.0,113.0){\rule[-0.200pt]{184.288pt}{0.400pt}}
\put(985.0,113.0){\rule[-0.200pt]{0.400pt}{118.764pt}}
\put(220.0,606.0){\rule[-0.200pt]{184.288pt}{0.400pt}}
\put(45,359){\makebox(0,0){$P(\tilde{n},\tilde{T})$}}
\put(602,23){\makebox(0,0){$\tilde{n}$}}
\put(220.0,113.0){\rule[-0.200pt]{0.400pt}{118.764pt}}
\put(228,114){\usebox{\plotpoint}}
\put(228,113.67){\rule{1.686pt}{0.400pt}}
\multiput(228.00,113.17)(3.500,1.000){2}{\rule{0.843pt}{0.400pt}}
\multiput(235.00,115.61)(1.579,0.447){3}{\rule{1.167pt}{0.108pt}}
\multiput(235.00,114.17)(5.579,3.000){2}{\rule{0.583pt}{0.400pt}}
\multiput(243.00,118.60)(1.066,0.468){5}{\rule{0.900pt}{0.113pt}}
\multiput(243.00,117.17)(6.132,4.000){2}{\rule{0.450pt}{0.400pt}}
\multiput(251.00,122.59)(0.569,0.485){11}{\rule{0.557pt}{0.117pt}}
\multiput(251.00,121.17)(6.844,7.000){2}{\rule{0.279pt}{0.400pt}}
\multiput(259.59,129.00)(0.485,0.569){11}{\rule{0.117pt}{0.557pt}}
\multiput(258.17,129.00)(7.000,6.844){2}{\rule{0.400pt}{0.279pt}}
\multiput(266.59,137.00)(0.488,0.692){13}{\rule{0.117pt}{0.650pt}}
\multiput(265.17,137.00)(8.000,9.651){2}{\rule{0.400pt}{0.325pt}}
\multiput(274.59,148.00)(0.488,0.758){13}{\rule{0.117pt}{0.700pt}}
\multiput(273.17,148.00)(8.000,10.547){2}{\rule{0.400pt}{0.350pt}}
\multiput(282.59,160.00)(0.488,0.824){13}{\rule{0.117pt}{0.750pt}}
\multiput(281.17,160.00)(8.000,11.443){2}{\rule{0.400pt}{0.375pt}}
\multiput(290.59,173.00)(0.485,1.026){11}{\rule{0.117pt}{0.900pt}}
\multiput(289.17,173.00)(7.000,12.132){2}{\rule{0.400pt}{0.450pt}}
\multiput(297.59,187.00)(0.488,0.824){13}{\rule{0.117pt}{0.750pt}}
\multiput(296.17,187.00)(8.000,11.443){2}{\rule{0.400pt}{0.375pt}}
\multiput(305.59,200.00)(0.488,0.890){13}{\rule{0.117pt}{0.800pt}}
\multiput(304.17,200.00)(8.000,12.340){2}{\rule{0.400pt}{0.400pt}}
\multiput(313.59,214.00)(0.485,0.950){11}{\rule{0.117pt}{0.843pt}}
\multiput(312.17,214.00)(7.000,11.251){2}{\rule{0.400pt}{0.421pt}}
\multiput(320.59,227.00)(0.488,0.758){13}{\rule{0.117pt}{0.700pt}}
\multiput(319.17,227.00)(8.000,10.547){2}{\rule{0.400pt}{0.350pt}}
\multiput(328.59,239.00)(0.488,0.692){13}{\rule{0.117pt}{0.650pt}}
\multiput(327.17,239.00)(8.000,9.651){2}{\rule{0.400pt}{0.325pt}}
\multiput(336.59,250.00)(0.488,0.626){13}{\rule{0.117pt}{0.600pt}}
\multiput(335.17,250.00)(8.000,8.755){2}{\rule{0.400pt}{0.300pt}}
\multiput(344.59,260.00)(0.485,0.569){11}{\rule{0.117pt}{0.557pt}}
\multiput(343.17,260.00)(7.000,6.844){2}{\rule{0.400pt}{0.279pt}}
\multiput(351.00,268.59)(0.569,0.485){11}{\rule{0.557pt}{0.117pt}}
\multiput(351.00,267.17)(6.844,7.000){2}{\rule{0.279pt}{0.400pt}}
\multiput(359.00,275.59)(0.671,0.482){9}{\rule{0.633pt}{0.116pt}}
\multiput(359.00,274.17)(6.685,6.000){2}{\rule{0.317pt}{0.400pt}}
\multiput(367.00,281.59)(0.821,0.477){7}{\rule{0.740pt}{0.115pt}}
\multiput(367.00,280.17)(6.464,5.000){2}{\rule{0.370pt}{0.400pt}}
\multiput(375.00,286.60)(0.920,0.468){5}{\rule{0.800pt}{0.113pt}}
\multiput(375.00,285.17)(5.340,4.000){2}{\rule{0.400pt}{0.400pt}}
\multiput(382.00,290.61)(1.579,0.447){3}{\rule{1.167pt}{0.108pt}}
\multiput(382.00,289.17)(5.579,3.000){2}{\rule{0.583pt}{0.400pt}}
\put(390,293.17){\rule{1.700pt}{0.400pt}}
\multiput(390.00,292.17)(4.472,2.000){2}{\rule{0.850pt}{0.400pt}}
\put(398,295.17){\rule{1.500pt}{0.400pt}}
\multiput(398.00,294.17)(3.887,2.000){2}{\rule{0.750pt}{0.400pt}}
\put(405,296.67){\rule{1.927pt}{0.400pt}}
\multiput(405.00,296.17)(4.000,1.000){2}{\rule{0.964pt}{0.400pt}}
\put(413,297.67){\rule{1.927pt}{0.400pt}}
\multiput(413.00,297.17)(4.000,1.000){2}{\rule{0.964pt}{0.400pt}}
\put(436,298.67){\rule{1.927pt}{0.400pt}}
\multiput(436.00,298.17)(4.000,1.000){2}{\rule{0.964pt}{0.400pt}}
\put(421.0,299.0){\rule[-0.200pt]{3.613pt}{0.400pt}}
\put(506,298.67){\rule{1.927pt}{0.400pt}}
\multiput(506.00,299.17)(4.000,-1.000){2}{\rule{0.964pt}{0.400pt}}
\put(444.0,300.0){\rule[-0.200pt]{14.936pt}{0.400pt}}
\put(529,297.67){\rule{1.927pt}{0.400pt}}
\multiput(529.00,298.17)(4.000,-1.000){2}{\rule{0.964pt}{0.400pt}}
\put(537,296.67){\rule{1.927pt}{0.400pt}}
\multiput(537.00,297.17)(4.000,-1.000){2}{\rule{0.964pt}{0.400pt}}
\put(545,295.17){\rule{1.500pt}{0.400pt}}
\multiput(545.00,296.17)(3.887,-2.000){2}{\rule{0.750pt}{0.400pt}}
\put(552,293.17){\rule{1.700pt}{0.400pt}}
\multiput(552.00,294.17)(4.472,-2.000){2}{\rule{0.850pt}{0.400pt}}
\put(560,291.17){\rule{1.700pt}{0.400pt}}
\multiput(560.00,292.17)(4.472,-2.000){2}{\rule{0.850pt}{0.400pt}}
\multiput(568.00,289.95)(1.355,-0.447){3}{\rule{1.033pt}{0.108pt}}
\multiput(568.00,290.17)(4.855,-3.000){2}{\rule{0.517pt}{0.400pt}}
\multiput(575.00,286.94)(1.066,-0.468){5}{\rule{0.900pt}{0.113pt}}
\multiput(575.00,287.17)(6.132,-4.000){2}{\rule{0.450pt}{0.400pt}}
\multiput(583.00,282.93)(0.821,-0.477){7}{\rule{0.740pt}{0.115pt}}
\multiput(583.00,283.17)(6.464,-5.000){2}{\rule{0.370pt}{0.400pt}}
\multiput(591.00,277.93)(0.821,-0.477){7}{\rule{0.740pt}{0.115pt}}
\multiput(591.00,278.17)(6.464,-5.000){2}{\rule{0.370pt}{0.400pt}}
\multiput(599.00,272.93)(0.492,-0.485){11}{\rule{0.500pt}{0.117pt}}
\multiput(599.00,273.17)(5.962,-7.000){2}{\rule{0.250pt}{0.400pt}}
\multiput(606.00,265.93)(0.569,-0.485){11}{\rule{0.557pt}{0.117pt}}
\multiput(606.00,266.17)(6.844,-7.000){2}{\rule{0.279pt}{0.400pt}}
\multiput(614.59,257.72)(0.488,-0.560){13}{\rule{0.117pt}{0.550pt}}
\multiput(613.17,258.86)(8.000,-7.858){2}{\rule{0.400pt}{0.275pt}}
\multiput(622.59,248.72)(0.488,-0.560){13}{\rule{0.117pt}{0.550pt}}
\multiput(621.17,249.86)(8.000,-7.858){2}{\rule{0.400pt}{0.275pt}}
\multiput(630.59,239.21)(0.485,-0.721){11}{\rule{0.117pt}{0.671pt}}
\multiput(629.17,240.61)(7.000,-8.606){2}{\rule{0.400pt}{0.336pt}}
\multiput(637.59,229.30)(0.488,-0.692){13}{\rule{0.117pt}{0.650pt}}
\multiput(636.17,230.65)(8.000,-9.651){2}{\rule{0.400pt}{0.325pt}}
\multiput(645.59,218.30)(0.488,-0.692){13}{\rule{0.117pt}{0.650pt}}
\multiput(644.17,219.65)(8.000,-9.651){2}{\rule{0.400pt}{0.325pt}}
\multiput(653.59,206.98)(0.485,-0.798){11}{\rule{0.117pt}{0.729pt}}
\multiput(652.17,208.49)(7.000,-9.488){2}{\rule{0.400pt}{0.364pt}}
\multiput(660.59,196.30)(0.488,-0.692){13}{\rule{0.117pt}{0.650pt}}
\multiput(659.17,197.65)(8.000,-9.651){2}{\rule{0.400pt}{0.325pt}}
\multiput(668.59,185.09)(0.488,-0.758){13}{\rule{0.117pt}{0.700pt}}
\multiput(667.17,186.55)(8.000,-10.547){2}{\rule{0.400pt}{0.350pt}}
\multiput(676.59,173.51)(0.488,-0.626){13}{\rule{0.117pt}{0.600pt}}
\multiput(675.17,174.75)(8.000,-8.755){2}{\rule{0.400pt}{0.300pt}}
\multiput(684.59,163.21)(0.485,-0.721){11}{\rule{0.117pt}{0.671pt}}
\multiput(683.17,164.61)(7.000,-8.606){2}{\rule{0.400pt}{0.336pt}}
\multiput(691.59,153.72)(0.488,-0.560){13}{\rule{0.117pt}{0.550pt}}
\multiput(690.17,154.86)(8.000,-7.858){2}{\rule{0.400pt}{0.275pt}}
\multiput(699.00,145.93)(0.494,-0.488){13}{\rule{0.500pt}{0.117pt}}
\multiput(699.00,146.17)(6.962,-8.000){2}{\rule{0.250pt}{0.400pt}}
\multiput(707.00,137.93)(0.671,-0.482){9}{\rule{0.633pt}{0.116pt}}
\multiput(707.00,138.17)(6.685,-6.000){2}{\rule{0.317pt}{0.400pt}}
\multiput(715.00,131.93)(0.581,-0.482){9}{\rule{0.567pt}{0.116pt}}
\multiput(715.00,132.17)(5.824,-6.000){2}{\rule{0.283pt}{0.400pt}}
\multiput(722.00,125.94)(1.066,-0.468){5}{\rule{0.900pt}{0.113pt}}
\multiput(722.00,126.17)(6.132,-4.000){2}{\rule{0.450pt}{0.400pt}}
\multiput(730.00,121.94)(1.066,-0.468){5}{\rule{0.900pt}{0.113pt}}
\multiput(730.00,122.17)(6.132,-4.000){2}{\rule{0.450pt}{0.400pt}}
\put(738,117.17){\rule{1.500pt}{0.400pt}}
\multiput(738.00,118.17)(3.887,-2.000){2}{\rule{0.750pt}{0.400pt}}
\put(745,115.67){\rule{1.927pt}{0.400pt}}
\multiput(745.00,116.17)(4.000,-1.000){2}{\rule{0.964pt}{0.400pt}}
\put(753,114.17){\rule{1.700pt}{0.400pt}}
\multiput(753.00,115.17)(4.472,-2.000){2}{\rule{0.850pt}{0.400pt}}
\put(514.0,299.0){\rule[-0.200pt]{3.613pt}{0.400pt}}
\put(769,112.67){\rule{1.686pt}{0.400pt}}
\multiput(769.00,113.17)(3.500,-1.000){2}{\rule{0.843pt}{0.400pt}}
\put(761.0,114.0){\rule[-0.200pt]{1.927pt}{0.400pt}}
\put(776.0,113.0){\rule[-0.200pt]{48.421pt}{0.400pt}}
\put(228,113){\usebox{\plotpoint}}
\put(228.00,113.00){\usebox{\plotpoint}}
\multiput(235,113)(20.756,0.000){0}{\usebox{\plotpoint}}
\put(248.76,113.00){\usebox{\plotpoint}}
\multiput(251,113)(20.756,0.000){0}{\usebox{\plotpoint}}
\multiput(259,113)(20.547,2.935){0}{\usebox{\plotpoint}}
\put(269.34,114.83){\usebox{\plotpoint}}
\multiput(274,116)(18.564,9.282){0}{\usebox{\plotpoint}}
\put(287.27,124.61){\usebox{\plotpoint}}
\multiput(290,127)(10.458,17.928){0}{\usebox{\plotpoint}}
\put(298.32,141.96){\usebox{\plotpoint}}
\multiput(305,157)(6.326,19.768){2}{\usebox{\plotpoint}}
\put(317.44,201.05){\usebox{\plotpoint}}
\multiput(320,212)(4.754,20.204){2}{\usebox{\plotpoint}}
\multiput(328,246)(4.276,20.310){2}{\usebox{\plotpoint}}
\multiput(336,284)(4.171,20.332){2}{\usebox{\plotpoint}}
\put(347.71,343.13){\usebox{\plotpoint}}
\multiput(351,361)(4.625,20.234){2}{\usebox{\plotpoint}}
\multiput(359,396)(5.034,20.136){2}{\usebox{\plotpoint}}
\put(371.59,444.07){\usebox{\plotpoint}}
\put(377.43,463.99){\usebox{\plotpoint}}
\put(383.97,483.67){\usebox{\plotpoint}}
\put(392.45,502.59){\usebox{\plotpoint}}
\put(402.81,520.56){\usebox{\plotpoint}}
\multiput(405,524)(14.676,14.676){0}{\usebox{\plotpoint}}
\put(417.55,534.84){\usebox{\plotpoint}}
\multiput(421,537)(18.564,9.282){0}{\usebox{\plotpoint}}
\multiput(429,541)(19.957,5.702){0}{\usebox{\plotpoint}}
\put(436.46,543.06){\usebox{\plotpoint}}
\multiput(444,544)(20.595,2.574){0}{\usebox{\plotpoint}}
\put(457.09,545.00){\usebox{\plotpoint}}
\multiput(460,545)(20.756,0.000){0}{\usebox{\plotpoint}}
\multiput(467,545)(20.756,0.000){0}{\usebox{\plotpoint}}
\put(477.85,545.00){\usebox{\plotpoint}}
\multiput(483,545)(20.756,0.000){0}{\usebox{\plotpoint}}
\multiput(490,545)(20.756,0.000){0}{\usebox{\plotpoint}}
\put(498.60,544.93){\usebox{\plotpoint}}
\multiput(506,544)(20.595,-2.574){0}{\usebox{\plotpoint}}
\put(519.03,541.56){\usebox{\plotpoint}}
\multiput(521,541)(19.434,-7.288){0}{\usebox{\plotpoint}}
\multiput(529,538)(17.601,-11.000){0}{\usebox{\plotpoint}}
\put(537.55,532.52){\usebox{\plotpoint}}
\put(551.67,517.43){\usebox{\plotpoint}}
\multiput(552,517)(10.878,-17.677){0}{\usebox{\plotpoint}}
\put(562.21,499.57){\usebox{\plotpoint}}
\put(570.70,480.67){\usebox{\plotpoint}}
\put(577.63,461.11){\usebox{\plotpoint}}
\multiput(583,445)(6.104,-19.838){2}{\usebox{\plotpoint}}
\put(595.66,401.52){\usebox{\plotpoint}}
\multiput(599,389)(4.435,-20.276){2}{\usebox{\plotpoint}}
\put(609.79,340.89){\usebox{\plotpoint}}
\multiput(614,323)(4.754,-20.204){2}{\usebox{\plotpoint}}
\multiput(622,289)(4.890,-20.171){2}{\usebox{\plotpoint}}
\put(633.64,239.89){\usebox{\plotpoint}}
\multiput(637,225)(5.896,-19.900){2}{\usebox{\plotpoint}}
\put(650.96,180.13){\usebox{\plotpoint}}
\put(657.94,160.58){\usebox{\plotpoint}}
\put(667.34,142.15){\usebox{\plotpoint}}
\multiput(668,141)(12.208,-16.786){0}{\usebox{\plotpoint}}
\put(680.39,126.16){\usebox{\plotpoint}}
\multiput(684,123)(16.889,-12.064){0}{\usebox{\plotpoint}}
\put(698.14,116.22){\usebox{\plotpoint}}
\multiput(699,116)(20.136,-5.034){0}{\usebox{\plotpoint}}
\multiput(707,114)(20.595,-2.574){0}{\usebox{\plotpoint}}
\put(718.56,113.00){\usebox{\plotpoint}}
\multiput(722,113)(20.756,0.000){0}{\usebox{\plotpoint}}
\multiput(730,113)(20.756,0.000){0}{\usebox{\plotpoint}}
\put(739.31,113.00){\usebox{\plotpoint}}
\multiput(745,113)(20.756,0.000){0}{\usebox{\plotpoint}}
\put(760.07,113.00){\usebox{\plotpoint}}
\multiput(761,113)(20.756,0.000){0}{\usebox{\plotpoint}}
\multiput(769,113)(20.756,0.000){0}{\usebox{\plotpoint}}
\put(780.82,113.00){\usebox{\plotpoint}}
\multiput(784,113)(20.756,0.000){0}{\usebox{\plotpoint}}
\multiput(792,113)(20.756,0.000){0}{\usebox{\plotpoint}}
\put(801.58,113.00){\usebox{\plotpoint}}
\multiput(807,113)(20.756,0.000){0}{\usebox{\plotpoint}}
\put(822.33,113.00){\usebox{\plotpoint}}
\multiput(823,113)(20.756,0.000){0}{\usebox{\plotpoint}}
\multiput(830,113)(20.756,0.000){0}{\usebox{\plotpoint}}
\put(843.09,113.00){\usebox{\plotpoint}}
\multiput(846,113)(20.756,0.000){0}{\usebox{\plotpoint}}
\multiput(854,113)(20.756,0.000){0}{\usebox{\plotpoint}}
\put(863.85,113.00){\usebox{\plotpoint}}
\multiput(869,113)(20.756,0.000){0}{\usebox{\plotpoint}}
\put(884.60,113.00){\usebox{\plotpoint}}
\multiput(885,113)(20.756,0.000){0}{\usebox{\plotpoint}}
\multiput(892,113)(20.756,0.000){0}{\usebox{\plotpoint}}
\put(905.36,113.00){\usebox{\plotpoint}}
\multiput(908,113)(20.756,0.000){0}{\usebox{\plotpoint}}
\multiput(915,113)(20.756,0.000){0}{\usebox{\plotpoint}}
\put(926.11,113.00){\usebox{\plotpoint}}
\multiput(931,113)(20.756,0.000){0}{\usebox{\plotpoint}}
\multiput(939,113)(20.756,0.000){0}{\usebox{\plotpoint}}
\put(946.87,113.00){\usebox{\plotpoint}}
\multiput(954,113)(20.756,0.000){0}{\usebox{\plotpoint}}
\put(967.62,113.00){\usebox{\plotpoint}}
\multiput(970,113)(20.756,0.000){0}{\usebox{\plotpoint}}
\put(977,113){\usebox{\plotpoint}}
\end{picture}

\caption{\label{voldep3} The distribution function
$P(\tilde{n},\tilde{T})$ at the critical point.
The critical isotherm $\tilde{T}=1$ is plotted for
$\tilde{\mu}^{\rm ext}=-17.36$ for two
different values of $\tilde{V}=20$ (solid line) and $60$ (dotted
line). Note the broad
distribution of densities, which remain broad even in
the thermodynamic limit.}
\end{center}
\end{figure}
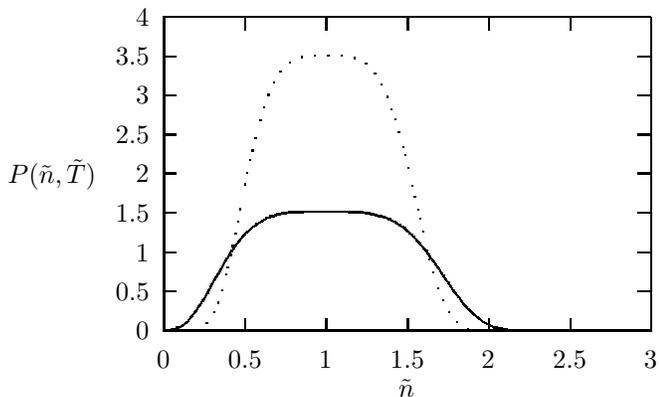

The size $\tilde{V}$ of the typical volume of a cell controls the
sharpness of the distribution function. In Fig. \ref{voldep1},
\ref{voldep2}, \ref{voldep3} we show the distribution function
$P(\tilde{n},\tilde{T}=0.85)$ for different values of $\tilde{V}$.
The fluctuations become much smaller as $\tilde{V}$ becomes larger,
consistent with the fact that in the thermodynamic limit fluctuations
vanish.  Another interesting property of the thermodynamic limit is
that the relative height of the to peaks in the distribution function
of the density increases as $\tilde{V}\rightarrow \infty$.  Eventually
only one of the two peaks survives. This is true, whenever the two
peaks are not equal in height. When they are exactly equal both spikes
coexist in the thermodynamic limit.

\begin{figure}[ht]
\begin{center}
\setlength{\unitlength}{0.240900pt}
\ifx\plotpoint\undefined\newsavebox{\plotpoint}\fi
\begin{picture}(1049,629)(0,0)
\font\gnuplot=cmr10 at 10pt
\gnuplot
\sbox{\plotpoint}{\rule[-0.200pt]{0.400pt}{0.400pt}}%
\put(220.0,113.0){\rule[-0.200pt]{184.288pt}{0.400pt}}
\put(220.0,113.0){\rule[-0.200pt]{0.400pt}{118.764pt}}
\put(220.0,113.0){\rule[-0.200pt]{4.818pt}{0.400pt}}
\put(198,113){\makebox(0,0)[r]{0}}
\put(965.0,113.0){\rule[-0.200pt]{4.818pt}{0.400pt}}
\put(220.0,195.0){\rule[-0.200pt]{4.818pt}{0.400pt}}
\put(198,195){\makebox(0,0)[r]{1}}
\put(965.0,195.0){\rule[-0.200pt]{4.818pt}{0.400pt}}
\put(220.0,277.0){\rule[-0.200pt]{4.818pt}{0.400pt}}
\put(198,277){\makebox(0,0)[r]{2}}
\put(965.0,277.0){\rule[-0.200pt]{4.818pt}{0.400pt}}
\put(220.0,360.0){\rule[-0.200pt]{4.818pt}{0.400pt}}
\put(198,360){\makebox(0,0)[r]{3}}
\put(965.0,360.0){\rule[-0.200pt]{4.818pt}{0.400pt}}
\put(220.0,442.0){\rule[-0.200pt]{4.818pt}{0.400pt}}
\put(198,442){\makebox(0,0)[r]{4}}
\put(965.0,442.0){\rule[-0.200pt]{4.818pt}{0.400pt}}
\put(220.0,524.0){\rule[-0.200pt]{4.818pt}{0.400pt}}
\put(198,524){\makebox(0,0)[r]{5}}
\put(965.0,524.0){\rule[-0.200pt]{4.818pt}{0.400pt}}
\put(220.0,606.0){\rule[-0.200pt]{4.818pt}{0.400pt}}
\put(198,606){\makebox(0,0)[r]{6}}
\put(965.0,606.0){\rule[-0.200pt]{4.818pt}{0.400pt}}
\put(220.0,113.0){\rule[-0.200pt]{0.400pt}{4.818pt}}
\put(220,68){\makebox(0,0){0}}
\put(220.0,586.0){\rule[-0.200pt]{0.400pt}{4.818pt}}
\put(348.0,113.0){\rule[-0.200pt]{0.400pt}{4.818pt}}
\put(348,68){\makebox(0,0){0.5}}
\put(348.0,586.0){\rule[-0.200pt]{0.400pt}{4.818pt}}
\put(475.0,113.0){\rule[-0.200pt]{0.400pt}{4.818pt}}
\put(475,68){\makebox(0,0){1}}
\put(475.0,586.0){\rule[-0.200pt]{0.400pt}{4.818pt}}
\put(603.0,113.0){\rule[-0.200pt]{0.400pt}{4.818pt}}
\put(603,68){\makebox(0,0){1.5}}
\put(603.0,586.0){\rule[-0.200pt]{0.400pt}{4.818pt}}
\put(730.0,113.0){\rule[-0.200pt]{0.400pt}{4.818pt}}
\put(730,68){\makebox(0,0){2}}
\put(730.0,586.0){\rule[-0.200pt]{0.400pt}{4.818pt}}
\put(858.0,113.0){\rule[-0.200pt]{0.400pt}{4.818pt}}
\put(858,68){\makebox(0,0){2.5}}
\put(858.0,586.0){\rule[-0.200pt]{0.400pt}{4.818pt}}
\put(985.0,113.0){\rule[-0.200pt]{0.400pt}{4.818pt}}
\put(985,68){\makebox(0,0){3}}
\put(985.0,586.0){\rule[-0.200pt]{0.400pt}{4.818pt}}
\put(220.0,113.0){\rule[-0.200pt]{184.288pt}{0.400pt}}
\put(985.0,113.0){\rule[-0.200pt]{0.400pt}{118.764pt}}
\put(220.0,606.0){\rule[-0.200pt]{184.288pt}{0.400pt}}
\put(45,359){\makebox(0,0){$P(\tilde{n},\tilde{T})$}}
\put(602,23){\makebox(0,0){$\tilde{n}$}}
\put(220.0,113.0){\rule[-0.200pt]{0.400pt}{118.764pt}}
\put(228,122){\usebox{\plotpoint}}
\multiput(228.59,122.00)(0.485,0.874){11}{\rule{0.117pt}{0.786pt}}
\multiput(227.17,122.00)(7.000,10.369){2}{\rule{0.400pt}{0.393pt}}
\multiput(235.59,134.00)(0.488,1.088){13}{\rule{0.117pt}{0.950pt}}
\multiput(234.17,134.00)(8.000,15.028){2}{\rule{0.400pt}{0.475pt}}
\multiput(243.59,151.00)(0.488,1.286){13}{\rule{0.117pt}{1.100pt}}
\multiput(242.17,151.00)(8.000,17.717){2}{\rule{0.400pt}{0.550pt}}
\multiput(251.59,171.00)(0.488,1.352){13}{\rule{0.117pt}{1.150pt}}
\multiput(250.17,171.00)(8.000,18.613){2}{\rule{0.400pt}{0.575pt}}
\multiput(259.59,192.00)(0.485,1.484){11}{\rule{0.117pt}{1.243pt}}
\multiput(258.17,192.00)(7.000,17.420){2}{\rule{0.400pt}{0.621pt}}
\multiput(266.59,212.00)(0.488,1.088){13}{\rule{0.117pt}{0.950pt}}
\multiput(265.17,212.00)(8.000,15.028){2}{\rule{0.400pt}{0.475pt}}
\multiput(274.59,229.00)(0.488,0.758){13}{\rule{0.117pt}{0.700pt}}
\multiput(273.17,229.00)(8.000,10.547){2}{\rule{0.400pt}{0.350pt}}
\multiput(282.00,241.59)(0.494,0.488){13}{\rule{0.500pt}{0.117pt}}
\multiput(282.00,240.17)(6.962,8.000){2}{\rule{0.250pt}{0.400pt}}
\multiput(290.00,249.60)(0.920,0.468){5}{\rule{0.800pt}{0.113pt}}
\multiput(290.00,248.17)(5.340,4.000){2}{\rule{0.400pt}{0.400pt}}
\multiput(305.00,251.95)(1.579,-0.447){3}{\rule{1.167pt}{0.108pt}}
\multiput(305.00,252.17)(5.579,-3.000){2}{\rule{0.583pt}{0.400pt}}
\multiput(313.00,248.93)(0.710,-0.477){7}{\rule{0.660pt}{0.115pt}}
\multiput(313.00,249.17)(5.630,-5.000){2}{\rule{0.330pt}{0.400pt}}
\multiput(320.00,243.93)(0.671,-0.482){9}{\rule{0.633pt}{0.116pt}}
\multiput(320.00,244.17)(6.685,-6.000){2}{\rule{0.317pt}{0.400pt}}
\multiput(328.00,237.93)(0.569,-0.485){11}{\rule{0.557pt}{0.117pt}}
\multiput(328.00,238.17)(6.844,-7.000){2}{\rule{0.279pt}{0.400pt}}
\multiput(336.00,230.93)(0.494,-0.488){13}{\rule{0.500pt}{0.117pt}}
\multiput(336.00,231.17)(6.962,-8.000){2}{\rule{0.250pt}{0.400pt}}
\multiput(344.59,221.69)(0.485,-0.569){11}{\rule{0.117pt}{0.557pt}}
\multiput(343.17,222.84)(7.000,-6.844){2}{\rule{0.400pt}{0.279pt}}
\multiput(351.00,214.93)(0.494,-0.488){13}{\rule{0.500pt}{0.117pt}}
\multiput(351.00,215.17)(6.962,-8.000){2}{\rule{0.250pt}{0.400pt}}
\multiput(359.00,206.93)(0.569,-0.485){11}{\rule{0.557pt}{0.117pt}}
\multiput(359.00,207.17)(6.844,-7.000){2}{\rule{0.279pt}{0.400pt}}
\multiput(367.00,199.93)(0.671,-0.482){9}{\rule{0.633pt}{0.116pt}}
\multiput(367.00,200.17)(6.685,-6.000){2}{\rule{0.317pt}{0.400pt}}
\multiput(375.00,193.93)(0.581,-0.482){9}{\rule{0.567pt}{0.116pt}}
\multiput(375.00,194.17)(5.824,-6.000){2}{\rule{0.283pt}{0.400pt}}
\multiput(382.00,187.93)(0.671,-0.482){9}{\rule{0.633pt}{0.116pt}}
\multiput(382.00,188.17)(6.685,-6.000){2}{\rule{0.317pt}{0.400pt}}
\multiput(390.00,181.93)(0.821,-0.477){7}{\rule{0.740pt}{0.115pt}}
\multiput(390.00,182.17)(6.464,-5.000){2}{\rule{0.370pt}{0.400pt}}
\multiput(398.00,176.94)(0.920,-0.468){5}{\rule{0.800pt}{0.113pt}}
\multiput(398.00,177.17)(5.340,-4.000){2}{\rule{0.400pt}{0.400pt}}
\multiput(405.00,172.94)(1.066,-0.468){5}{\rule{0.900pt}{0.113pt}}
\multiput(405.00,173.17)(6.132,-4.000){2}{\rule{0.450pt}{0.400pt}}
\multiput(413.00,168.95)(1.579,-0.447){3}{\rule{1.167pt}{0.108pt}}
\multiput(413.00,169.17)(5.579,-3.000){2}{\rule{0.583pt}{0.400pt}}
\multiput(421.00,165.95)(1.579,-0.447){3}{\rule{1.167pt}{0.108pt}}
\multiput(421.00,166.17)(5.579,-3.000){2}{\rule{0.583pt}{0.400pt}}
\put(429,162.17){\rule{1.500pt}{0.400pt}}
\multiput(429.00,163.17)(3.887,-2.000){2}{\rule{0.750pt}{0.400pt}}
\put(436,160.17){\rule{1.700pt}{0.400pt}}
\multiput(436.00,161.17)(4.472,-2.000){2}{\rule{0.850pt}{0.400pt}}
\put(444,158.17){\rule{1.700pt}{0.400pt}}
\multiput(444.00,159.17)(4.472,-2.000){2}{\rule{0.850pt}{0.400pt}}
\put(452,156.67){\rule{1.927pt}{0.400pt}}
\multiput(452.00,157.17)(4.000,-1.000){2}{\rule{0.964pt}{0.400pt}}
\put(460,155.67){\rule{1.686pt}{0.400pt}}
\multiput(460.00,156.17)(3.500,-1.000){2}{\rule{0.843pt}{0.400pt}}
\put(297.0,253.0){\rule[-0.200pt]{1.927pt}{0.400pt}}
\put(498,155.67){\rule{1.927pt}{0.400pt}}
\multiput(498.00,155.17)(4.000,1.000){2}{\rule{0.964pt}{0.400pt}}
\put(506,156.67){\rule{1.927pt}{0.400pt}}
\multiput(506.00,156.17)(4.000,1.000){2}{\rule{0.964pt}{0.400pt}}
\put(514,158.17){\rule{1.500pt}{0.400pt}}
\multiput(514.00,157.17)(3.887,2.000){2}{\rule{0.750pt}{0.400pt}}
\put(521,159.67){\rule{1.927pt}{0.400pt}}
\multiput(521.00,159.17)(4.000,1.000){2}{\rule{0.964pt}{0.400pt}}
\put(529,161.17){\rule{1.700pt}{0.400pt}}
\multiput(529.00,160.17)(4.472,2.000){2}{\rule{0.850pt}{0.400pt}}
\multiput(537.00,163.61)(1.579,0.447){3}{\rule{1.167pt}{0.108pt}}
\multiput(537.00,162.17)(5.579,3.000){2}{\rule{0.583pt}{0.400pt}}
\multiput(545.00,166.61)(1.355,0.447){3}{\rule{1.033pt}{0.108pt}}
\multiput(545.00,165.17)(4.855,3.000){2}{\rule{0.517pt}{0.400pt}}
\multiput(552.00,169.61)(1.579,0.447){3}{\rule{1.167pt}{0.108pt}}
\multiput(552.00,168.17)(5.579,3.000){2}{\rule{0.583pt}{0.400pt}}
\multiput(560.00,172.60)(1.066,0.468){5}{\rule{0.900pt}{0.113pt}}
\multiput(560.00,171.17)(6.132,4.000){2}{\rule{0.450pt}{0.400pt}}
\multiput(568.00,176.60)(0.920,0.468){5}{\rule{0.800pt}{0.113pt}}
\multiput(568.00,175.17)(5.340,4.000){2}{\rule{0.400pt}{0.400pt}}
\multiput(575.00,180.59)(0.821,0.477){7}{\rule{0.740pt}{0.115pt}}
\multiput(575.00,179.17)(6.464,5.000){2}{\rule{0.370pt}{0.400pt}}
\multiput(583.00,185.59)(0.821,0.477){7}{\rule{0.740pt}{0.115pt}}
\multiput(583.00,184.17)(6.464,5.000){2}{\rule{0.370pt}{0.400pt}}
\multiput(591.00,190.59)(0.671,0.482){9}{\rule{0.633pt}{0.116pt}}
\multiput(591.00,189.17)(6.685,6.000){2}{\rule{0.317pt}{0.400pt}}
\multiput(599.00,196.59)(0.581,0.482){9}{\rule{0.567pt}{0.116pt}}
\multiput(599.00,195.17)(5.824,6.000){2}{\rule{0.283pt}{0.400pt}}
\multiput(606.00,202.59)(0.671,0.482){9}{\rule{0.633pt}{0.116pt}}
\multiput(606.00,201.17)(6.685,6.000){2}{\rule{0.317pt}{0.400pt}}
\multiput(614.00,208.59)(0.569,0.485){11}{\rule{0.557pt}{0.117pt}}
\multiput(614.00,207.17)(6.844,7.000){2}{\rule{0.279pt}{0.400pt}}
\multiput(622.00,215.59)(0.569,0.485){11}{\rule{0.557pt}{0.117pt}}
\multiput(622.00,214.17)(6.844,7.000){2}{\rule{0.279pt}{0.400pt}}
\multiput(630.00,222.59)(0.492,0.485){11}{\rule{0.500pt}{0.117pt}}
\multiput(630.00,221.17)(5.962,7.000){2}{\rule{0.250pt}{0.400pt}}
\multiput(637.00,229.59)(0.671,0.482){9}{\rule{0.633pt}{0.116pt}}
\multiput(637.00,228.17)(6.685,6.000){2}{\rule{0.317pt}{0.400pt}}
\multiput(645.00,235.59)(0.671,0.482){9}{\rule{0.633pt}{0.116pt}}
\multiput(645.00,234.17)(6.685,6.000){2}{\rule{0.317pt}{0.400pt}}
\multiput(653.00,241.59)(0.581,0.482){9}{\rule{0.567pt}{0.116pt}}
\multiput(653.00,240.17)(5.824,6.000){2}{\rule{0.283pt}{0.400pt}}
\multiput(660.00,247.61)(1.579,0.447){3}{\rule{1.167pt}{0.108pt}}
\multiput(660.00,246.17)(5.579,3.000){2}{\rule{0.583pt}{0.400pt}}
\multiput(668.00,250.61)(1.579,0.447){3}{\rule{1.167pt}{0.108pt}}
\multiput(668.00,249.17)(5.579,3.000){2}{\rule{0.583pt}{0.400pt}}
\put(467.0,156.0){\rule[-0.200pt]{7.468pt}{0.400pt}}
\put(684,251.17){\rule{1.500pt}{0.400pt}}
\multiput(684.00,252.17)(3.887,-2.000){2}{\rule{0.750pt}{0.400pt}}
\multiput(691.00,249.93)(0.821,-0.477){7}{\rule{0.740pt}{0.115pt}}
\multiput(691.00,250.17)(6.464,-5.000){2}{\rule{0.370pt}{0.400pt}}
\multiput(699.00,244.93)(0.569,-0.485){11}{\rule{0.557pt}{0.117pt}}
\multiput(699.00,245.17)(6.844,-7.000){2}{\rule{0.279pt}{0.400pt}}
\multiput(707.59,236.72)(0.488,-0.560){13}{\rule{0.117pt}{0.550pt}}
\multiput(706.17,237.86)(8.000,-7.858){2}{\rule{0.400pt}{0.275pt}}
\multiput(715.59,226.74)(0.485,-0.874){11}{\rule{0.117pt}{0.786pt}}
\multiput(714.17,228.37)(7.000,-10.369){2}{\rule{0.400pt}{0.393pt}}
\multiput(722.59,214.89)(0.488,-0.824){13}{\rule{0.117pt}{0.750pt}}
\multiput(721.17,216.44)(8.000,-11.443){2}{\rule{0.400pt}{0.375pt}}
\multiput(730.59,201.47)(0.488,-0.956){13}{\rule{0.117pt}{0.850pt}}
\multiput(729.17,203.24)(8.000,-13.236){2}{\rule{0.400pt}{0.425pt}}
\multiput(738.59,186.26)(0.485,-1.026){11}{\rule{0.117pt}{0.900pt}}
\multiput(737.17,188.13)(7.000,-12.132){2}{\rule{0.400pt}{0.450pt}}
\multiput(745.59,172.68)(0.488,-0.890){13}{\rule{0.117pt}{0.800pt}}
\multiput(744.17,174.34)(8.000,-12.340){2}{\rule{0.400pt}{0.400pt}}
\multiput(753.59,158.89)(0.488,-0.824){13}{\rule{0.117pt}{0.750pt}}
\multiput(752.17,160.44)(8.000,-11.443){2}{\rule{0.400pt}{0.375pt}}
\multiput(761.59,146.51)(0.488,-0.626){13}{\rule{0.117pt}{0.600pt}}
\multiput(760.17,147.75)(8.000,-8.755){2}{\rule{0.400pt}{0.300pt}}
\multiput(769.59,136.45)(0.485,-0.645){11}{\rule{0.117pt}{0.614pt}}
\multiput(768.17,137.73)(7.000,-7.725){2}{\rule{0.400pt}{0.307pt}}
\multiput(776.00,128.93)(0.671,-0.482){9}{\rule{0.633pt}{0.116pt}}
\multiput(776.00,129.17)(6.685,-6.000){2}{\rule{0.317pt}{0.400pt}}
\multiput(784.00,122.93)(0.821,-0.477){7}{\rule{0.740pt}{0.115pt}}
\multiput(784.00,123.17)(6.464,-5.000){2}{\rule{0.370pt}{0.400pt}}
\multiput(792.00,117.95)(1.579,-0.447){3}{\rule{1.167pt}{0.108pt}}
\multiput(792.00,118.17)(5.579,-3.000){2}{\rule{0.583pt}{0.400pt}}
\put(800,114.67){\rule{1.686pt}{0.400pt}}
\multiput(800.00,115.17)(3.500,-1.000){2}{\rule{0.843pt}{0.400pt}}
\put(807,113.67){\rule{1.927pt}{0.400pt}}
\multiput(807.00,114.17)(4.000,-1.000){2}{\rule{0.964pt}{0.400pt}}
\put(815,112.67){\rule{1.927pt}{0.400pt}}
\multiput(815.00,113.17)(4.000,-1.000){2}{\rule{0.964pt}{0.400pt}}
\put(676.0,253.0){\rule[-0.200pt]{1.927pt}{0.400pt}}
\put(823.0,113.0){\rule[-0.200pt]{37.099pt}{0.400pt}}
\put(228,113){\usebox{\plotpoint}}
\put(228.00,113.00){\usebox{\plotpoint}}
\multiput(235,115)(15.620,13.668){0}{\usebox{\plotpoint}}
\multiput(243,122)(6.819,19.604){2}{\usebox{\plotpoint}}
\multiput(251,145)(3.412,20.473){2}{\usebox{\plotpoint}}
\multiput(259,193)(1.929,20.666){4}{\usebox{\plotpoint}}
\multiput(266,268)(1.798,20.677){4}{\usebox{\plotpoint}}
\multiput(274,360)(1.879,20.670){4}{\usebox{\plotpoint}}
\multiput(282,448)(2.461,20.609){3}{\usebox{\plotpoint}}
\multiput(290,515)(3.962,20.374){2}{\usebox{\plotpoint}}
\put(302.47,551.68){\usebox{\plotpoint}}
\put(310.55,534.66){\usebox{\plotpoint}}
\multiput(313,527)(3.335,-20.486){2}{\usebox{\plotpoint}}
\multiput(320,484)(3.156,-20.514){3}{\usebox{\plotpoint}}
\multiput(328,432)(3.156,-20.514){2}{\usebox{\plotpoint}}
\multiput(336,380)(3.344,-20.484){2}{\usebox{\plotpoint}}
\multiput(344,331)(3.335,-20.486){3}{\usebox{\plotpoint}}
\put(355.29,268.70){\usebox{\plotpoint}}
\multiput(359,252)(5.519,-20.008){2}{\usebox{\plotpoint}}
\put(371.95,208.76){\usebox{\plotpoint}}
\put(379.16,189.30){\usebox{\plotpoint}}
\put(388.41,170.78){\usebox{\plotpoint}}
\multiput(390,168)(12.208,-16.786){0}{\usebox{\plotpoint}}
\put(400.60,154.02){\usebox{\plotpoint}}
\multiput(405,149)(16.604,-12.453){0}{\usebox{\plotpoint}}
\put(416.46,140.84){\usebox{\plotpoint}}
\multiput(421,138)(19.434,-7.288){0}{\usebox{\plotpoint}}
\put(435.30,132.30){\usebox{\plotpoint}}
\multiput(436,132)(20.136,-5.034){0}{\usebox{\plotpoint}}
\multiput(444,130)(20.136,-5.034){0}{\usebox{\plotpoint}}
\put(455.48,127.57){\usebox{\plotpoint}}
\multiput(460,127)(20.547,-2.935){0}{\usebox{\plotpoint}}
\multiput(467,126)(20.756,0.000){0}{\usebox{\plotpoint}}
\put(476.13,126.00){\usebox{\plotpoint}}
\multiput(483,126)(20.756,0.000){0}{\usebox{\plotpoint}}
\put(496.88,126.00){\usebox{\plotpoint}}
\multiput(498,126)(20.595,2.574){0}{\usebox{\plotpoint}}
\multiput(506,127)(20.595,2.574){0}{\usebox{\plotpoint}}
\put(517.48,128.50){\usebox{\plotpoint}}
\multiput(521,129)(20.136,5.034){0}{\usebox{\plotpoint}}
\multiput(529,131)(19.434,7.288){0}{\usebox{\plotpoint}}
\put(537.38,134.14){\usebox{\plotpoint}}
\multiput(545,137)(18.021,10.298){0}{\usebox{\plotpoint}}
\put(555.86,143.41){\usebox{\plotpoint}}
\multiput(560,146)(15.620,13.668){0}{\usebox{\plotpoint}}
\put(571.22,157.14){\usebox{\plotpoint}}
\multiput(575,162)(12.208,16.786){0}{\usebox{\plotpoint}}
\put(583.50,173.87){\usebox{\plotpoint}}
\put(593.40,192.09){\usebox{\plotpoint}}
\put(601.30,211.24){\usebox{\plotpoint}}
\multiput(606,226)(6.104,19.838){2}{\usebox{\plotpoint}}
\put(618.75,270.99){\usebox{\plotpoint}}
\multiput(622,284)(4.503,20.261){2}{\usebox{\plotpoint}}
\multiput(630,320)(3.412,20.473){2}{\usebox{\plotpoint}}
\multiput(637,362)(3.713,20.421){2}{\usebox{\plotpoint}}
\multiput(645,406)(3.556,20.449){2}{\usebox{\plotpoint}}
\multiput(653,452)(3.412,20.473){2}{\usebox{\plotpoint}}
\multiput(660,494)(4.754,20.204){2}{\usebox{\plotpoint}}
\put(670.86,535.51){\usebox{\plotpoint}}
\put(682.13,550.53){\usebox{\plotpoint}}
\put(690.82,533.45){\usebox{\plotpoint}}
\put(694.98,513.12){\usebox{\plotpoint}}
\multiput(699,493)(2.836,-20.561){3}{\usebox{\plotpoint}}
\multiput(707,435)(2.425,-20.613){4}{\usebox{\plotpoint}}
\multiput(715,367)(2.095,-20.650){3}{\usebox{\plotpoint}}
\multiput(722,298)(2.656,-20.585){3}{\usebox{\plotpoint}}
\multiput(730,236)(3.344,-20.484){2}{\usebox{\plotpoint}}
\multiput(738,187)(4.070,-20.352){2}{\usebox{\plotpoint}}
\put(748.29,143.78){\usebox{\plotpoint}}
\put(757.74,125.48){\usebox{\plotpoint}}
\multiput(761,121)(17.601,-11.000){0}{\usebox{\plotpoint}}
\put(774.56,114.41){\usebox{\plotpoint}}
\multiput(776,114)(20.595,-2.574){0}{\usebox{\plotpoint}}
\multiput(784,113)(20.756,0.000){0}{\usebox{\plotpoint}}
\put(795.20,113.00){\usebox{\plotpoint}}
\multiput(800,113)(20.756,0.000){0}{\usebox{\plotpoint}}
\multiput(807,113)(20.756,0.000){0}{\usebox{\plotpoint}}
\put(815.95,113.00){\usebox{\plotpoint}}
\multiput(823,113)(20.756,0.000){0}{\usebox{\plotpoint}}
\put(836.71,113.00){\usebox{\plotpoint}}
\multiput(838,113)(20.756,0.000){0}{\usebox{\plotpoint}}
\multiput(846,113)(20.756,0.000){0}{\usebox{\plotpoint}}
\put(857.46,113.00){\usebox{\plotpoint}}
\multiput(861,113)(20.756,0.000){0}{\usebox{\plotpoint}}
\multiput(869,113)(20.756,0.000){0}{\usebox{\plotpoint}}
\put(878.22,113.00){\usebox{\plotpoint}}
\multiput(885,113)(20.756,0.000){0}{\usebox{\plotpoint}}
\put(898.97,113.00){\usebox{\plotpoint}}
\multiput(900,113)(20.756,0.000){0}{\usebox{\plotpoint}}
\multiput(908,113)(20.756,0.000){0}{\usebox{\plotpoint}}
\put(919.73,113.00){\usebox{\plotpoint}}
\multiput(923,113)(20.756,0.000){0}{\usebox{\plotpoint}}
\multiput(931,113)(20.756,0.000){0}{\usebox{\plotpoint}}
\put(940.48,113.00){\usebox{\plotpoint}}
\multiput(946,113)(20.756,0.000){0}{\usebox{\plotpoint}}
\put(961.24,113.00){\usebox{\plotpoint}}
\multiput(962,113)(20.756,0.000){0}{\usebox{\plotpoint}}
\multiput(970,113)(20.756,0.000){0}{\usebox{\plotpoint}}
\put(977,113){\usebox{\plotpoint}}
\end{picture}

\caption{\label{voldep2} The distribution function
$P(\tilde{n},\tilde{T})$ along the isotherm $\tilde{T}=0.8971$ for two
different values of $\tilde{V}=20$ (solid line) and $60$ (dotted
line). The chemical potential is that
of coexistence, $\tilde{\mu}^{\rm ext}=-14.8971$. Because the heights
are equal, as the typical volume $\tilde{V}$ of the cells increases,
they remain equal at later times. In the thermodynamic limit two
spikes localized at the gas and liquid densities remain. }
\end{center}
\end{figure}
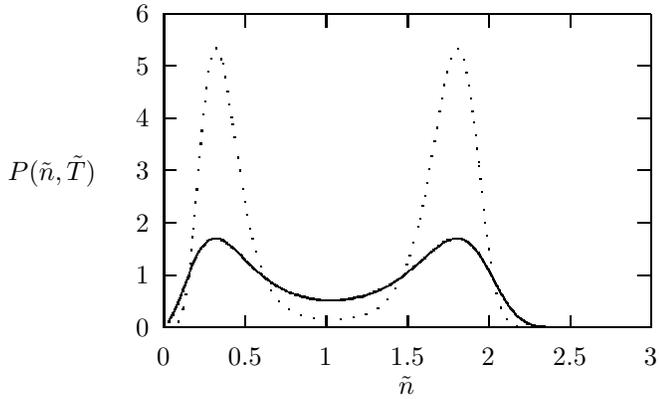

\section{Discussion}
\label{discusion}
We have presented a Lagrangian finite Voronoi volume discretization of
continuum hydrodynamics and have shown that the discrete equations
have almost the GENERIC structure. The GENERIC structure can be
restored by simply adding a small term into the reversible part of the
dynamics. Therefore, the obtained equations conserve mass, momentum,
and energy, and the entropy is an strictly increasing function of time
in the absence of fluctuations. Thermal fluctuations are consistently
included which lead to the strict increase of the entropy functional
and to the correct Einstein distribution function. The size of the
fluctuations is given by the typical size of the volumes of the
particles, arguably scaling like the square root of this volume. The
need of incorporating thermal fluctuations in a particular system will
be determined by the external length scales that need to be
resolved. For example, if sub-micron colloidal particles are
considered, we need to resolve the size of the colloidal particle with
fluid particles of size, say, an order of magnitude or two smaller
than the diameter of the colloidal particle. For these small volumes,
fluctuations are important and lead to the Brownian motion of the
particle. A ping-pong ball, on the other hand requires fluid particles
much larger, for which thermal fluctuations are negligible. Of course,
one could use a very large number of small fluid particles to deal
with the ping-pong ball, but in this case the (large) thermal
fluctuations on each fluid particle average out among the (large)
number of fluid particles.

The original formulations of Dissipative Particle Dynamics lack this
effect of switching-off thermal fluctuations depending on the size of
the fluid particles. This is due to the fact that early formulations
did not include the volume and/or the mass of the particles as a
relevant dynamical variable. In this paper, we have extended the range
of variables and have shown that a thermodynamically consistent
Dissipative Particle Dynamics model can be formulated. This model has
the same reversible part as the finite Voronoi volume model. Hence,
the usual conservative forces between dissipative particles are
substituted by truly pressure forces. Therefore, prescribed
thermodynamics can be given, without the need of ``reverse inference''
(find out which conservative force will produce the desired equation
of state \cite{Groot97}). Even though the present DPD model has clear
advantages over previous DPD models, it is yet inferior to the finite
Voronoi volume model presented in this paper. Note that the finite
volume is closely related to a discretization of Navier-Stokes
equations, whereas the DPD model in this paper does not converge to
the Navier-Stokes equations, even though it displays correct
hydrodynamic behaviour. This means that in order to simulate a fluid
of a given viscosity, one has to tune the viscosity parameter of the
DPD model by previous simulation runs. With the finite volume model
the true viscosity is given as input.

We have shown also that the Smoothed Particle Dynamics method has the
GENERIC structure. However, we have pointed out several problems that
favour the use of the finite volume method of the present paper. For
example, in the Smoothed Particle Dynamics method we have noted that
remnant forces appear even for the situation in which all particles
are at rest with identical pressures. This effect can be made
arbitrarily small by increasing the overlapping coefficient. But this
has two drawbacks. The first one is of practical nature: when the
overlapping coefficient is increased the number of interacting
neighbours increases and, therefore, the computational time increases
too. A second, more conceptual drawback is that in the large
overlapping limit which, strictly speaking, is needed in order to
``deduce'' the SPH equations for Navier-Stokes, the volume (or
density) of the particles becomes a constant and the equations start
loosing its sense. The pressures, which depend essentially on the
volumes, become a constant and the resulting equations cannot be
claimed to be a faithful discretization of Navier-Stokes equations.
Even though hydrodynamic behaviour is displayed (momentum is conserved
locally) we expect that the thermodynamics and the transport
properties of the simulated fluid differ from the actual desired
values.

Concerning the GENERIC finite volume model of fluid particles
presented in this paper, one of the most surprising realizations has
been the fact that the mass of the fluid particles changes due to the
reversible part of the dynamic. In Refs. \cite{Espanol-prl99},
\cite{Flekkoy99} this was overlooked. Of course, we could, in
principle, formulate a model with ${\bf c}_{\mu\nu}= 0$ in the final
discrete equations (\ref{REVER}). This would result in a great
simplification of the equations without losing the desired GENERIC
properties. This approximation would imply that the mass of the
particles are constant. This spoils the equilibrium distribution
function that would not be given by the Einstein distribution function
(\ref{ein}) but rather by

\begin{eqnarray}
\rho^{\rm eq}(x)&=& 
\frac{1}{\Omega}\prod_\mu^M
\delta(M_\mu(x)-M_\mu^0)\delta(E(x)-E_0)\delta({\bf P}(x)-{\bf P}_0),
\nonumber\\
&\times &\exp\{k_B^{-1} S(x)\}
\label{einmd0}
\end{eqnarray}
where $M_\mu^0$ is the initial value of the mass of particle $\mu$.
In this case, not only total mass but also the individual masses are
dynamical invariants of the discrete equations when ${\bf
c}_{\mu\nu}=0$.  We note that we could not follow the same arguments
that lead to (\ref{pmsap}) and which describe the liquid-gas
coexistence of a van der Waals fluid.  This very same argument can be
applied to the case of the SPH model discussed in section
\ref{sphmodel} which renders the SPH model unsuitable to discuss
gas-liquid coexistence. From a theoretical point of view, the
possibility of simulating liquid-gas coexistence with a particle model
represents a stringent condition on the thermodynamic consistency of
the model proposed. Actually, we have learnt that the mass, and not
the volume, of the fluid particles should be included as an
independent dynamic variable if coexistence is to be described.

Finally, we would like to note that the model presente does not
exibit the phenomena of surface tension. Surface tension can
be easily included as an extra conservative contribution to the
energy function \cite{Preprint}.

\section*{Acknowledgments}
We are grateful to E.G. Flekkoy and P.V. Coveney for exchange of
preprints and useful comments. Helpful conversations and comments by
H.C. \"Ottinger are greatly acknowledged. Discussions with R. Delgado
and M. Ripoll are appreciated. This work has been partially supported
by DGYCIT PB97-0077.

\section{Appendix: Voronoi properties}
\label{ap-vol}
In this appendix we explicitly compute the derivative of the volume of
cell $\mu$ with respect to the position ${\bf R}_\nu$ of cell $\nu$,
this is

\begin{equation}
{\bf G}_{\mu\nu}=\frac{\partial{{\cal V}_\mu}}{{\partial {\bf R}_\nu}}
=\frac{1}{\sigma^2}
\int_{V_T}d{\bf R}
\chi_\mu({\bf R})(\delta_{\mu\nu}-\chi_\nu({\bf R}))({\bf R}-{\bf R}_\nu).
\label{omvor}
\end{equation}
Its worth considering the cases $\mu\neq \nu$ and $\mu=\nu$ explicitly.
\begin{eqnarray}
\mbox{\boldmath $G$}_{\mu\nu}
&=&-\frac{1}{\sigma^2}
\int_{V_T}d{\bf R}
\chi_\mu({\bf R})\chi_\nu({\bf R})({\bf R}-{\bf R}_\nu),
\quad\quad \nu\neq \mu
\nonumber\\
\mbox{\boldmath $G$}_{\mu\mu}
&=&
\frac{1}{\sigma^2}
\int_{V_T}d{\bf R}
\chi_\mu({\bf R})(1-\chi_\mu({\bf R}))({\bf R}-{\bf R}_\mu)
\nonumber\\
&=&
\sum_{\nu\neq \mu}\frac{1}{\sigma^2}
\int_{V_T}d{\bf R}
\chi_\mu({\bf R})\chi_\nu({\bf R})({\bf R}-{\bf R}_\mu)
\nonumber\\
&=&-\sum_{\nu\neq\mu}\mbox{\boldmath $G$}_{\nu\mu},
\label{omvor2}
\end{eqnarray}

It is convenient to rewrite Eqn. (\ref{omvor2}) for $\nu\neq\mu$ as

\begin{eqnarray}
\mbox{\boldmath $G$}_{\mu\nu}
&=&-\frac{1}{\sigma^2}
\int_{V_T}d{\bf R}
\chi_\mu({\bf R})\chi_\mu({\bf R})
\left({\bf R}-\frac{{\bf R}_\mu+{\bf R}_\nu}{2}\right)
\nonumber\\
&&-{\bf R}_{\mu\nu}\frac{1}{2\sigma^2}
\int_{V_T}d{\bf R}
\chi_\mu({\bf R})\chi_\mu({\bf R})
\label{omvor3}
\end{eqnarray}
with ${\bf R}_{\mu\nu}={\bf R}_\mu-{\bf R}_\nu$.  The first task is to
compute the limit $\sigma\rightarrow0$ for these two integrals. For
this reason, it is instructive to work out the actual forms of
$\chi_\mu({\bf R})$ and $\chi_\mu({\bf R})\chi_\nu({\bf R})$ for the
case that only two particles are present in the system, as has been
done by Flekkoy and Coveney \cite{Flekkoy99}. Simple algebra leads to
\begin{eqnarray}
\chi_\mu({\bf R})
&=&\frac{1}{1+\exp\{-{\bf R}_{\mu\nu}
\!\cdot\!({\bf R}-({\bf R}_\mu+{\bf R}_\nu)/2)/\sigma^2\}}
\nonumber\\
\chi_\mu({\bf R})\chi_\nu({\bf R})
&=&\frac{1}{4\cosh^2({\bf R}_{\mu\nu}
\!\cdot\!({\bf R}-({\bf R}_\mu+{\bf R}_\nu)/2)/2\sigma^2\})}
\nonumber\\
&&\label{explicit}
\end{eqnarray}
Note that $\chi_{\mu}({\bf R})\chi_{\nu}({\bf R})$ is different from
zero only around the boundary of the Voronoi cells of particles
$\mu,\nu$. In the limit of small $\sigma$ this is even more true. The
integrals in (\ref{omvor3}) therefore can be performed not over the
full volume ${V}_T$ but only over a region $\partial_{\mu\nu}$
``around'' the boundary of the $\mu,\nu$ cells. In this region, we can
further substitute the expression of $\chi_{\mu}({\bf
r})\chi_{\nu}({\bf R})$, which depend on the positions of all the
center cells, by Eqn. (\ref{explicit}), which depends only on the
position of the centers of cells $\mu,\nu$. Actually, we can make a
translation from ${\bf R}$ to ${\bf R}'= {\bf R}-({\bf R}_\mu+{\bf
R}_\nu)/2$ (we put the origin exactly at the boundary between
cells). We can also make a rotation in such a way that the $x$ axis is
along the line joining the cell centers. In this way, we can write
\begin{eqnarray}
&&\frac{1}{\sigma^2}
\int_{V_T} d{\bf R} \chi_\mu({\bf R})\chi_\nu({\bf R})
\nonumber\\
&=&
\frac{1}{4\sigma^2}\int_{\partial_{\mu\nu}}d{\bf R}'
\frac{1}{\cosh^2({\bf R}'\!\cdot\!{\bf R}_{\mu\nu}/2\sigma^2)}
\nonumber\\
&=&
\frac{1}{4\sigma^2}A_{\mu\nu}\int_{-\infty}^{\infty}dx
\frac{1}{\cosh^2(x R_{\mu\nu}/2\sigma^2)}
\nonumber\\
&=&\frac{A_{\mu\nu}}{R_{\mu\nu}}
\label{area1}
\end{eqnarray}
Note that $\int_0^\infty \cosh^{-2}(x)dx=1$.  Here, $A_{\mu\nu}$ is
the actual area of the boundary between Voronoi cells of particles
$\mu,\nu$.

In a similar way, one computes the first integral in Eqn. (\ref{omvor3})
with the result

\begin{equation}
\frac{1}{\sigma^2}
\int_{V_T}d{\bf R}
\chi_\mu({\bf R})\chi_\mu({\bf R})
\left({\bf R}-\frac{{\bf R}_\mu+{\bf R}_\nu}{2}\right)
= \frac{A_{\mu\nu}}{R_{\mu\nu}}{\bf c}_{\mu\nu}
\label{area2}
\end{equation}
where the vector ${\bf c}_{\mu\nu}$ is, by definition, the position of
the center of mass of the face between Voronoi cells $\mu,\nu$ with
respect to the point $({\bf R}_\mu+{\bf R}_\nu)/2$. Collecting
Eqns. (\ref{omvor3}),(\ref{area2}) and (\ref{area1}) one finally
obtains

\begin{equation}
\mbox{\boldmath $G$}_{\mu\nu} =-A_{\mu\nu}\left(\frac{{\bf
c}_{\mu\nu}}{R_{\mu\nu}}+\frac{{\bf e}_{\mu\nu}}{2}\right)
\label{fin1}
\end{equation}
where 
\begin{eqnarray}
{\bf e}_{\mu\nu}& =& \frac{{\bf R}_{\mu\nu}}{R_{\mu\nu}}
\nonumber\\
{\bf R}_{\mu\nu}&=&{\bf R}_{\mu}-{\bf R}_\nu
\nonumber\\
R_{\mu\nu}&=& |{\bf R}_{\mu\nu}|
\label{defis3}
\end{eqnarray}

Note that $\sum_\nu {\bf G}_{\mu\nu}=0$ due to Eqn. (\ref{omvor2}).
Also $\sum_\mu {\bf G}_{\mu\nu}=0$ because of Eqn. (\ref{omvor})
and the fact that total volume is a constant, independent of positions.

\section{Appendix: Dependent vs. independent variables in GENERIC}
\label{ap-dep}
What happens if in the description of the state of a system one
introduces, perhaps without knowing it, variables which are not
independent of each other?  We provide in this appendix the 
answer to this question.

Let us assume that the state of a system is described by
a set of {\em independent} variables $x$. Assume, for the
sake of simplicity, that the dynamics is purely
reversible. The case $M\neq 0$ follows along a similar way as
the one presented here for the reversible part.

Now assume that the basic building blocks of GENERIC have
the following structure

\begin{eqnarray}
E(x) &=& \overline{E}(x,y(x))
\nonumber\\
S(x) &=& \overline{S}(x,y(x))
\nonumber\\
L(x) &=& \overline{L}(x,y(x))
\end{eqnarray}
where $y(x)$ is a prescribed (vector) function of the sate $x$.
From the chain rule one has

\begin{eqnarray}
\nabla E(x) = \nabla_x\overline{E}(x,y(x))+J^T(x)\nabla_y
\overline{E}(x,y(x))
\label{derove}
\end{eqnarray}
and similarly for $S(x)$ and $L(x)$. Here, the matrix $J(x)$ is
defined by $J(x)=\partial y/\partial x$.

The GENERIC reversible part of the dynamics is given by
\begin{eqnarray}
\dot{x}&=&L(x)\nabla E(x)
\nonumber\\
&=&
\overline{L}(x,y(x))\nabla_x\overline{E}(x,y(x))
\nonumber\\
&+&\overline{L}(x,y(x))J^T(x)\nabla_y\overline{E}(x,y(x))
\label{eq1}
\end{eqnarray}
We can consider the time derivative of the functions $y(x(t))$
which, again through the chain rule, is given by

\begin{equation}
\dot{y}(x) = J(x)\dot{x}
\label{eq2}
\end{equation}
By using now Eqn. (\ref{eq1}) into (\ref{eq2}) we can finally
group both equation into the form

\begin{equation}
\left(
\begin{array}{c}
\dot{x}\\
\\
\dot{y}\\
\end{array}\right)
=
\left(
\begin{array}{ccc}
\overline{L}&& \overline{L}J^T  \\
\\
J\overline{L}&& J\overline{L}J^T  \\
\end{array}\right)
\left(
\begin{array}{c}
\nabla_x\overline{E}\\
\\
\nabla_y\overline{E}\\
\end{array}\right)
\label{compact0}
\end{equation}
or, by introducing the  obvious notation $z=(x,y(x))$,
\begin{equation}
\dot{z}={\cal L}(z)\nabla_z E(z)
\label{compact}
\end{equation}
A very striking nicety is that the matrix ${\cal L}(z)$ {\em has all
the required properties for being a proper GENERIC reversible
matrix}. This is
\begin{eqnarray}
{\cal L}^T(z)&=& -{\cal L}(z)
\nonumber\\
{\cal L}(z)\nabla_z S(z) &=& 0
\nonumber\\
\nabla_z[{\cal L}(z)\nabla_z E(z)] &=& 0
\label{allprop}
\end{eqnarray}
as can be easily checked from the properties of $L(x)$.

Now, let us assume that we start describing the state of
a system by a vector $z=(x,y)$ and that the matrix $L(z)$
has the structure

\begin{equation}
\left(
\begin{array}{ccc}
\overline{L}(x,y)&& \overline{L}(x,y)J^T(x)  \\
\\
J(x)\overline{L}(x,y)&& J(x)\overline{L}(x,y)J^T(x)  \\
\end{array}\right)
\label{compact2}
\end{equation}
with $J(x) = \partial h/\partial x$ for some set of functions
 $h(x)$. Then, we will have

\begin{equation}
\frac{d}{dt}y 
= \frac{\partial h}{\partial x}(x) \dot{x}=\frac{d}{dt}h(x)
\label{148}
\end{equation}
and, therefore, $y=h(x)$. Therefore, if in a given model the matrix
 $L(z)$ happens to have the structure given in (\ref{compact2}), then,
 necessarily, the variables $y$ are dependent on the variables $x$.

In our formulation in Ref. \cite{Espanol-prl99} we have assumed that 
the volume evolves according to
\begin{equation}
\dot{\cal V}_\mu = \sum_\nu{\bf \Omega}_{\mu\nu}\dot{\bf R}_\nu
\end{equation}
which for most reasonable forms of ${\bf \Omega}_{\mu\nu}$ has
the form of Eqn. (\ref{148}). Therefore, the dynamical equation
for the volume is nothing else than application of the chain
rule and the volume cannot be considered as a truly independent
variable of positions.

\section{Appendix: Molecular ensemble}
\label{ap-mol}
In this appendix we want to compute explicitly the following integral

\begin{eqnarray}
\Phi({\bf P}_0,E_0,M)&=&\int d^{DM}{\bf P}
\,\,\delta\!\left(\sum_\mu^M \frac{{\bf P}_\mu^2}{2M_\mu}-E_0\right)
\nonumber\\
&\times&
\delta^D\!\left(\sum_\mu^M{\bf P}_\mu-{\bf P}_0\right)
\nonumber\\
&=&
\prod_\mu^M(2M_\mu)^{D/2}\int d^{DM}{\bf P}
\,\,\delta\!\left(\sum_\mu^M {\bf P}_\mu^2-E_0\right)
\nonumber\\
&\times&
\delta^D\!\left(\sum_\mu^M(2M_\mu)^{1/2}
{\bf P}_\mu-{\bf P}_0\right)
\label{4}
\end{eqnarray}
which appears repeatedly when computing molecular averages.  The
equation $\sum_\mu^M(2M_\mu)^{1/2}{\bf P}_\mu={\bf P}_0$ are actually D
equations (one for each component of the momentum) which define D
planes in $R^{DM}$. The integral in (\ref{4}) is actually over a
submanifold which is the intersection of the D planes with the surface
of a $DM$ dimensional sphere of radius $E_0^{1/2}$. This
intersection will be also a sphere, which will be now of smaller
radius and also of smaller dimension, $D(M-1)$.

In order to compute (\ref{4}), we change to the following notation

\begin{eqnarray}
\mbox{\boldmath ${\cal P}$}
&=&(p_1^x,\ldots,p_M^x,p_1^y,\ldots,p_M^y,p_1^z,\ldots,p_M^z)
\nonumber\\
\mbox{\boldmath ${\cal C}$}_x &=&((2M_1)^{1/2},\ldots,(2M_M)^{1/2},0,\ldots,0,0,\ldots,0)
\nonumber\\
\mbox{\boldmath ${\cal C}$}_y &=&(0,\ldots,0,(2M_1)^{1/2},\ldots,(2M_M)^{1/2},0,\ldots,0)
\nonumber\\
\mbox{\boldmath ${\cal C}$}_z &=&(0,\ldots,0,0,\ldots,0,(2M_1)^{1/2},\ldots,(2M_M)^{1/2})
\label{defis}
\end{eqnarray}
Note that these vectors satisfy
 $\mbox{\boldmath ${\cal C}$}_x\!\cdot\!\mbox{\boldmath ${\cal C}$}_y=0$
 $\mbox{\boldmath ${\cal C}$}_y\!\cdot\!\mbox{\boldmath ${\cal C}$}_z=0$
 $\mbox{\boldmath ${\cal C}$}_z\!\cdot\!\mbox{\boldmath ${\cal C}$}_x=0$. 
With these vectors so defined, Eqn. (\ref{4}) becomes
\begin{eqnarray}
\Phi({\bf P}_0,E_0,M)
&=&\int d^{DM}\mbox{\boldmath ${\cal P}$}
\delta\!\left(\mbox{\boldmath ${\cal P}$}^2-E\right)
\,\,\delta\!\left(
\mbox{\boldmath ${\cal C}$}_x\!\cdot\!
\mbox{\boldmath ${\cal P}$}-P_0^x\right)
\nonumber\\
&\times&
\,\,\delta\!\left(
\mbox{\boldmath ${\cal C}$}_y\!\cdot\!
\mbox{\boldmath ${\cal P}$}-P_0^y\right)
\,\,\delta\!\left(
\mbox{\boldmath ${\cal C}$}_z\!\cdot\!
\mbox{\boldmath ${\cal P}$}-P_0^z\right)
\nonumber\\
&\times& \prod_\mu^M(2M_\mu)^{D/2}
\label{4b}
\end{eqnarray}

Now we consider the following change of variables

\begin{equation}
\mbox{\boldmath ${\cal P}$}'=\mbox{\boldmath ${\cal P}$}-
\left(
\frac{\mbox{\boldmath ${\cal C}$}_x}{|\mbox{\boldmath ${\cal C}$}_x|^2}
 P_0^x
+
\frac{\mbox{\boldmath ${\cal C}$}_y}{|\mbox{\boldmath ${\cal C}$}_y|^2} P_0^y
+
\frac{\mbox{\boldmath ${\cal C}$}_z}{|\mbox{\boldmath ${\cal C}$}_z|^2} P_0^z
\right)
\label{tra}
\end{equation}
which is simply a translation and has unit Jacobian. Simple algebra 
leads to

\begin{eqnarray}
\Phi({\bf P}_0,E_0,M)&=& \prod_\mu^M(2M_\mu)^{D/2}
\int d^{DM}\mbox{\boldmath ${\cal P}$}'
\,\,\delta\!\left(
\frac{{\mbox{\boldmath ${\cal P}$'}^2}}{2m}
-U_0\right)
\nonumber\\
&\times&
\delta\!\left(\mbox{\boldmath ${\cal
C}$}_x\!\cdot\!\mbox{\boldmath ${\cal P}$}'\right)
\,\,\delta\!\left(\mbox{\boldmath ${\cal
C}$}_y\!\cdot\!\mbox{\boldmath ${\cal P}$}'\right)
\,\,\delta\!\left(\mbox{\boldmath ${\cal
C}$}_z\!\cdot\!\mbox{\boldmath ${\cal P}$}'\right)
\label{4c}
\end{eqnarray}
where we have introduced the total internal energy

\begin{equation}
U_0=\left(E_0-\frac{{\bf P}_0^2}{2{\cal M}_0}\right)
\label{u}
\end{equation}
where ${\cal M}_0=\sum_\mu M_\mu$ is the total mass.

We now consider a second change of variables $\mbox{\boldmath ${\cal
P}$}''={\bf \Lambda}\!\cdot\!\mbox{\boldmath ${\cal P}$}'$ through a
rotation ${\bf \Lambda}$ such that

\begin{eqnarray}
{\bf \Lambda}\mbox{\boldmath ${\cal C}$}_x 
&=&(2{\cal M}_0)^{1/2}(1,\ldots,0,0,\ldots,0,0,\ldots,0)
\nonumber\\
{\bf \Lambda}\mbox{\boldmath ${\cal C}$}_y 
&=&(2{\cal M}_0)^{1/2}(0,\ldots,0,1,\ldots,0,0,\ldots,0)
\nonumber\\
{\bf \Lambda}\mbox{\boldmath ${\cal C}$}_z 
&=&(2{\cal M}_0)^{1/2}(0,\ldots,0,0,\ldots,0,1,\ldots,0)
\label{defis2b}
\end{eqnarray}
It is always possible to find a matrix ${\bf \Lambda}$ that satisfies
Eqns. (\ref{defis2b}).  For example, consider a block diagonal matrix
made of three identical blocks of size $M\times M$. Then assume that
each block is the same orthogonal matrix which transforms the vector
$((2M_1)^{1/2},(2M_2)^{1/2},\ldots,(2M_M)^{1/2})$ into $(2{\cal
M}_0)^{1/2}(1,0,\ldots,0)$. After the rotation (which has unit
Jacobian and leaves the modulus of a vector invariant) the integral
(\ref{4c}) becomes

\begin{eqnarray}
\Phi&=&
\prod_\mu^M(2M_\mu)^{D/2}
\int d^{DM}\mbox{\boldmath ${\cal P}$}''
\delta\!\left(\mbox{\boldmath ${\cal P}$}''^2-U_0\right)
\nonumber\\
&\times&
\,\,\delta((2{\cal M}_0)^{1/2}p_1''^x)
\,\,\delta((2{\cal M}_0)^{1/2}p_1''^y)
\,\,\delta((2{\cal M}_0)^{1/2}p_1''^z)
\nonumber\\
&=&\prod_\mu^M(2M_\mu)^{D/2}
\int d^{D(M-1)}\mbox{\boldmath ${\cal P}$}''
\delta\!\left(\mbox{\boldmath ${\cal P}$}''^2-U_0\right).
\label{10}
\end{eqnarray}

We compute now the integral over the sphere in Eqn. (\ref{10}) by
using that the integral of an arbitrary function $F({\bf x}) = f(|{\bf
x}|) $ that depends on ${\bf x}$ only through its modulus $|{\bf x}|$
can be computed by changing to polar coordinates
\begin{equation}
\int F({\bf x})d^M{\bf x}=\omega_M\int_0^{\infty} f(r) r^{M-1}dr.
\label{0}
\end{equation}
The numerical factor $\omega_M$, which comes from the integration of
the angles, can be computed by considering the special case when
$f(r)$ is a Gaussian. The result is
\begin{equation}
\omega_M = 2 \frac{\pi^{M/2}}{\Gamma(M/2)}
\label{1b}
\end{equation}
By using Eqn. (\ref{0}), Eqn. (\ref{10}) becomes

\begin{eqnarray}
\Phi({\bf P}_0,E_0,M)
&=& \prod_\mu^M(2M_\mu)^{D/2}\omega_{D(M-1)}
\nonumber\\
&\times&\int dp \,\,p^{D(M-1)-1}
\,\,\delta\!\left(p^2-U\right)
\label{10b}
\end{eqnarray}
We need now the property

\begin{equation}
\delta(f(x))=\sum_\mu\frac{\delta(x-x_\mu)}{|f'(x_\mu)|}
\end{equation}
where $x_\mu$ are the zeros of $f(x_\mu)=0$. For the case of Eqn. (\ref{10b}) we
have $f(x)=x^2-U$, $x_\mu =\pm (U)^{1/2}$ and $f'(x)=2x$. Therefore,

\begin{equation}
\Phi({\bf P}_0,E_0,M) = \frac{1}{2}\omega_{D(M-1)}
U_0^{\frac{D(M-1)-2}{2}}\prod_\mu^M(2M_\mu)^{D/2}
\label{11}
\end{equation}

\section{Appendix: van der Waals fluid}
\label{ap-vdW}
The free energy $F$ of a system has as natural variables the number
of particles $N$, the temperature $T$ and the volume $V$, this
is $F=F(N,T,V)$. Its derivatives are given by the pressure, the
entropy and the chemical potential
\begin{eqnarray}
-P &=& \left. \frac{\partial F}{\partial V}\right|_{T,N}
\nonumber\\
-S &=& \left. \frac{\partial F}{\partial T}\right|_{V,N}
\nonumber\\
\mu  &=& \left. \frac{\partial F}{\partial N}\right|_{T,V}
\label{if}
\end{eqnarray}
Because the free energy is a first order function of the
extensive variables $N,V$, we have
\begin{equation}
F(N,T,V) = V f(n,T)
\label{fex}
\end{equation}
where $n$ is the number density and $f(n,T)$ is the free energy
density. The property (\ref{fex}) implies in Eqns. (\ref{if})
\begin{eqnarray}
-P(n,T) &=& f(n,T) -\mu n
\nonumber\\
s&=&-\frac{\partial f}{\partial T}(n,T)
\nonumber\\
\mu &=& \frac{\partial f}{\partial n}(n,T)
\label{ifd}
\end{eqnarray}
where $s=S/V$ is the entropy density.

For a van der Waals gas, the free energy density is given by
\begin{equation}
f(n,T) = -k_BTn\left(1+\ln\left(\frac{1-nb}{n\Lambda^D(T)}\right)\right)-an^2
\label{fnt}\end{equation}
where the thermal wavelength is defined by

\begin{equation}
\Lambda(T) = \frac{h}{(2\pi m_0 k_BT)^{1/2}}
\label{lamb}\end{equation}
Here, $h$ is Planck's constant, $m_0$ is the mass of a molecule, and
$k_B$ is Boltzmann constant. The constants $a,b$ are the attraction parameter
and the excluded volume, respectively.

With Eqns. (\ref{ifd}) plus the well-known relationship $f=\epsilon - Ts$
where $\epsilon$ is the internal energy density, we easily arrive at
the following relations

\begin{eqnarray}
\epsilon(n,s) &=& \frac{D}{2}k_B T(n,s)n - an^2
\nonumber\\
k_BT(n,s) &=& \left(\frac{h^2}{2\pi m_0}\right)
 \left(\frac{n}{1-nb}\right)^{2/D}\exp \left\{\frac{2s}{Dk_B n}
-\frac{D+2}{D}\right\}
\nonumber\\
P(n,s) &=& \frac{k_BT n}{1-nb}-an^2
\nonumber\\
\mu(n,s)&=&
k_BT\left(\frac{nb}{1-nb}+ \ln\left(\frac{nb}{1-nb}\right)\right)
\nonumber\\
&-&k_BT\ln\left( \frac{b}{\Lambda^D(T)}\right)-2an
\label{finden}
\end{eqnarray}
which give the fundamental equation $\epsilon(n,s)$ and the three
equations of state in terms of the variables $n,s$.

As it is customary for the van der Waals fluid, we introduce reduced variables
\begin{eqnarray}
\tilde{T}&=& \frac{27 b}{8a} k_B T,
\nonumber\\
\tilde{P}&=& \frac{27 b^2}{a}P.
\nonumber\\
\tilde{\mu}&=& \frac{27 b}{8a}\mu.
\nonumber\\
\tilde{n} &=& 3bn,
\nonumber\\
\tilde{s}&=& \frac{3bs}{k_B}
\nonumber\\
\tilde{\epsilon}&=& \frac{27 b}{8a}3b\epsilon
\label{redapp}
\end{eqnarray}

In reduced units we have that the fundamental equation becomes

\begin{equation}
\tilde{\epsilon}(\tilde{n},\tilde{s}) 
= \frac{D}{2}\tilde{T}(\tilde{n},\tilde{s}) \tilde{n}-\frac{9}{8}\tilde{n}^2
\label{ep}
\end{equation}
and the three equations of state are
\begin{eqnarray}
\tilde{T}(\tilde{n},\tilde{s}) &=& c \exp\frac{2}{D}
\left\{\frac{\tilde{s}}{\tilde{n}}-\frac{D+2}{2}+
\ln\left(\frac{\tilde{n}}{3-\tilde{n}}\right)\right\}
\nonumber\\
\tilde{P}(\tilde{n},\tilde{s})
&=& 8\frac{\tilde{T}\tilde{n}}{3-\tilde{n}}-3\tilde{n}^2
\nonumber\\
\tilde{\mu}(\tilde{n},\tilde{T})
&=&\tilde{T}\left(\ln\left(\frac{\tilde{n}}{3-\tilde{n}}\right)+
\frac{\tilde{n}}{3-\tilde{n}}\right)-\frac{9}{4}\tilde{n}
-\tilde{T}\ln\left(\frac{\tilde{T}}{c}\right)^{D/2}
\label{eqest}
\end{eqnarray}
In this equation, we have introduced the the constant $c$ which
 depends on microscopic parameters through

\begin{equation}
c=\frac{27 b^{(D-2)/D}h^2}{16\pi m_0 a}=4.836 \times 10^{-5}
\label{c}
\end{equation}
where we have used the values corresponding to water in $D=3$
dimensions, $a = 1.5262 \times 10^{-48} m^5 kg/s^2$, $b= 5.0622 \times
10^{-29} m^3$, $m_0=2.991\times 10^{-26} kg $, $h=1.054 \times
10^{-34} Js$ \cite{handbook}. This constant appears in the the
dimensionless quantity
\begin{equation}
\frac{b}{\Lambda^D(T)} = \left(\frac{\tilde{T}}{c}\right)^{D/2}
\label{l}
\end{equation}

Note that in reduced units we still have the relationships
\begin{eqnarray}
\frac{\partial\tilde{\epsilon}(\tilde{n},\tilde{s})}{\partial \tilde{n}}
&=& \tilde{\mu}(\tilde{n},\tilde{s})
\nonumber\\
\frac{\partial\tilde{\epsilon}(\tilde{n},\tilde{s})}{\partial \tilde{s}}
&=& \tilde{T}(\tilde{n},\tilde{s})
\label{intred}
\end{eqnarray}

Let us focus now on the distribution function (\ref{pden})
which in reduced units becomes

\begin{equation}
P(\tilde{n},\tilde{s})=
\frac{1}{Z}
\exp \{\tilde{V}\left\{
\tilde{s}-\tilde{\beta}^{\rm ext}(\tilde{\epsilon}(\tilde{n},\tilde{s})-
\tilde{\mu}^{\rm ext}\tilde{n}\right\}
\label{apptil}
\end{equation}
where $\tilde{V}=\overline{V}/3b$, $\tilde{\beta}^{\rm
ext}=8a\beta/27b$, is the reduced inverse external temperature and
$\tilde{\mu}^{\rm ext}= \overline{\lambda}m_0 27b/8a$ is the reduced
external chemical potential.

Due to the particular functional form of the fundamental equation for
the van der Waals gas, it is more convenient to study the distribution
function of $\tilde{n},\tilde{T}$ instead of $\tilde{n},\tilde{s}$.
The Jacobian of the transformation is $\frac{D\tilde{n}}{2\tilde{T}}$ and,
therefore, the new distribution function is
\begin{equation}
P(\tilde{n},\tilde{T})=\frac{\tilde{n}}{Z\tilde{T}}
\exp\tilde{V}\{\tilde{s}(\tilde{n},\tilde{s})-
\tilde{\beta}^{\rm ext}
(\tilde{\epsilon}(\tilde{n},\tilde{s})-\tilde{\mu}^{\rm ext}\tilde{n})\}
\label{appnt}
\end{equation}

\end{document}